\documentclass[prd,aps,10pt,superscriptaddress,twocolumn,amsmath,amssymb,nofootinbib,longbibliography, floatfix]{revtex4-1}

\usepackage{ragged2e}

\usepackage[english]{babel}
\usepackage[version=4]{mhchem}

\usepackage{natbib}
\usepackage[colorlinks]{hyperref}
\hypersetup{linkcolor=blue,citecolor=blue,filecolor=blue,urlcolor=blue}

\usepackage{amssymb,amsmath,multirow}
\usepackage{bm}
\usepackage{dsfont}
\usepackage{orcidlink}
\usepackage{color}
\usepackage{array}

\usepackage[utf8]{inputenc}
\usepackage{times}
\usepackage{setspace}
\usepackage{float}
\usepackage{caption}
\usepackage{booktabs}
\usepackage{multirow}
\usepackage{siunitx}

\usepackage{calligra}

\DeclareMathAlphabet{\mathcalligra}{T1}{calligra}{m}{n}
\DeclareFontShape{T1}{calligra}{m}{n}{<->s*[2.2]callig15}{}
\newcommand{\scriptr}{\mathcalligra{r}\,}

\usepackage[normalem]{ulem}
\usepackage{comment}

\usepackage{subfig,hyperref}

\usepackage{graphicx}
\usepackage{dcolumn}
\usepackage{bm}
\usepackage{wasysym}
\usepackage[
   starfontserif 
    ]{starfont}
\usepackage{amsmath}

\usepackage{mathtools}

\definecolor{darkgreen}{rgb}{0,0.5,0}
\definecolor{midgreen}{rgb}{0.0,0.675,0.0}

\DeclareSymbolFont{starfontsym}{OT1}{sts}{m}{n}
\DeclareMathSymbol{\mathTerra}{\mathord}{starfontsym}{76}

\begin{document}

\title{Gravitational Ionization by Schwarzschild Primordial Black Holes}

\author{Alexandra P. Klipfel \orcidlink{0000-0002-1907-7468}}
 \email{aklipfel@mit.edu}
\affiliation{%
Department of Physics, Massachusetts Institute of Technology, Cambridge, MA 02139, USA}
 
\author{David I.~Kaiser \orcidlink{0000-0002-5054-6744}}
 \email{dikaiser@mit.edu}

\affiliation{%
Department of Physics, Massachusetts Institute of Technology, Cambridge, MA 02139, USA}

\date{\today}

\begin{abstract}

Primordial black holes (PBHs) are theorized to form from the collapse of overdensities in the very early Universe. PBHs in the asteroid-mass range $10^{17} \, {\rm g}\lesssim M \lesssim 10^{23} \, {\rm g}$ could serve as all or most of the dark matter today, but are particularly difficult to detect due to their modest rates of Hawking emission and sub-micron Schwarzschild radii. We consider whether the steep gradients of a PBH's gravitational field could generate tidal forces strong enough to disrupt atoms and nuclei. Such phenomena may yield new observables that could uniquely distinguish a PBH from a macroscopic object of the same mass. We first consider the gravitational ionization of ambient neutral hydrogen and evaluate prospects for detecting photon radiation from the recombination of ionized atoms. During the present epoch, this effect would be swamped by Hawking radiation---which would itself be difficult to detect for PBHs at the upper end of the asteroid-mass window. We then consider the gravitational ionization and heating of neutral hydrogen immediately following recombination at $z\simeq1090$, and identify a broad class of PBH distributions with typical mass $5\times10^{21}\,{\rm g}\lesssim M \lesssim 10^{23}\, {\rm g}$ within which gravitational interactions would have been the dominant form of energy deposition to the medium. We also identify conditions under which tidal forces from a transiting PBH could overcome the strong nuclear force, either by dissociating deuterons, which would be relevant during big bang nucleosynthesis (BBN), or by inducing fission of heavy nuclei. We find that gravitational dissociation of deuterons dominates photodissociation rates due to Hawking radiation for PBHs with masses $10^{14}\,{\rm g}\lesssim M \lesssim 10^{16}\,{\rm g}$. We additionally identify the phenomenon of gravitationally induced fission of heavy nuclei via tidal deformation.
\end{abstract}

\maketitle

\section{Introduction}

Primordial black holes (PBHs) \cite{zeldovichHypothesisCoresRetarded1966,hawkingGravitationallyCollapsedObjects1971,carrBlackHolesEarly1974} are of significant interest as curious theoretical objects in their own right, and as candidates that could account for most or all of the dark matter abundance today \cite{Khlopov:2008qy,carrPrimordialBlackHoles2020,greenPrimordialBlackHoles2021,carrObservationalEvidencePrimordial2024,escrivaPrimordialBlackHoles2024}. Remarkably, in his first paper on PBHs, Hawking considered the implications of ionization by extremely light \emph{charged} black holes with $M\lesssim 10^3 \, {\rm g}$, and estimated detection rates for a population of such objects with bubble chambers \cite{hawkingGravitationallyCollapsedObjects1971}. However, physically realistic populations of PBHs,  which would be consistent with modern constraints and models, would likely be {\it uncharged} and are expected to form with exponentially larger masses, $M > 5.34 \times 10^{14} \, {\rm g}$, if they are to persist to today \cite{Klipfel:2025bvh}. For PBHs with such masses, transit rates through the Earth---let alone through a detector---would be vanishingly small. 

Present constraints on the PBH dark matter fraction from microlensing and from measured fluxes of high-energy cosmic rays leave an open window of PBH masses within which PBHs could constitute all of the dark matter: $10^{17} \, {\rm g} \lesssim  M \lesssim 10^{23} \, {\rm g}$, dubbed the ``asteroid-mass range'' \cite{carrConstraintsPrimordialBlack2016,Smyth:2019whb,carrConstraintsPrimordialBlack2021,gortonHowOpenAsteroidmass2024,DelaTorreLuque:2024qms}. The Schwarzschild radius of an asteroid-mass PBH with vanishing spin and charge, $r_s=2GM/c^2$, would be sub-micron scale, with $10^{-13} \,{\rm m} \lesssim r_s \lesssim 10^{-7} \, {\rm m}$. This range is centered around an angstrom: PBHs within the mass range of interest would be comparable to the size of an atom, making direct detection particularly challenging. 

Various gravitational signatures of individual asteroid-mass PBHs in the present epoch have been proposed, such as perturbations to the motions of well-tracked visible objects \cite{Dror:2019twh,Li:2022oqo,tranCloseEncountersPrimordial2024,cuadrat-grzybowskiProbingPrimordialBlack2024,Brown:2025awt} or signatures that could be measured in next-generation gravitational-wave detectors \cite{DeLorenci:2025wbn,Thoss:2025yht}. In each of these scenarios, the signal would depend on the perturber's mass and velocity, making it difficult to distinguish a PBH transit from the effects of more mundane macroscopic objects (such as actual asteroids) that happen to have similar mass and motion. 

In this paper we consider a novel gravitational signature arising from PBHs' interactions with matter, which we call ``gravitational ionization.'' We assume 
only Standard Model particles and interactions. The very small spatial size of PBHs within the asteroid-mass range suggests that such a PBH traversing a neutral gas could exert a strong tidal gravitational force on nearby atoms, thus overwhelming the electrostatic attraction between the atomic constituents and ionizing the atoms. Recombination of the ionized electrons with nearby ions would then emit photons, which might be detectable at Earth if produced by a nearby PBH transit today. Gravitational ionization followed by recombination might also yield measurable effects from a large population of PBHs in the early universe. Such electromagnetic radiation would be caused by {\it strictly gravitational interactions} by (uncharged) Schwarzschild PBHs. We consider gravitational ionization observables from Schwarzschild PBHs transiting through the Solar System today and at various epochs in the early universe. 

\begin{table*}
\caption{\label{tab:Mdefs}\justifying Definitions of various PBH mass scales. } 
\begin{ruledtabular}
\begin{tabular}{lll}
Symbol & Definition \\ 
\hline
$M_{\rm peak}^{ \{ \rm H, D, A \} }$ & Mass that maximizes ionization rate of hydrogen atoms (H), dissociation rate of deuterons (D), or fission of nuclei (A)\\
$M_{\rm max}^{ \{ {\rm H, D, A}\} }$ & Maximum mass that can ionize hydrogen atoms (H), dissociate deuterons (D), or induce fission of nuclei (A)\\
$M_{\rm trans}$ & Mass at the transition scale above which we can take $b_{\rm max}^{\rm H}=b_{\rm th}^{\rm H}$\\
$M_c$ & Mass below which a PBH in a medium at temperature $T_b$ is a net Hawking emitter\\
$M_{E_1}$ & Mass for which the primary Hawking photon emission spectrum peaks at the hydrogen ground-state energy $E_1$. \\ 
$M_i$ & Mass at time of PBH formation  \\
$M_*$ & Mass at formation time that would complete evaporation today \\
$\bar{M}$ & Mass at peak of PBH number distribution 
\end{tabular}
\end{ruledtabular}
\end{table*}

Unlike the gravitational signatures identified in Refs.~\cite{Dror:2019twh,Li:2022oqo,tranCloseEncountersPrimordial2024,cuadrat-grzybowskiProbingPrimordialBlack2024,Brown:2025awt,DeLorenci:2025wbn,Thoss:2025yht}, gravitational ionization depends on the perturber's mass, velocity, \emph{and} spatial size, and hence would be {\it unique} to small-mass PBHs---distinguishing a transiting PBH from a transiting asteroid.

The expected emission rate of photons from gravitational ionization depends on the number density of neutral atoms through which a PBH would pass. Given typical neutral-gas number densities throughout the Universe today, we find that the gravitational ionization photon emission rate would be swamped by the primary photon Hawking emission rate at comparable frequencies from the same passing PBH. On the other hand, for a population of PBHs, energy deposition to a medium by gravitational scattering could have exceeded total energy deposition from Hawking emission during at least one special epoch in cosmic history: immediately following recombination, at redshift $z_{\rm rec} \simeq 1090$, when the isotropic number density of neutral hydrogen was at its maximum. 

PBHs at and below the lower end of the asteroid-mass range could comparably disrupt nuclear matter. For example, gravitational tidal forces could dissociate bound deuterons. If the rates for such gravitationally induced dissociation were large enough during the very early Universe, they could have impacted big bang nucleosynthesis (BBN). Meanwhile heavy, unstable nuclei could also be affected by the gravitational tidal forces of a passing PBH, which could induce nuclear fission by deforming nuclei sufficiently such that their nuclear deformation energy surpasses the fission barrier.

In Section~\ref{sec:GravIon} we introduce the basic mechanism of gravitational ionization and compute expected event rates for various scenarios. Section~\ref{sec:HawkingPhotons} briefly reviews the formalism for Hawking emission of photons from black holes, and compares the gravitational ionization rate from a single PBH transit to the Hawking emission rate of photons with comparable energies from that same PBH. In Section~\ref{sec:GravIonDominates} we consider gravitational ionization from a population of PBHs around the epoch of recombination, at $z_{\rm rec} = 1090$. We find that across the asteroid-mass range, re-ionization of the medium would be dominated by Hawking emission rather than by gravitational ionization. On the other hand, we find wide regions of parameter space within which total energy deposited in the medium via gravitational scattering from the PBHs would exceed the energy deposition from Hawking radiation. Section~\ref{sec:GravNuclear} investigates gravitational effects on nuclear matter from a transiting PBH---ranging from dissociation of deuterons to deformation-induced fission of uranium nuclei. We identify a range of masses near but below the lower end of asteroid range mass in which gravitational tidal forces from transiting PBHs could be more effective at dissociating deuterons during BBN than photodissociation via Hawking emission. We also identify a regime of PBH masses within the asteroid-mass range in which fission via tidal deformation of heavy nuclei like \ce{^{235}_{92}U} is possible. Concluding remarks follow in Section~\ref{sec:Discussion}. 

We adopt ``natural'' units in which $c = \hbar = k_B = 1$. In these units, Newton's gravitational constant $G$ may be related to the (reduced) Planck mass $M_{\rm pl}$ as $G = 1 / (8 \pi M_{\rm pl}^2)$, with $M_{\rm pl}  =2.43 \times 10^{18} \, {\rm GeV} = 4.33 \times 10^{-6} \, {\rm g}$. Throughout the paper, we introduce several different PBH masses. For convenience, we collect and define these various masses in Table~\ref{tab:Mdefs}.

\section{Gravitational Ionization }
\label{sec:GravIon}

In this section, we consider radiation resulting from the recombination of gravitationally ionized neutral hydrogen following a PBH transit and evaluate the prospects for PBH detection with this new observable. We assume the PBH has no charge and no spin. PBHs form from the collapse of (scalar) curvature perturbations, so they typically begin with little or no spin. Similarly, although scenarios have been identified in which PBHs much smaller than the asteroid-mass range could have formed with large initial charge \cite{alonso-monsalvePrimordialBlackHoles2024}, such short-lived objects are expected to discharge very rapidly (if charged only under Standard Model gauge groups  \cite{Baker:2025zxm,Santiago:2025rzb}). Meanwhile, following their formation, black holes that happened to form with charge and/or spin will preferentially emit particles to reduce those quantities \cite{carterChargeParticleConservation1974,gibbonsVacuumPolarizationSpontaneous1975,pageParticleEmissionRates1976a}. Furthermore, PBHs within the asteroid-mass range will not spin up over time due to negligible accretion rates \cite{chibaSpinDistributionPrimordial2017,Jaraba:2021ces,delucaEvolutionPrimordialBlack2020,chongchitnanExtremeValueStatisticsSpin2021}. For the cosmological epochs we consider here, we may therefore restrict attention to Schwarzschild black holes.

PBHs that form in the asteroid-mass range have relatively low Hawking temperatures and thus lose very little mass via Hawking emission over cosmological timescales \cite{Klipfel:2025bvh}. In addition, their very small radii yield highly inefficient accretion over those same timescales \cite{Rice:2017avg,delucaEvolutionPrimordialBlack2020}. Hence we can assume $M (t) = M_i$ for all $M_i \gtrsim 10^{17}\, {\rm g}$ (where $M_i$ is the mass at formation time), implying that the gravitational effects of such PBHs are the same over cosmological history. What would change over cosmological time, however, is the composition of the Universe, and hence the various media that such PBHs would encounter.

As we will see, the rate of gravitational ionization depends on the relative velocity between PBHs and baryonic matter. Throughout different cosmological epochs, the relative velocity of dark matter to baryonic matter is known. Today, for example, the dark matter velocity distribution in the Milky Way galaxy is set by the virial velocity, and is typically parameterized by a truncated Maxwellian distribution with $\langle v \rangle \simeq 246 \, {\rm km /s}$ in the neighborhood of the Solar System \cite{Klipfel:2025bvh}. Typical relative velocities between dark matter and baryonic matter at earlier epochs, such as cosmic recombination, were about an order of magnitude smaller \cite{Tseliakhovich_2010}. If PBHs comprise a significant fraction of the dark matter, then those same estimates for relative velocities would apply to PBHs encountering baryonic matter. We use those velocity estimates when evaluating gravitational ionization rates.

\subsection{Energy Transfer from a PBH Transit}
\label{sec:EnTransfer}

Energy transfer to the components of an atom by gravitational scattering is distinct from the well-known case of Coulomb scattering, which is responsible for conventional ionization by charged particles. For the case of ionization of a medium by Coulomb scattering, one neglects the nuclei and only considers the energy transfer to electrons. In contrast, energy transfer to the (more massive) nuclei dominates for gravitational ionization. We briefly quantify the distinction in this subsection.

Given a central potential of the form
$U (r) = k/r$, if mass $m_1$ with velocity $v_{\rm rel}\ll 1$ is incident upon mass $m_2$, then the energy transfer in the rest frame of $m_2$ is (see, e.g., Chap.~13 of Ref.~\cite{Jackson})
\begin{equation}
\varepsilon =  \frac{2 v_{\rm rel}^2\mu^2}{m_2}\frac{1}{ 1 + \left( b / b_{\perp}\right)^2} , 
\label{EnergyTransfer1}
\end{equation}
where $\mu \equiv  m_1 m_2 / (m_1 + m_2)$ is the reduced mass and $b_{\perp} = k / ( \mu v_{\rm rel}^2)$ is the impact parameter for which the scattering angle in the center-of-mass frame is $90^\circ$. In the limit of small scattering angle ($b \gg b_{\perp})$, which is particularly applicable in a diffuse medium, Eq.~(\ref{EnergyTransfer1}) reduces to
\begin{equation}
    \varepsilon\simeq \frac{ 2 k^2}{m_2 v_{\rm rel}^2 b^2} .
    \label{EnergyTransfer2}
\end{equation}
In the case of Coulomb scattering, $k=z_1 z_2 \alpha_{\rm EM}$, where $\alpha_{\rm EM} = e^2 / (4 \pi \epsilon_0 ) \simeq 1/137$ is the fine structure constant, and the energy transfer is
\begin{equation}
\varepsilon_C = 2 \left( \frac{ z_1 z_2 \alpha_{\rm EM}}{v_{\rm rel} b} \right)^2 \frac{1}{m_2} .
\label{EnergyTransferCoulomb}
\end{equation}
For an identical scattering geometry, the ratio of energy transfer to a proton versus the transfer to an electron is given by
\begin{equation}
\frac{ \varepsilon_{C, p}}{\varepsilon_{C, e}} = \frac{ m_e}{m_p} .
\label{EnergyTransferCoulombEP}
\end{equation}
Hence calculations of ionization energy losses in media neglect Coulomb scattering off nuclei 
\cite{Landau:216256, osti_4311507}.

On the other hand, for the case of gravitational scattering we have $k = G m_1 m_2$. The energy transfer is
\begin{equation}
    \varepsilon_G = 2 \left( \frac{ G m_1}{v_{\rm rel} b} \right)^2 m_2,
    \label{EnergyTransferG}
\end{equation}
and the ratio of energy transfer to a proton versus the transfer to an electron becomes
\begin{equation}
\frac{ \varepsilon_{G, p}}{\varepsilon_{G, e}} = \frac{ m_p}{m_e}.
\label{EnergyTransferGravEP}
\end{equation}
Unlike electromagnetic scattering, the energy transfer to the proton now dominates---as expected, given the large hierarchy in mass, $m_p \gg m_e$, and the fact that gravitation couples to mass. Thus, energy losses by gravitational scattering for a Schwarzschild PBH incident upon a neutral medium are governed by its interactions with the nuclei of the material, while the electrons may be neglected. 

\begin{figure}[t]
    \centering
    \includegraphics[width=0.9\linewidth]{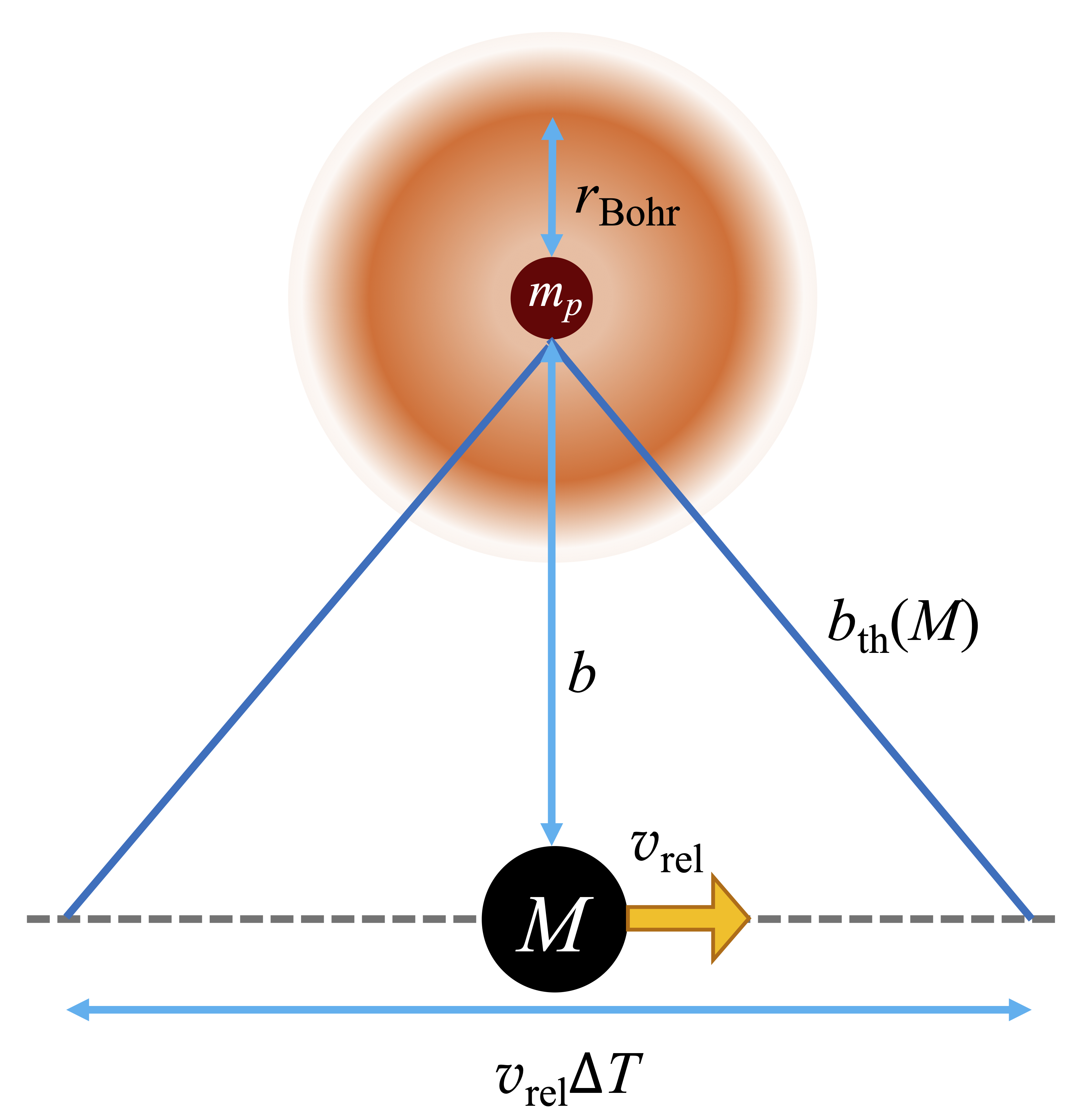}
    \caption{\justifying Schematic diagram of the gravitational ionization of a hydrogen atom  by a passing PBH of mass $M$ and relative velocity $v_{\rm rel}$ in the hydrogen atom rest frame. The hydrogen atom $1s$ ground state is represented by the orange cloud and the transiting PBH by the black circle. The Bohr radius and PBH radius would be drawn to relative scale ($r_s(M) = r_{\rm Bohr}/2$) if we took $M=1.78\times10^{19} \, {\rm g}$. The distances, however, are not shown to scale for such a system, which would have $b_{\rm th}(M) = 3.3\times10^{-9} \, {\rm m} \simeq 62 \, r_{\rm Bohr}$. The interaction timescale $\Delta T$, given by Eq.~(\ref{dT}), is set by the system parameters $b$, $M$, and $v_{\rm rel}$. }
    \label{fig:Schematic}
\end{figure}

\subsection{Ionization by Gravitational Tidal Forces}

Consider a hydrogen atom that consists of a proton of mass $m_p$ at point $\boldsymbol{r}_p$, a distance $b$ from a passing PBH, and an electron of mass $m_e$, with probability to be at point $\boldsymbol{r}_e$ given by its ground state wave function. See Fig.~\ref{fig:Schematic}. We restrict attention to physically motivated scenarios in which the PBH moves at nonrelativistic speeds, as expected for dark matter. Note that in this section we work in the PBH rest frame, with the PBH fixed at the origin to simplify the geometry, while Fig.~\ref{fig:Schematic} shows the hydrogen atom rest frame.

If the gradient of the gravitational field of a passing PBH of mass $M$ is large enough, we cannot consider the electron and proton to be in the same freely falling frame. Instead, they will experience different accelerations, set by their relative proximity to the PBH, $\Delta r \equiv|\boldsymbol{r}_e|-|\boldsymbol{r}_p|$.

As discussed in Section \ref{sec:EnTransfer}, we neglect energy transfer to the electron due to the mass hierarchy $m_p \gg m_e$. Thus we only consider energy transfer for the case of net acceleration on the proton, which occurs for $|\boldsymbol{r}_e|-|\boldsymbol{r}_p| > 0$. We impose this condition with a Heaviside theta function in Eq.~(\ref{AccelGradient}), which sets $|\Delta a| = 0$ for $|\boldsymbol{r}_e|-|\boldsymbol{r}_p| < 0$.

The net tidal acceleration of the proton relative to that of the electron will therefore be
\begin{equation}
\begin{split}
    \vert \Delta a \vert & = \frac{ 2 G M}{b^3} \Delta r \,\Theta(\Delta r)\\ & \simeq \frac{ 2 G M}{b^3} |\scriptr| \cos \theta \,\Theta(\pi/2 - \theta),
\end{split}
\label{AccelGradient}
\end{equation}
where $\boldsymbol{\scriptr} \equiv \boldsymbol{r}_e-\boldsymbol{r}_p$ and $\theta$ is the polar angle in a spherical coordinate system centered at $\boldsymbol{r}_p$. We take a spatial average over the spherical $1s$ ground state orbital to find
\begin{equation}
    \langle |\Delta a|\rangle = \frac{ 2 G M}{b^3} \langle \Delta r\rangle,
\end{equation}
where 
\begin{equation}
    \label{DeltaR1}
    \begin{split}
    \langle \Delta r\rangle & \simeq \langle |\scriptr| \cos \theta \, \Theta(\pi/2 - \theta)\rangle \\&
    = \langle |\scriptr| \rangle \,\langle \cos \theta \, \Theta(\pi/2 - \theta)\rangle.
    \end{split}
\end{equation}
For the radial integral of the average, the expectation value of the electron's distance from the proton is given by (see, e.g., Chap.~11 of Ref.~\cite{Zwiebach})
\begin{equation}
\langle |\scriptr| \rangle = \frac{1}{2} r_{\rm Bohr} \left(3 n^2 - \ell (\ell + 1 ) \right),
\label{rExpectation}
\end{equation}
for an electron in state $\{ n, \ell \}$, where $r_{\rm Bohr} \equiv 1 / (\alpha_{\rm EM \, }m_e) \simeq 0.53$ \AA \> is the Bohr radius. Hence we take $\langle |\scriptr| \rangle = \frac{3}{2}r_{\rm Bohr}$ for the ground state. The integral over $\theta$ then picks up a factor of $1/2\pi$, so Eq.~(\ref{DeltaR1}) becomes
\begin{equation}
\langle \Delta r \rangle \simeq \frac{\langle |\scriptr| \rangle}{2\pi} = \frac{3}{ 4 \pi} r_{\rm Bohr} .
\label{DeltaR}
\end{equation}

We assume that a necessary but not sufficient condition for gravitational ionization by a PBH of mass $M$ is $b\leq b_{\rm th}(M)$, where the \emph{threshold impact parameter} $b_{\rm th}(M)$ is set by the force balance 
\begin{equation}
\frac{ 2 G M m_p}{b_{\rm th}^3} \langle \Delta r\rangle = \frac{ \alpha_{\rm EM}}{\langle \Delta r\rangle^2 }.
\label{GravIonCondition}
\end{equation}
For the hydrogen atom geometry shown in Fig.~\ref{fig:Schematic}, we find 
\begin{equation}
b_{\rm th}^{\rm H} (M) = \left( \frac{ 2 G M m_p}{\alpha_{\rm EM}} \right)^{1/3} \langle \Delta r\rangle.
\label{bmaxH1}
\end{equation}

We can estimate the energy transfer to the proton in terms of the net tidal acceleration from Eq.~(\ref{AccelGradient}) and the timescale of the interaction, 
\begin{equation}
    \label{dT}
    \Delta T(b, M, v_{\rm rel}) = \frac{2}{v_{\rm rel}}\sqrt{b_{\rm th}^{\rm H}(M)^2-b^2}.
\end{equation}
Thus the energy transfer is given by
\begin{equation}
    \label{dE}
    \begin{split}
    \Delta E(b,& M, v_{\rm rel}) \\& \simeq m_p |\Delta a|^2 (\Delta T)^2 \\ &= \frac{4m_p}{v_{\rm rel}^2}\left(\frac{2 G M}{b^3} \langle \Delta r\rangle\right)^2 \left(b_{\rm th}^{\rm H}(M)^2-b^2 \right).
    \end{split}
\end{equation}

\begin{figure}[t!]
    \centering
    \includegraphics[width=1.0\linewidth]{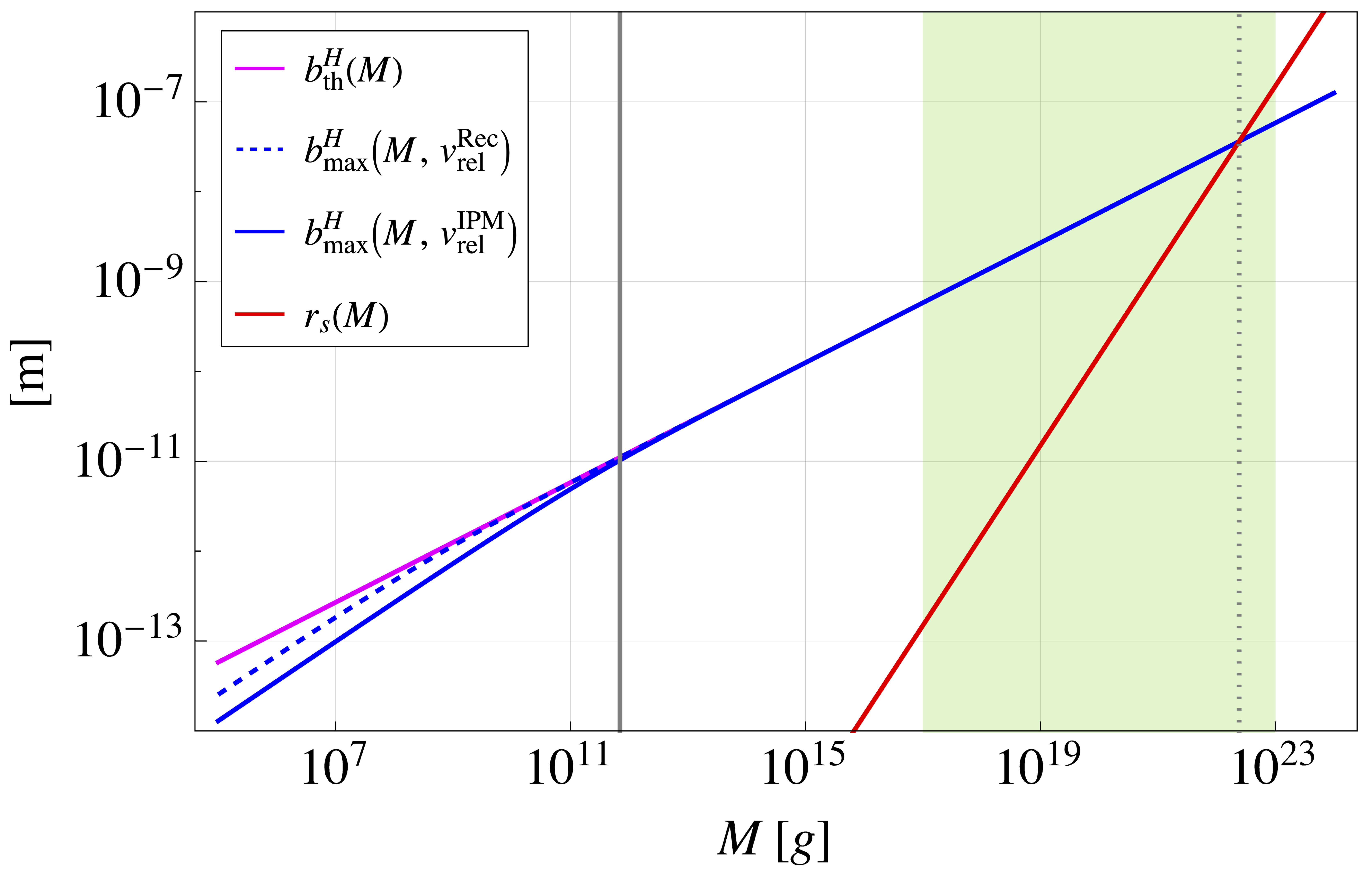}\caption{\justifying  A comparison of the maximum impact parameter $b_{\rm max}^{\rm H} (M, v_{\rm rel})$ from Eq.~(\ref{bMax}), the threshold impact parameter $b_{\rm th}^{\rm H} (M)$ from Eq.~(\ref{bmaxH1}), and the Schwarzschild radius $r_s (M)$. Note that $M_{\rm trans} = 6.93\times10^{11} \, {\rm g}$ (vertical, gray) marks the mass at which the scaling of $b_{\rm max}^{\rm H} $ with $M$ changes. The regime in which $b_{\rm max}^{\rm H}  \simeq b_{\rm th}^{\rm H}$ covers the whole asteroid-mass range (green, shaded). The maximum PBH mass which can gravitationally ionize hydrogen is $M_{\rm max} = 2.43 \times 10^{22} \, {\rm g}$ (vertical, dotted), as in Eq.~(\ref{Mmax}). The maximum impact parameter $b_{\rm max}^{\rm H}$ is shown for PBH transits through neutral hydrogen atoms at the epoch of recombination (Rec) and through the present-day interplanetary medium (IPM), with appropriate values for $v_{\rm rel}$ taken from Table~\ref{tab:medium}. }  
    \label{fig:bComp}
\end{figure}

We define the condition for gravitational ionization 
as: $\Delta E=E_1 \equiv 13.6 \, {\rm eV}$, where $E_1$ is the ground-state binding energy of a hydrogen atom. Using this condition, we define the \emph{maximum impact parameter} for which a transiting PBH can ionize a hydrogen atom as the value $b_{\rm max}^{\rm H}$ which satisfies
\begin{equation}
    \label{GIcriterion}
    E_1 = \Delta E(b_{\rm max}^{\rm H} , M, v_{\rm rel}).
\end{equation}
To solve Eq.~(\ref{GIcriterion}) exactly, we recast it in terms of two dimensionless parameters $B$ and $\xi$, which are related via
\begin{equation}
    \label{GIcriterionND}
    1 = \frac{B}{\xi^6}(1-\xi^2),
\end{equation}
where
\begin{equation}
    \label{xi}
    \xi \equiv \frac{b_{\rm max}^{\rm H}}{b_{\rm th}^{\rm H}(M)}
\end{equation}
is the ratio of the impact parameter $b$ to the threshold impact parameter, and
\begin{equation}
    \label{B}
    B \equiv \frac{4}{E_1 m_p v_{\rm rel}^2}\left(\frac{2 G M m_p\langle \Delta r\rangle}{b_{\rm th}^{\rm H}(M)^2} \right)^2.
\end{equation}
The quantity $B$ compares the atom's ground-state energy $E_1$ and its kinetic energy (in the PBH rest frame) $\frac{1}{2} m_p v_{\rm rel}^2$ to the (square of the) gravitational potential energy between the PBH and the proton $\Delta U \sim GM m_p \langle \Delta r \rangle /(b_{\rm th}^{\rm H})^2$. The exact solution to Eq.~(\ref{GIcriterionND}) is
\begin{equation}
\begin{split}
\xi_{\rm max} (B) &= \left( \frac{B}{2 \left( 1 + \sqrt{1 + 4B/27} \right)}\right)^{\frac{1}{6}} \\
& \quad\times  \left[\left( 1 + \sqrt{1 + \frac{4B}{27}}\right)^{\frac{2}{3}} - \left( \frac{4B}{27} \right)^{\frac{1}{3}}\right]^{\frac{1}{2}}
\end{split}
\label{ximaxB}
\end{equation} 
which is governed by the dimensionless parameter $B$, which separates the solution into two regimes with different scalings with $M$, depending on whether $B\ll1$ or $B\gg 1$. The transition scale where $B=1$ occurs for $M_{\rm trans}=6.93\times10^{11} \, {\rm g}$ if $v_{\rm rel}=246 \, {\rm km/s}$. See Fig.~\ref{fig:bComp}. 

We can now define 
\begin{equation}
    \label{bMax}
    b_{\rm max}^{\rm H} (M, v_{\rm rel}) = b_{\rm th}^{\rm H}(M) \,\xi_{\rm max}(B(M, v_{\rm rel})).
\end{equation}
The asymptotic scalings of $b_{\rm max}^{\rm H} $ with $M$ can be easily found by using $\xi_{\rm max}\sim 1$ for $B\gg 1$ and $\xi_{\rm max}\sim B^{1/6}$ for $B\ll 1$ from Eq.~(\ref{ximaxB}):
\begin{equation}
    \label{bMaxScaling}
    b_{\rm max}^{\rm H}  \rightarrow  \begin{cases} 
      b_{\rm th}^{\rm H} \sim M^{1/3}, & M \gg 10^{11} \, {\rm g}\\
      b_{\rm th}^{\rm H}\,B^{1/6} \sim M^{4/9}v_{\rm rel}^{-1/3}, &  M \ll 10^{11} \, {\rm g}\\
   \end{cases}
\end{equation}

Even if $\Delta E > E_1$, in order for the PBH to actually ionize the atom, the proton's position should change faster than the electron's wavefunction would be able to adjust in response; otherwise the electron could remain bound to the proton and the impulse from the PBH transit would simply result in center-of-mass motion. We thus want to ensure that the PBH transit results in a \emph{non-adiabatic perturbation} to the hydrogen atom potential, which can be quantified by comparing the interaction timescale for an energy transfer $\Delta E = E_1$,
\begin{equation}
    \begin{split}
    \tau &\equiv \Delta T(b_{\rm max}^{\rm H} ) \\ &= \frac{2}{v_{\rm rel}}b_{\rm th}^{\rm H}(M)\sqrt{1 - \xi_{\rm max}(M, v_{\rm rel})^2} ,
    \end{split}
    \label{tau}
\end{equation}
to the typical timescale of the electron's motion. In its ground state, the electron's timescale is set by $E_1^{-1} \sim 3.2 \times 10^{-16} \, {\rm s}$. We expect gravitational ionization to be effective only if $\tau < E_1^{-1}$. 

\begin{figure}[t!]
    \centering
    \includegraphics[width=1.0\linewidth]{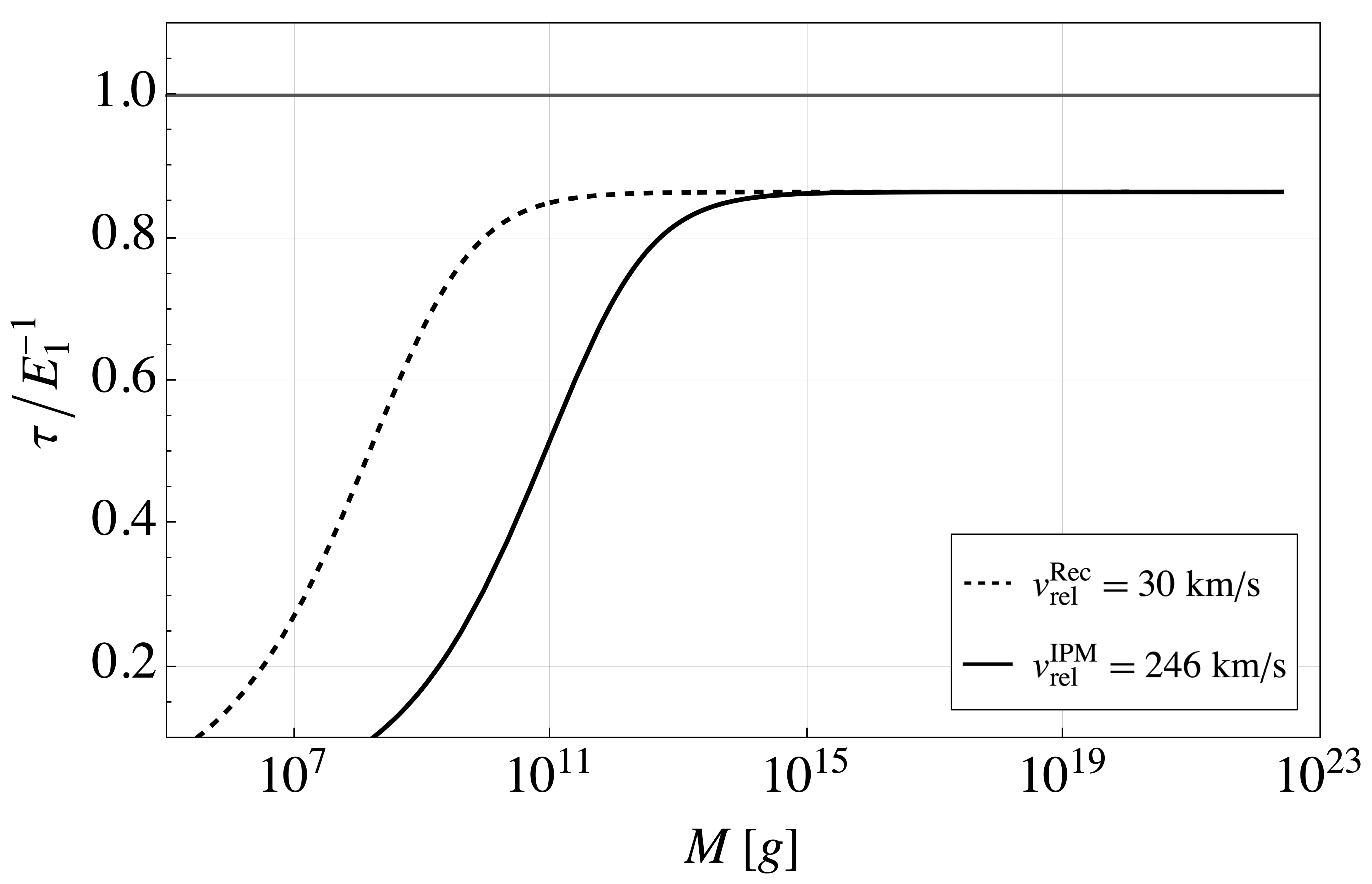}\caption{\justifying Comparison of the electron timescale $E_1^{-1}$ to the energy-transfer timescale $\tau$ for gravitational ionization of a hydrogen atom by a PBH of mass $M$. Here $E_1 = 13.6 \, {\rm eV}$ is the hydrogen ground-state energy.    We find that $\tau/E_1^{-1}<1$ for all $M \leq M_{\rm max}^{\rm H}$, which implies that energy transfer can always be treated as instantaneous and that the encounter effectively results in a non-adiabatic perturbation to the hydrogen atom potential.}
    \label{fig:TimeScales}
\end{figure}

Fig.~\ref{fig:TimeScales} shows $\tau/E_1^{-1}$ as a function of PBH mass $M$ for different values of $v_{\rm rel}$. We find that in the asteroid-mass range $(B \gg 1, M \gg 10^{11} \, {\rm g})$, 
\begin{equation}
    \label{tauRatio}
    \frac{\tau}{E_1^{-1}} \longrightarrow 0.87 \,\,<1 \,\quad {\rm for} \>\> M \gg 10^{11} \, {\rm g}.
\end{equation}
From Eq.~(\ref{tauRatio}) we conclude that, over the entire asteroid-mass range, the energy transfer from the transiting PBH to the proton is effectively instantaneous. In that limit, Eq.~(\ref{bMaxScaling}) indicates that we may take $b_{\rm max}^{\rm H} (M,v_{\rm rel}) \rightarrow b_{\rm max}^{\rm H} (M)= b_{\rm th}^{\rm H}(M)$, which is independent of the PBH relative velocity, consistent with Fig.~\ref{fig:bComp}.

Across the asteroid-mass range, we therefore find the scaling between the maximum impact parameter and the PBH Schwarzschild radius to be
\begin{equation}
\frac{ b_{\rm max}^{\rm H} (M)}{r_s (M)} \propto M^{-2/3} .
\label{bmaxrs}
\end{equation}
The fact that $b_{\rm max}^{\rm H}  / r_s$ falls with $M$ makes sense: we do not expect large astrophysical black holes to spontaneously ionize their surroundings simply due to gravitational gradients. Moreover, the scaling in Eq.~(\ref{bmaxrs}) implies that there should be some largest mass $M_{\rm max}$ capable of gravitationally ionizing a gas of neutral hydrogen atoms, determined by the condition $b_{\rm max}^{\rm H} (M_{\rm max}) = r_s (M_{\rm max})$. From Eq.~(\ref{bmaxH1}), this yields
\begin{equation}
    \label{Mmax}
    M_{\rm max}^{\rm H} = \left(\frac{m_p \langle \Delta r\rangle^3}{4 \alpha_{\rm EM}G^2} \right)^{1/2} = 2.43\times10^{22} \, {\rm g} ,
\end{equation}
which lies 
the top of the asteroid-mass range. Remarkably, we find that gravitational ionization would only occur for PBHs with masses that are already of prime interest as dark matter candidates. 

\begin{table}
    \begin{tabular}{c | c | c | c }
    Medium & $n_{\rm H} \> [ {\rm m}^{-3} ]$ &  $v_{\rm rel} \> [ {\rm km /s} ]$ & $\Gamma_{\rm max} \> [ {\rm s}^{-1} ]$ \\
    \hline
    IPM & $10^5$ & $246$ & $3.9\times10^{-5}$\\
    ISM & $2 \times 10^5$ & $246$ & $7.8\times10^{-5}$\\
    Rec. & $2.3 \times 10^8$ & $30$ & $1.1\times10^{-2}$\\
    \hline       
    \end{tabular}
    \vspace{5 pt}
    \caption{\justifying Number density $n_{\rm H}$ of neutral hydrogen atoms in the interplanetary medium (IPM) \cite{holzerNeutralHydrogenInterplanetary1977,swaczynaInterstellarNeutralHydrogen2024}, interstellar medium (ISM) \cite{hayakawaRadiationInterstellarHydrogen1961,swaczynaDensityNeutralHydrogen2020}, and immediately following cosmological recombination (Rec.) \cite{Lynch:2024gmp,planckcollaborationPlanck2018Results2020}, together with the relative velocity $v_{\rm rel}$ between dark matter particles (here assumed to be PBHs) and hydrogen gas \cite{cerdenoParticleDarkMatter2010,choiImpactDarkMatter2014,Tseliakhovich_2010}, and the maximum achievable gravitational ionization rate $\Gamma_{\rm max}$, which occurs for $M_{\rm peak} = 1.07 \times 10^{22} \, {\rm g}$.
    } 
    \label{tab:medium}
\end{table}

We may generalize Eq.~(\ref{bmaxH1}) to an arbitrary neutral atom with atomic number $Z$, atomic mass $A$, and typical electron distance $\langle\Delta r\rangle_{A,Z}$:
\begin{equation}
    \label{bmaxAZ1}
    \begin{split}
        b_{\rm max} &(M \vert Z, A) \\ &= \left(\frac{2 G M}{Z \alpha_{\rm EM}} (Z m_p + (A-Z)m_n) \right)^{1/3} \langle\Delta r\rangle_{A,Z}.
    \end{split}
\end{equation}
Taking $m_p \approx m_n$, this yields
\begin{equation}
    \label{bmaxAZ2}
        b_{\rm max} (M \vert Z, A) = \left[ \left( \frac{ A}{Z} \right)^{1/3} \frac{ \langle\Delta r\rangle_{A,Z}}{ \langle\Delta r\rangle } \right] b_{\rm max}^{\rm H} (M).
\end{equation}
For all atoms with $A > 1$ and $A/Z > 1$, regardless of the typical scale for the atomic radius one uses (Van der Waals, covalent, and so on), $\langle |\scriptr| \rangle_{A,Z}> \frac{3}{2} r_{\rm Bohr} \implies \langle\Delta r\rangle_{A,Z} > \langle\Delta r\rangle$ for valence electrons \cite{SlaterRadii}, so we find
\begin{equation}
\frac{ b_{\rm max} (M \vert Z, A)}{b_{\rm max}^{\rm H} (M)} > 1 \>\> \forall A > 1 ,
\label{bmax3}
\end{equation}
which implies that gravitational ionization of a medium by a PBH would be \emph{more effective} for heavier atoms than for hydrogen atoms. Thus, for all $A > 1$, the corresponding value of $M_{\rm max}$ would be larger and the ionization rates discussed in the following section would higher.

\begin{figure}[t!]
    \centering
    \includegraphics[width=1.0\linewidth]{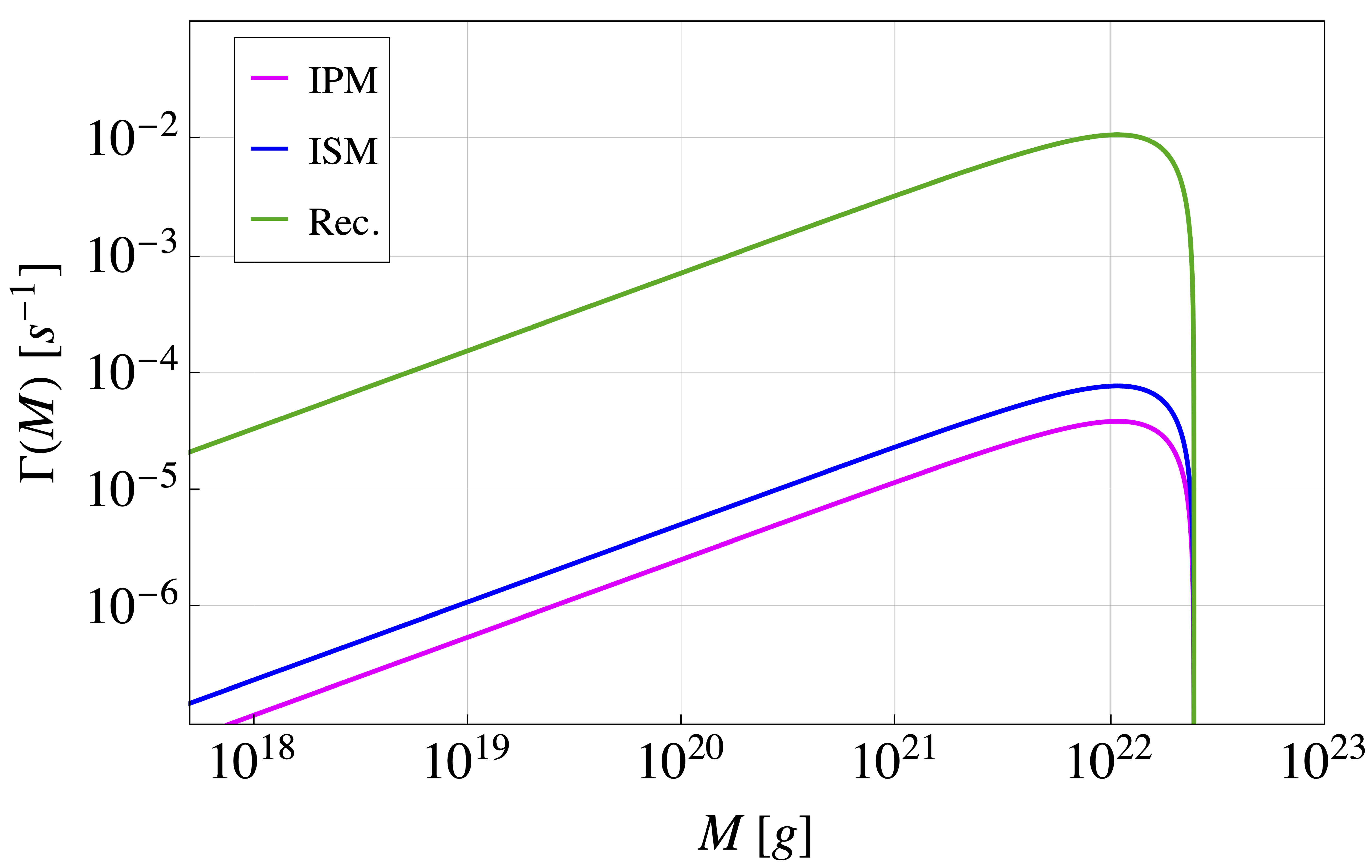}
    \caption{\justifying Gravitational ionization rates $\Gamma(M)$ for a PBH of mass $M$ transiting through the three different neutral hydrogen media listed in Table \ref{tab:medium}. Note that $\Gamma$ is maximized at $M_{\rm peak}^{\rm H} =1.07\times10^{22} \, {\rm g}$, as given in Eq.~(\ref{Msigma}), and cuts off sharply at $M_{\rm max}^{\rm H} =2.43\times10^{22} \, {\rm g}$, as in Eq.~(\ref{Mmax}). 
    }
    \label{fig:IonRates}
\end{figure}

\subsection{Gravitational Ionization Rates}

Given that we can take $b_{\rm max}^{\rm H} (M) = b_{\rm th}^{\rm H}(M)$ as the maximum ionization impact parameter for asteroid-mass PBHs, we may estimate the ionization rate for a Schwarzschild PBH with relative velocity $v_{\rm rel}$ traveling through neutral hydrogen atoms with number density $n_{\rm H}$.

The ionization rate is given by
\begin{equation}
    \Gamma = n_{\rm H} \,\sigma\, v_{\rm rel},
    \label{GammaIonize1}
\end{equation}
where $v_{\rm rel}$ and $n_{\rm H}$ can vary depending on the physical system and cosmological epoch being considered. We take the cross section $\sigma$ to be an annulus with outer radius $b_{\rm max}^{\rm H}(M)$ and inner radius $r_s$,
\begin{equation}
    \sigma (M) = \pi \left[ b^{\rm H}_{\rm max} (M)^2 - r_s^2 (M) \right] ,
    \label{sigmaM}
\end{equation}
which yields an ionization rate for atomic hydrogen of
\begin{equation}
\Gamma (M) = \pi n_{\rm H} v_{\rm rel} \left[ b_{\rm max}^{\rm H} (M )^2 - r_s^2 (M) \right] .
\label{GammaIonize2}
\end{equation}

For a cloud of neutral hydrogen atoms, the gravitational ionization cross section (and hence also the ionization rate $\Gamma$) is maximized for a PBH mass $M_{\rm peak}$, given by
\begin{equation}
    \label{Msigma}
    M_{\rm peak}^{\rm H} = \frac{1}{3^{3/4}}\left(\frac{m_p \langle \Delta r\rangle^3}{4 \alpha_{\rm EM}G^2} \right)^{1/2} = 1.07\times10^{22} \, {\rm g} .
\end{equation}
We may use Eq.~(\ref{GammaIonize2}) to compute the gravitational ionization rate as a function of PBH mass for different cosmological epochs and different astrophysical media. We consider three scenarios: the interplanetary medium (IPM), the interstellar medium (ISM), and the early universe immediately after recombination (Rec). See Table~\ref{tab:medium}. (To compute $n_{\rm H}$ at cosmological recombination, we set $n_{\rm H} (z) = 0.76 (\Omega_b  / m_p) \rho_{c, 0} (1 + z)^3$, where $0.76$ is the fraction of baryons in hydrogen nuclei following big bang nucleosynthesis, $\rho_{c, 0}$ is the present value of the critical density, and we take $z_{\rm rec} \simeq 1090$ \cite{Lynch:2024gmp,planckcollaborationPlanck2018Results2020}.) Fig.~\ref{fig:IonRates} shows ionization rates for each scenario as a function of PBH mass. 

The rate $\Gamma$ in Eq.~(\ref{GammaIonize2}) concerns ionization events, but we are ultimately interested the rate of photon emission from subsequent \emph{recombination} events. In future sections we take the gravitational ionization photon emission rate to be $\Gamma(M)$, which implicitly assumes that each gravitational ionization event yields (on average) one recombination event. This is a two-fold assumption that (1) each ionized electron does not go on to ionize any other electrons and (2) only one photon is emitted on average during recombination. Making the first assumption results in underestimating the photon emission rate, though we expect the difference to be minimal in diffuse media. The second assumption is well-motivated. For example, in the case of ionization of the diffuse interplanetary medium with ambient temperature $T \sim {\cal O} (10^2 \, {\rm K})$, the dominant radiation from recombination consists of single-photon emission of Lyman-$\alpha$ photons (from the $2p \rightarrow 1s$ transition), with $\lambda_\alpha = 1215$ \AA ,  corresponding to $E_\alpha = 10.2 \, {\rm eV}$. Across the energy range of interest, the cross sections for the next-leading emission processes follow roughly as $\sigma_{ H\alpha} / \sigma_{ {\rm Ly} \alpha}  \simeq 0.7$, $\sigma_{P\alpha}  / \sigma_{ {\rm Ly} \alpha} \simeq 0.4$, and $\sigma_{2h\nu} / \sigma_{ {\rm Ly}\alpha}  \simeq 0.09$, where $H\alpha$ refers to single-photon emission with $\lambda_{H \alpha} = 6563$ \AA \>\,($E_{H\alpha} = 1.9 \, {\rm eV}$); $P\alpha$ refers to single-photon emission with $\lambda_{P\alpha} = 18651$ \AA \>\,($E_{P\alpha} = 0.7 \, {\rm eV}$); and $2h\nu$ refers to two-photon emission from the transition $2s \rightarrow 1s$, with each photon at $\lambda_\nu \gtrsim 1215$ \AA \>\,($E_\nu \lesssim 10.2 \, {\rm eV}$) \cite{hayakawaRadiationInterstellarHydrogen1961}. Henceforth, in future sections we take the gravitational ionization photon emission rate to be $\Gamma(M)$ in Eq.~(\ref{GammaIonize2}).

\begin{table*}
    \begin{tabular}{| c | c | c | c | c |}
    \hline  
    Epoch & $T_b$ [eV] &  $M_c$ [g] &  $r_s(M_c)$ [m] & $T_H(M_c)$ [eV] \\
    \hline
    BBN & $7.0\times10^4$ & $4.21 \times 10^{18}$ & $6.27 \times 10^{-12}$  &  $2.50 \times 10^3$ \\
    Rec. & $0.256$ & $1.15 \times 10^{24}$  & $1.72 \times 10^{-6}$ & $9.14 \times 10^{-3}$ \\
    $t_{0, \, {\rm ISM}}$ & $8.62 \times 10^{-3}$ & $3.42 \times 10^{25}$ & $5.09 \times 10^{-5}$ &  $3.08\times 10^{-4}$ \\
    $t_{0,\, {\rm CMB}}$ & $2.35\times10^{-4}$ & $1.26\times10^{27}$ & $1.87\times10^{-3}$ & $8.39\times10^{-6}$\\
    \hline       
    \end{tabular}
    \vspace{5 pt}
    \caption{\justifying Criteria for a black hole to be a net emitter in a given medium with temperature $T_b$. A black hole with mass $M \leq M_c$, as defined in Eq.~(\ref{EmitterCondition2}), will have a Schwarzschild radius $r_s$ smaller than the typical wavelength of surrounding radiation and will be a net Hawking emitter according to Eq.~(\ref{Hawk1}) with temperature $T_{H}$. We assume that black holes with masses $M\gtrsim M_c$ behave as net absorbers of radiation from their surrounding medium and do not significantly emit Hawking radiation until the time when $T_b$ cools sufficiently due to Hubble expansion. We consider the time of deuteron formation during big bang nucleosynthesis (BBN) ($t_{\rm D} \simeq 120 \, {\rm s}$) \cite{Pospelov:2010hj}, the time of recombination 
    ($t_{\rm rec} \simeq 380,000 \, {\rm yr}$) \cite{Planck:2018vyg},
    and two media at the present time ($t_0 \simeq 13.8 \, {\rm Gyr}$): the ISM, with dominant cold neutral atomic hydrogen fraction at $T_{\rm ISM} = 100 \, {\rm K}$ \cite{BasuISM,PatraISM}, and the CMB at $T_{\rm CMB} = 2.726 \, {\rm K}$ \cite{FixsenCMBT0}.
    } 
    \label{tab:EmissionMasses}
\end{table*}

Photon emission from recombination is a separate event from the gravitational ionization of neutral atoms. In our case, the properties of a hydrogen cloud, including its density, temperature, and ionization fraction, determine the time delay between when a PBH passes through the medium and when an electron from an ionized hydrogen atom recombines with an ${\rm H}^+$ ion \cite{hayakawaRadiationInterstellarHydrogen1961}.

If gravitational ionization rates in the IPM were sufficiently large so as to be detectable by ground- or space-based telescopes, the resulting signature from a local PBH transit in the IPM would be distinct. The signature would appear as a point source of photons sharply peaked at $E_{\alpha}$ moving on a linear trajectory across the sky with a signal-duration timescale on the order of days or weeks (given expected PBH speeds $v \sim {\cal O} (250 \, {\rm km/s})$ within the Solar System \cite{tranCloseEncountersPrimordial2024,Klipfel:2025bvh}). Such a signature, in combination with gravitational perturbations to the motions of Solar System bodies \cite{tranCloseEncountersPrimordial2024,cuadrat-grzybowskiProbingPrimordialBlack2024}, would uniquely distinguish a PBH transit from an asteroid of similar mass.  However, as we will show in later sections, the gravitational ionization rates in the present-day IPM are swamped by Hawking emission rates, which themselves would likely remain undetectable for asteroid-mass PBH transits at reasonable impact parameters from Earth, as discussed in detail in Ref.~\cite{KlipfelEMsignatures}.

\section{Hawking Emission of Photons}
\label{sec:HawkingPhotons}
The gravitational ionization rates are quite low, especially during the late universe. Another possible electromagnetic signature of an asteroid-mass PBH transit is the emission of photons via Hawking radiation. As mentioned above, there would be a time delay between emission of Hawking radiation photons and those produced via recombination of gravitationally ionized electrons and ${\rm H}^+$ ions. In this section, we introduce the Hawking radiation formalism and investigate whether there are any PBH masses for which the gravitational ionization rate surpasses the Hawking emission rate. In Sec.~\ref{sec:GravIonDominates} we determine whether there are any realistic PBH number distributions for which, at a certain cosmological epoch, the dominant energy-deposition mechanism from the PBHs to the ambient medium is due to gravitational interactions rather than Hawking radiation. As in the previous section, we restrict attention to Schwarzschild black holes.

\subsection{Criterion for a PBH to be a Net Emitter}
\label{sec:EmissionCriteria}

When two blackbodies of different temperatures are brought into proximity there will be a heat flow between them, governed by the Stefan-Boltzmann law; the colder object will be a net absorber, rather than emitter, until local equilibrium is established. We are interested in PBHs with high Hawking temperatures that are immersed within hot media with thermal spectra, so we assume a comparable criterion should apply for when the PBHs will be net emitters of Hawking radiation. 

A black hole with a high enough temperature can emit all Standard Model particles (as well as any additional particles, beyond the Standard Model, which might exist \cite{Baker:2021btk,Baker:2022rkn,Baker:2025ffi}). The Hawking temperature of a Schwarzschild black hole of mass $M$ is given by
\begin{equation}
    T_H (M) \equiv \frac{1}{8 \pi G M }. 
    \label{TH}
\end{equation}
We assume that a black hole in a thermal bath of photons at temperature $T_b$ will be a net emitter if its Schwarzschild radius $r_s$ is smaller than the wavelength of the typical surrounding radiation. For a thermal bath of photons at temperature $T_b$, the Planck distribution peaks at $E_{\rm peak} (T_b) = 2.82 \, T_b$, so a PBH will be a net emitter if 
\begin{equation}
    \label{EmitterCondition1}
    r_s(M) < \lambda_{\rm peak}(T_b) =\frac{2 \pi}{2.82 \, T_b }.
\end{equation}
(For similar considerations, see, e.g., Refs.~\cite{Rice:2017avg,Barrau:2022bfg,Loeb:2024gga}; cf. Ref.~\cite{Chatterjee:2025wnt}.)
Eq.~(\ref{EmitterCondition1}) amounts to a condition on the PBH mass: 
\begin{equation}
    \label{EmitterCondition2}
    M<M_c(T_b) \equiv \frac{\pi}{2.82}\frac{1}{G \, T_b}, 
\end{equation}
where $M_c(T_b)$ is the critical mass for net emission.
Using Eq.~(\ref{TH}), we can define an equivalent condition on the black hole temperature for net emission:
\begin{equation}
    \label{EmitterCondition3}
    T_{H} > T_H(M_c) = \frac{2.82}{8 \pi^2}T_b.
\end{equation}
See Table \ref{tab:EmissionMasses} for a list of several cosmological epochs and their corresponding background temperatures and critical black hole masses, temperatures, and radii. As we will see, in each scenario of interest here, we find $M_c (T_b)$ greater than the range of PBH masses of interest, and so in practice the criterion of Eq.~(\ref{EmitterCondition1}) is satisfied. We leave additional consideration of this general criterion to further research.

\subsection{Hawking Emission Formalism}
\label{sec:HawkingFormalism}

The primary Hawking emission spectrum for a Standard Model particle species $j$ may be parameterized as \cite{pageParticleEmissionRates1976, macgibbonQuarkGluonjetEmission1990}:
\begin{equation}
    \frac{ d^2 N_j^{(1)}}{dt dE} = g_j \frac{ \Gamma_{s_j}(M,E)}{2 \pi} \left[ \exp \left( \frac{ E}{T_H(M)}\right) - (-1)^{2 s_j} \right]^{-1},
    \label{Hawk1}
\end{equation}
where $g_j$ is the number of degrees of freedom associated with that species, $s_j$ is the particle's spin, $\Gamma_{s_j}$ is the greybody factor \cite{pageParticleEmissionRates1976, macgibbonQuarkGluonjetEmission1990, teukolskyPerturbationsRotatingBlack1973, teukolskyPerturbationsRotatingBlack1974}, 
$E$ is the energy of the emitted particle, and $M$ is the black hole mass.

We use \texttt{BlackHawk v2.2} to calculate the full energy spectra of all emitted particles for a black hole of mass $M$ \cite{arbeyBlackHawkV20Public2019, arbeyPhysicsStandardModel2021a}. For $M \gtrsim 5 \times 10^{17} \, {\rm g}$---which includes most of the open asteroid-mass range---the PBHs are too cold to emit electrons and positrons or any other more-massive particles \cite{Klipfel:2025bvh}, so the output would consist only of photons and neutrinos. 

\begin{figure}[t]
    \centering 
    \includegraphics[width=1.0\linewidth]{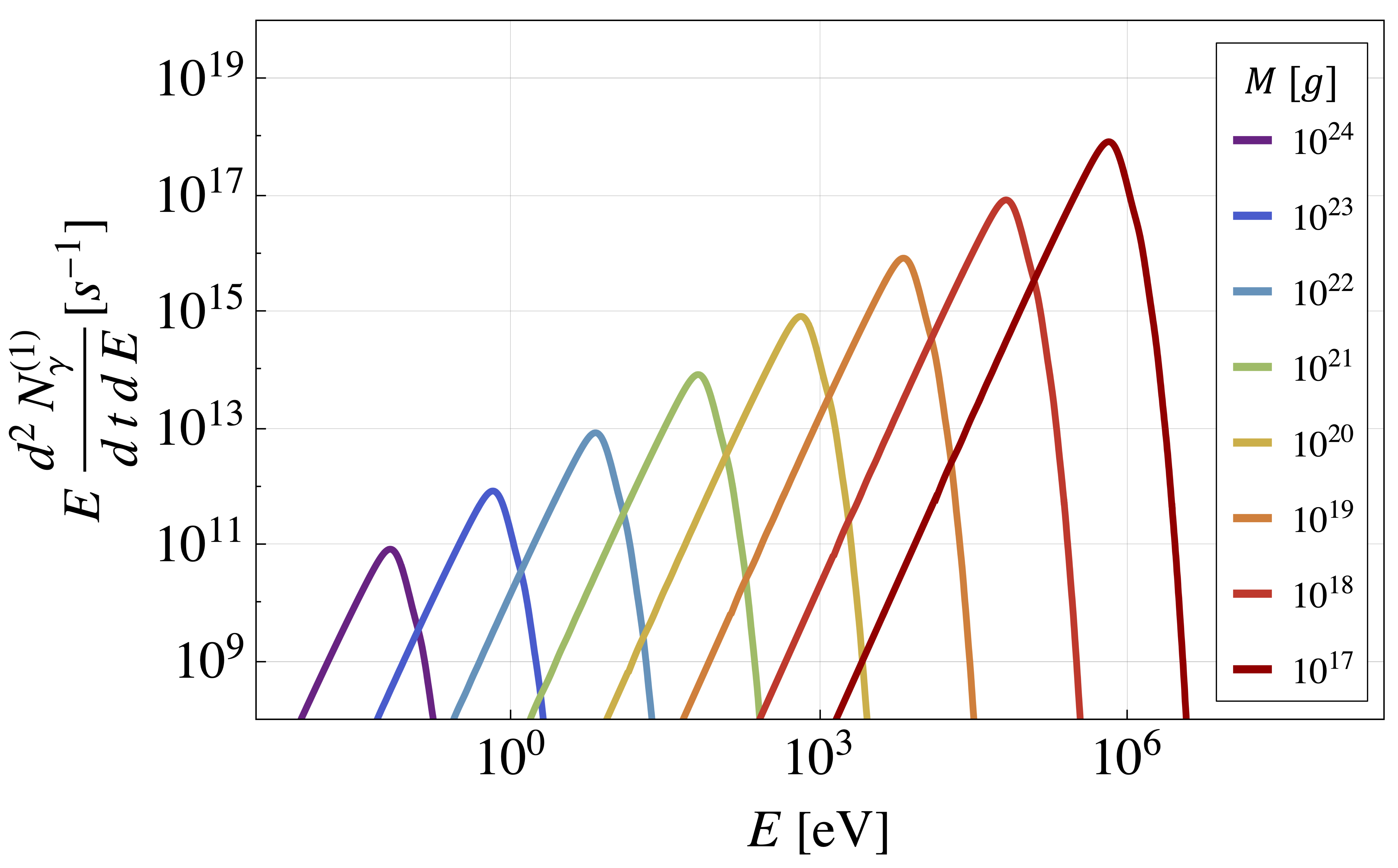}
    \caption{\justifying Primary Hawking emission of photons from a Schwarzschild black hole of mass $M$ within the range $10^{17} \, {\rm g} \leq M \leq 10^{24} \, {\rm g}$.  The spectra are sharply peaked, thus $E \frac{d^2N_{\gamma}^{(1)}}{dtdE}\big|_{E_{\rm peak}} \approx \frac{dN_\gamma^{(1)}}{dt}$, the total primary-emission rate for photons. Plot prepared using \texttt{BlackHawk v2.2} \cite{arbeyBlackHawkV20Public2019, arbeyPhysicsStandardModel2021a}.}
    \label{fig:HawkPrimary}
\end{figure}

For the case of interest---namely, emission of photons---we have $g_j = 2$ for the number of distinct polarization states, and $s_j = 1$ for the photon's spin. The greybody factor for photons may be parameterized as \cite{macgibbonQuarkGluonjetEmission1990}
\begin{equation}
    \Gamma_\gamma = \frac{ \sigma_s (E, M)}{ \pi} E^2 .
    \label{greybody1}
\end{equation}
The cross sections $\sigma_s (E, M)$ must be evaluated numerically for arbitrary $E$; their analytic forms are only known for $E \rightarrow 0$ and $E \rightarrow \infty$. In general the values of $\sigma_s (E, M)$ arise from solving for the transmission coefficients of modes scattering in the Regge-Wheeler effective potential of a Schwarzschild black hole. (See Ref.~\cite{grayGreybodyFactorsSchwarzschild2018} for a helpful discussion.) Primary photon emission spectra for asteroid-mass PBHs are shown in Fig.~\ref{fig:HawkPrimary}.

In Fig.~\ref{fig:HawkPrimary} we can see that the primary photon emission spectra are sharply peaked. The photon energy at the peak of the primary emission spectrum for species $j$ is given by \cite{macgibbonQuarkGluonjetEmission1991}
\begin{equation}
E_{\rm peak}^{\rm Hawk}(M) = \beta_{s_j} \, T_H(M) ,
\label{EpeakHawk}
\end{equation}
where $T_H$ is the Hawking temperature as given in Eq.~(\ref{TH}) and $\beta_{1}=6.04$ \cite{macgibbonQuarkGluonjetEmission1991}. 

The integrated photon emission rate obeys
\begin{equation}
\begin{split}
\frac{ dN_{\gamma}^{(1)}}{dt}  & \equiv \int_0^\infty dE  \frac{ d^2 N_{\gamma}^{(1)}}{dt dE} \\ & = 5.97 \times 10^{17} \, {\rm s}^{-1} \left( \frac{10^{17} \, {\rm g}}{ M } \right) ,
\end{split}
\label{dNdtHawk}
\end{equation}
scaling inversely with PBH mass $M$. As expected, smaller-mass PBHs are hotter and hence emit more particles, with a greater typical energy per particle, than larger-mass PBHs. 

For black holes with masses $M\lesssim 5 \times 10^{17} \, {\rm g}$, accurate emission spectra of long-lived particles require one to account for the formation and hadronization of quark-gluon jets and decays of unstable particles \cite{macgibbonQuarkGluonjetEmission1990, macgibbonQuarkGluonjetEmission1991}. We use \texttt{BlackHawk v2.2} \cite{arbeyBlackHawkV20Public2019, arbeyPhysicsStandardModel2021a} with \texttt{PYTHIA} \cite{Sjostrand:2014zea}, and \texttt{hazma} \cite{Coogan:2019qpu} to compute the secondary Hawking emission spectrum for photons, which is defined as
\begin{equation}
    \label{HawkSecondary}
    \frac{d^2N_{\gamma}^{(2)}}{dtdE} = \int_0^{\infty} \sum_j \frac{d^2N^{(1)}_{j}}{dtdE'} \frac{dN_{\gamma}^j}{dE}dE',
\end{equation}
where $dN_{\gamma}^j(E',E)/dE$ are the differential branching ratios \cite{arbeyBlackHawkV20Public2019}. Note that for $M\gtrsim 5 \times 10^{17} \, {\rm g}$, $d^2N_{\gamma}^{(2)}/dtdE = d^2N_{\gamma}^{(1)}/dtdE$, since the PBHs are too cold to emit any particles other than photons or neutrinos. 

\subsection{Comparison with Gravitational Ionization}
\label{sec:CompareGravIonHawking}

\begin{figure}[t]
    \centering
    \includegraphics[width=1.0\linewidth]{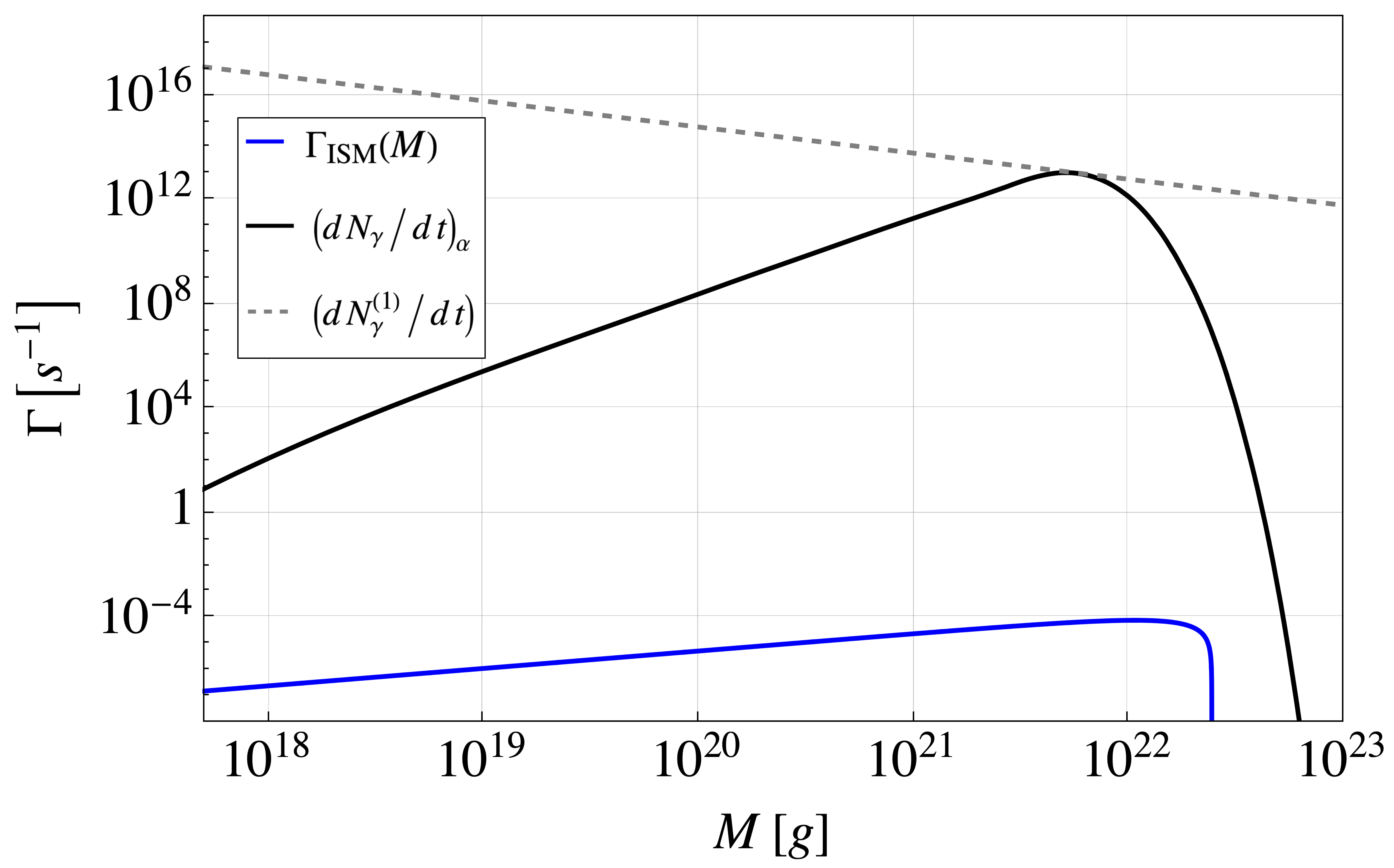}
    \caption{\justifying 
    Comparison of the total integrated Hawking photon emission rate from Eq.~(\ref{dNdtHawk}) (dashed), Hawking Lyman-$\alpha$ emission rate from Eq.~(\ref{LymanHawk}) (black), and the gravitational ionization rate $\Gamma (M)$ for a PBH transiting the ISM in the present epoch from Eq.~(\ref{GammaIonize2}) (blue).}
    \label{fig:PhotonComp}
\end{figure}

In this section, we compare the gravitational ionization rate $\Gamma(M)$ to the integrated Hawking emission rate from Eq.~(\ref{dNdtHawk}) and the Hawking emission rate of photons at the Lyman-$\alpha$ energy $E_\alpha$, for a single PBH of mass $M$ transiting through a gas of neutral hydrogen atoms. We define the Hawking Lyman-$\alpha$ emission rate for a PBH of mass $M$ as:
\begin{equation}
\label{LymanHawk}
\left(\frac{dN_{\gamma}}{dt}\right)_{\alpha} \equiv E_{\alpha}\frac{d^2N_{\gamma}^{(1)}}{dtdE}\Bigg|_{E=E_{\alpha}}.
\end{equation}
Figure \ref{fig:PhotonComp} compares the integrated photon emission rate, Hawking Lyman-$\alpha$ emission rate, and gravitational ionization rate for a PBH of mass $M$ transiting through neutral hydrogen in the ISM during the present epoch.

Table~\ref{tab:EmissionMasses} indicates that the cutoff mass $M_c (T_b)$ of Eq.~(\ref{EmitterCondition2}) for a PBH traversing the present-day ISM is $M_c \simeq 3.42 \times 10^{25} \, {\rm g}$. Given that only PBHs with $M \leq M_{\rm max}^{\rm H} = 2.43 \times 10^{22} \, {\rm g}$ can gravitationally ionize neutral hydrogen, the hierarchy $M_{\rm max}^{\rm H} \ll M_c (T_{\rm ISM})$ ensures that all PBHs with $M \leq M_{\rm max}^{\rm H}$ that traverse the ISM are net Hawking emitters.

The energy spectra from both types of photon emission differ considerably, with $E_{\rm peak}^{\rm GI} \simeq E_\alpha =  10.2 \, {\rm eV}$ from gravitational ionization, whereas $E_{\rm peak}^{\rm Hawk}$ for Hawking emission in Eq.~(\ref{EpeakHawk}) depends on the Hawking temperature of the PBH, as  shown in Fig.~\ref{fig:HawkPrimary}. As indicated by the dashed line in Fig.~\ref{fig:PhotonComp}, across the entire asteroid-mass range $(10^{17} \, {\rm g} \leq M \leq 10^{23} \, {\rm g})$, the total integrated Hawking photon emission would utterly dominate the photon emission rate from gravitational ionization within a present-day medium such as the ISM. In the following section, we search for scenarios in which either photon emission from gravitational ionization or total energy transfer to the medium from gravitational interactions could have dominated primary Hawking photon emission. 

\section{Regimes in which Gravitational Interactions Dominate}
\label{sec:GravIonDominates}

We investigate whether there exists a class of realistic PBH number distributions such that gravitational interactions could have been the dominant mode of energy deposition to the surrounding medium when integrated over the whole PBH population at some time in cosmological history. We consider two sources of energy deposition from gravitational interactions: (1) gravitational ionization of neutral hydrogen atoms, as described in Sec.~\ref{sec:GravIon}, and (2) energy transfer to hydrogen atoms from gravitational scattering as PBHs traverse the medium, regardless of whether such interactions induce gravitational ionization.

In order to maximize the gravitational ionization rate $\Gamma$ in Eq.~(\ref{GammaIonize1}) during some epoch, we need to maximize the product $n_{\rm H} v_{\rm rel}$ and the population of PBHs with mass $M\simeq M^{\rm H}_{\rm peak}$. The density of neutral hydrogen is maximized immediately following recombination ($z_{\rm rec} \simeq 1090, \, t_{\rm rec} \simeq 380,000 \, {\rm yr}$) and the population of PBHs with the optimal mass for gravitational ionization can be maximized by peaking the PBH number distribution around $M\simeq M_{\rm peak}^{\rm H}$. 

In Section \ref{sec:MassFunc} we introduce the generalized critical collapse PBH number distribution. Then in Section \ref{sec:Comparison}, we consider two ratios of interest: (1) the ratio of power deposited in the medium from gravitational ionization compared to the power deposited from ionizing Hawking radiation, and (2) the ratio of total power deposited in the medium from gravitational scattering (independent of possible ionization) to the total power deposited via Hawking emission. As we will see from the first ratio $R_{p}^{\rm ion}$, gravitational ionization rates always remain subdominant to photoionization rates from Hawking radiation across the entire asteroid-mass range. On the other hand, the second ratio $R_p^{\rm tot}$ indicates that there do exist robust regions of PBH parameter space within which total power deposited from gravitational interactions surpasses total power deposited from Hawking emission.

\subsection{PBH Number Distribution}
\label{sec:MassFunc}

At the time $t_i$ of PBH formation, the normalized PBH differential \emph{number distribution function} is defined as \cite{mosbechEffectsHawkingEvaporation2022,cang21cmConstraintsSpinning2022,Klipfel:2025bvh}
\begin{equation}
    \label{eqn:PsiDef}
    \phi(M_i, t_i) \equiv \frac{1}{n_{{\rm PBH},i}}\frac{dn_{ { \rm PBH}}}{dM_i} \simeq \frac{\bar{M}}{M}\psi(M_i, t_i) ,
\end{equation}
where $\phi(M_i, t_i)$ satisfies $\int dM_i \, \phi(M_i, t_i) = 1$ and the more commonly used quantity $\psi(M_i, t_i)$ is the \emph{mass function}. Here $n_{ {\rm PBH}, i}$ is the comoving PBH number density at formation time, and $\bar{M}$ is the mass at the peak of the number distribution. 

We assume PBHs form via \textit{critical collapse} of primordial overdensities in the aftermath of inflation~\cite{Choptuik:1992jv, Evans:1994pj, Niemeyer:1999ak, Gundlach:2007gc,Musco:2008hv}. This formation mechanism gives rise to mass functions with a sharp peak at a typical mass $\bar{M}$ and generic, physically motivated features---most notably, a power-law tail that extends to arbitrarily small masses $M < \bar{M}$ and an exponential cutoff for $M>\bar{M}$. The power-law tail guarantees the existence of a small sub-population of extremely light PBHs at all cosmological times, if the peak of the distribution is located a some $\bar{M} > M_*$. Here 
\begin{equation}
M_* = 5.34 \times 10^{14} \, {\rm g}
\label{Mstardef}
\end{equation}
is the PBH mass at the time of formation which has a lifetime equal to the current age of the Universe, $t_0$ \cite{Klipfel:2025bvh}. Meanwhile, causality restricts PBHs to form with masses no larger than the Hubble mass at the time of formation, which gives rise to the large-mass exponential cutoff feature \cite{Klipfel:2025bvh}.

The authors of Ref.~\cite{Gow:2020cou} find that these features of the PBH formation process are most accurately captured by the generalized critical collapse (GCC) parameterization of the PBH distribution. The normalized GCC number distribution function is defined as
\begin{equation}
    \label{eqn:GCCinit}
    \begin{split}
    \phi & (M_i|\bar{M},\alpha, \beta) \\
    &= \frac{C (\alpha,\beta)}{\bar{M}}\left( \frac{M_i}{\bar{M}}\right)^{\alpha-1}\exp\left[ -\frac{(\alpha-1)}{\beta} \left(\frac{M_i}{\bar{M}}\right)^{\beta}\right],
    \end{split}
\end{equation}
where $\bar{M}$ is the value of $M_i$ which maximizes $\phi$, $\alpha > 1$ controls the power-law scaling of the low-mass tail, and $\beta > 0$ controls the high-mass exponential cutoff. The dimensionless coefficient is given by 
\begin{equation}
    C (\alpha, \beta) = \frac{\beta}{\Gamma (\alpha / \beta)} \left( \frac{\alpha - 1}{\beta} \right)^{\alpha/\beta} .
\end{equation}
The original critical-collapse function corresponds to $\alpha = \beta = \nu^{-1}$, where $\nu = 0.36$ for PBH formation within a fluid with a radiation-dominated equation of state \cite{Niemeyer:1997mt,Green:1999xm,niemeyerDynamicsPrimordialBlack1999}.  

We note that unlike other analyses of particle emission from PBH populations \cite{Klipfel:2025bvh, Klipfel:2025jql}, which account for the time-evolution of the GCC mass function from formation until the epoch of interest, we can neglect time-evolution here because the population of interest at the peak of the distribution have masses $M \gg M_*$. We may therefore assume that $M=M_i$ and take
\begin{equation}
    \phi(M, t|\alpha, \beta,\bar{M}) =\phi(M_i|\alpha, \beta,\bar{M})
\end{equation}
for all $t\leq t_0$, so we suppress the parameter $t$ going forward.

Critical collapse gives a direct relationship between the typical PBH mass at formation time (the mass $\bar{M}$ that maximizes $\phi$) and the time of formation (the time when the overdensity that will trigger gravitational collapse crosses back inside the Hubble radius) \cite{Carr:1975qj, Niemeyer:1997mt, Green:1999xm, Kuhnel:2015vtw}:
\begin{equation}
    \label{eqn:Mbar}
    \bar{M}(t_i) \simeq 0.2 M_H(t_i).
\end{equation}
Here $M_H (t_i)$ is the mass enclosed within the Hubble radius at time $t_i$. For $\bar{M} (t_i) \gg M_*$, we may consider $\bar{M}$ to be effectively independent of time.

For this analysis, we treat the number distribution as a function of three model parameters: $\alpha$, $\beta$, and $\bar{M}$. Gravitational ionization rates can be maximized by letting $\bar{M} \simeq M_{\rm peak}^{\rm H}$ and forcing $\phi$ to be sharply peaked by tuning $\alpha$ and $\beta$. In the following subsection, we consider various scenarios for energy deposition from a population of PBHs into the medium around the epoch of recombination as functions of $\{ \alpha, \beta, \bar{M} \}$.

\subsection{Energy Deposition Comparisons}
\label{sec:Comparison}

In the epoch immediately following recombination, when $n_{\rm H}\simeq 2.3 \times10^8 \, {\rm m}^{-3}$, a PBH population with comoving number density $dn_{{\rm PBH}}/dM$ will deposit energy to the surrounding medium via two distinct mechanisms: photon Hawking emission and gravitational scattering. The latter will only ionize neutral hydrogen atoms if $b \leq b_{\rm th}^{\rm H} (M)$, where $b_{\rm th}^{\rm H} (M)$ is given in Eq.~(\ref{bmaxH1}). We compute the deposited power per comoving volume for each mechanism and identify regions of PBH model parameter space $\{ \alpha, \beta , \bar{M} \}$ in which gravitational scattering is the dominant mechanism of energy deposition. 

We take the end of recombination to occur at $z_{\rm rec}=1090$ \cite{Planck:2018vyg}, which corresponds to a background temperature $T_b (t_{\rm rec}) = T_{\rm CMB} (t_0) \times (1 + z_{\rm rec}) = 0.259 \, {\rm eV}$. This is significantly below the binding energy of neutral hydrogen atoms, $E_1=13.6 \, {\rm eV}$, because high-energy photons in the tail of the blackbody distribution continue to ionize significant fractions of the medium until $T_b \ll E_1$. (See, e.g., Ref.~\cite{Lynch:2024gmp}.)

The maximum black hole mass that can be considered a net emitter at $z_{\rm rec}=1090$ is found via Eq.~(\ref{EmitterCondition2}): $M_c(t_{\rm rec}) = 1.15\times10^{24} \, {\rm g}$. (See Table~\ref{tab:EmissionMasses}.) The fact that $M_c(t_{\rm rec}) > M_{\rm max}^{\rm H} = 2.43\times10^{22} \, {\rm g}$ implies that all PBHs capable of gravitational ionization will also be net Hawking emitters at $t_{\rm rec}$. Additionally, we note that the PBH mass for which photon emission peaks at $E=E_1$ is $M_{E_1}= 6.04/ (8 \pi G E_1) = 4.67\times10^{21} \, {\rm g} < M_{\rm max}^{\rm H}$. Thus, in this section we first investigate whether there can exist a PBH population such that gravitational ionization dominates photoionization by Hawking radiation immediately after cosmic recombination. Due to the low gravitational ionization (GI) rates compared to Hawking emission (see Fig.~\ref{fig:PhotonComp}), we find that photoionization by Hawking radiation will always dominate at $t=t_{\rm tec}$---even for sharply peaked (nearly monochromatic) PBH distributions with $M_{E_1}<\bar{M}<M_{\rm max}^{\rm H}$, for which most Hawking-emitted photons have $E<E_1$. We then show, however, that \emph{total energy transfer} to the medium via gravitational scattering from PBHs at $t_{\rm rec}$ \emph{does} dominate total Hawking power for a relatively large class of realistic PBH distributions peaked in the asteroid-mass range.

A single PBH with velocity $v_{\rm rec}$ in the rest frame of neutral hydrogen with density $n_{\rm H}$ loses energy via gravitational scattering at a rate: 
\begin{equation}
    \begin{split}
        \dot{E}(M, \Lambda_b) & = 2 \pi n_{\rm H} v_{\rm rec}\int_{r_s}^{\Lambda_{\rm b}} db \,\varepsilon(M, b, v_{\rm rec}) \,b \\
        & = 4\pi v_{\rm rec}^3n_{\rm H} m_p \int_{r_s(M)}^{\Lambda_{\rm b}} db \frac{b}{1 + (b/b_{\perp})^2} \\
        & = \frac{2 \pi n_{\rm H} m_p G^2 M^2}{v_{\rm rec}}\log\left[\frac{\Lambda_{\rm b}^2+b_{\perp}^2}{r_s^2+b_{\perp}^2} \right],
    \end{split}
    \label{dEdt1}
\end{equation}
where we used $\varepsilon(M, b, v_{\rm rec})$ from Eq.~(\ref{EnergyTransfer1}) and consider some large-$b$ cutoff $\Lambda_{\rm b}$.

From Eq.~(\ref{eqn:PsiDef}) we define the differential comoving PBH number density at $t_{\rm rec}$ as
\begin{equation}
    \frac{dn_{\rm PBH}}{dM} = n_{\rm PBH, i} \, \phi(M|\bar{M},\alpha, \beta),
\end{equation}
which holds exactly for $\bar{M} \gg M_*$.
Thus, for a population of PBHs with number distribution $\phi(M|\bar{M},\alpha, \beta)$, the total deposited power per comoving volume via the gravitational scattering of neutral hydrogen is given by
\begin{equation}
    \begin{split}
    p_{\rm GI}&(\Lambda_b, \Lambda_M, \bar{M},\alpha, \beta) \\ &\equiv n_{{\rm PBH}, i}\int_0^{\Lambda_M}dM \, \dot{E}(M,\Lambda_b) \,\phi(M|\bar{M}, \alpha, \beta),
    \end{split}
    \label{pGI}
\end{equation}
where the upper integration bound $\Lambda_M$ assumes different values for each of the cases considered below.

\begin{figure*}[t]
    \centering

    \subfloat{\includegraphics[width=0.48\linewidth]{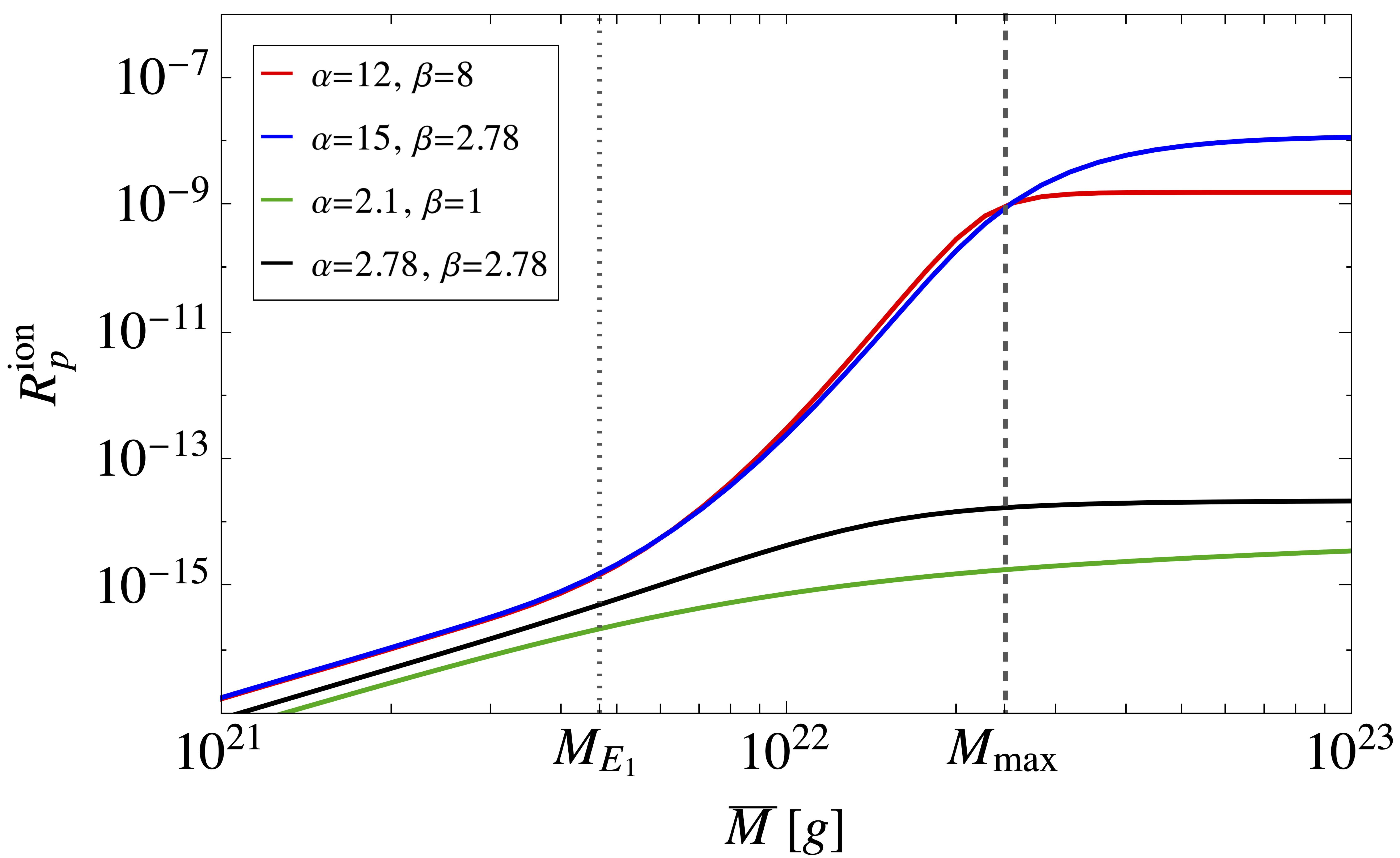}}
    \subfloat{\includegraphics[width=0.48\linewidth]{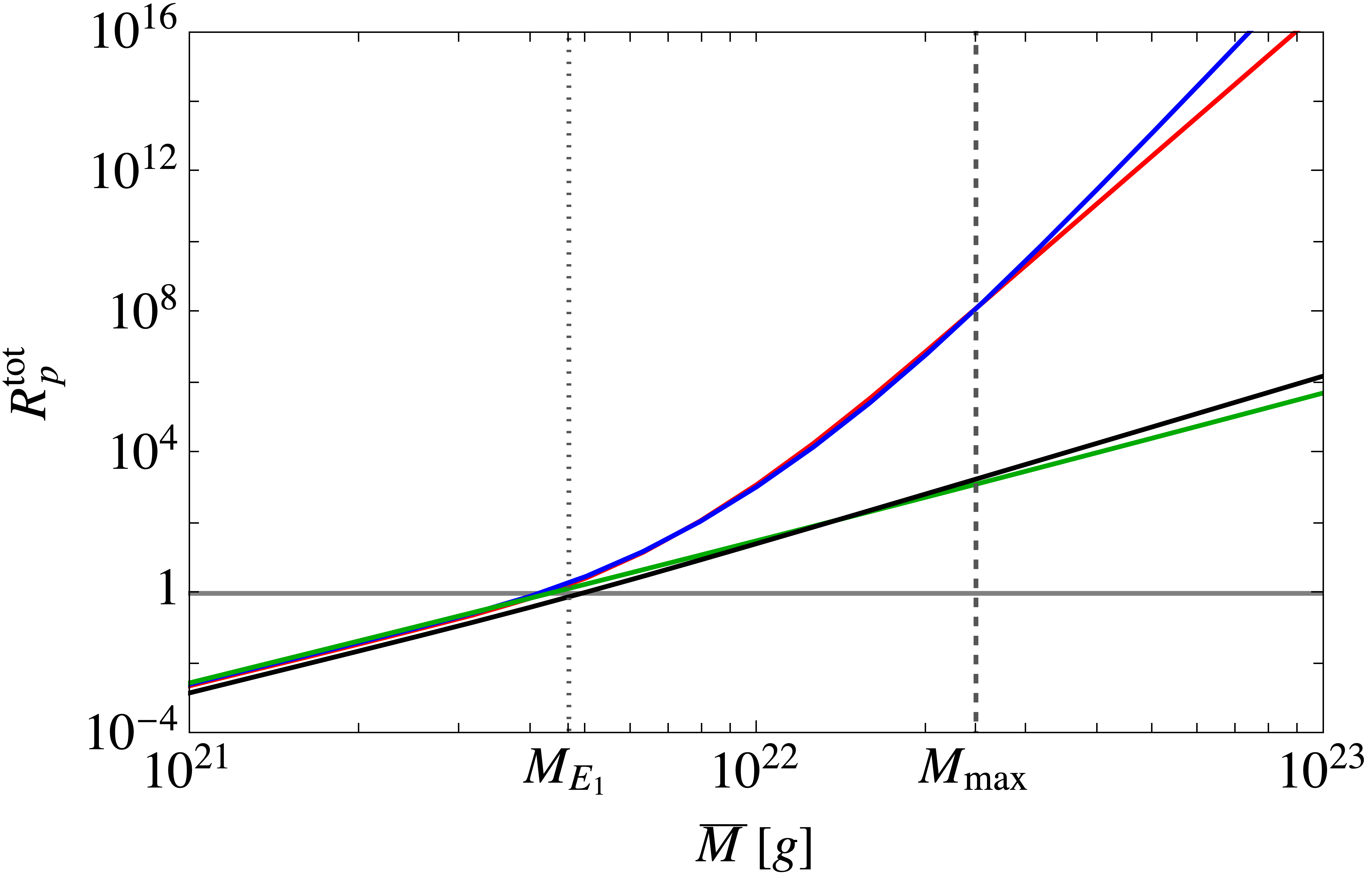}}
    \caption{\justifying  (\textit{Left}) The ionizing power ratio $R_p^{\rm ion}$ defined in Eq.~(\ref{RpIon}) at time $t_{\rm rec}$ immediately following recombination. Gravitational ionization never dominates ionizing Hawking radiation, even for large values of $\alpha$ and $\beta$, which force the PBH number distribution function to be sharply peaked at $\bar{M}$. (\textit{Right}) The total power ratio $R_p^{\rm tot}$ defined in Eq.~(\ref{RpTotAnalytical}) at time $t_{\rm rec}$. We find that the total  power deposited in the medium from gravitational scattering of neutral hydrogen can significantly exceed the power deposited via Hawking radiation for $M_{E_1}\lesssim M \lesssim M_{\rm max}$ across a wide range of values of the GCC model parameters $\{\alpha, \beta \}$, including the standard critical collapse parameters $\alpha=\beta=\nu^{-1}=2.78$. Note that both figures are truncated at $\bar{M}=10^{23}\, {\rm g}$, which is considered to be the upper bound on the asteroid-mass range, and the largest value of $\bar{M}$ such that a monochromatic PBH distribution can have a dark matter fraction $f_{\rm PBH}=1$. 
    }
    \label{fig:ParamSpace}
\end{figure*}

To calculate the power deposited in the medium by Hawking photons from the PBH population, we first note that photons with $E < E_1$ are effectively free-streaming and will only scatter elastically from the hydrogen atoms, and hence will not transfer any energy to the medium. So we restrict attention to $E \geq E_1$ when integrating over the Hawking emission spectra. For ionizing photons, we take the mean free path to be $\ell_\gamma (E) = 1 / [n_H \, \sigma (E) ]$, where $\sigma (E) = \sigma_L ( E / E_1 )^{-3}$ is the photoionziation cross section for photons with $E \geq E_1$. (Here $\sigma_L = 6.3 \times 10^{-22} \, {\rm m}^2$.) Then the power deposited in the medium per comoving volume by photon Hawking radiation from a PBH population is
\begin{equation}
\begin{split}
    &p_{\rm Hawk} (\bar{M}, \alpha, \beta) \\ & \equiv n_{{\rm PBH},i} \, n_{\rm H}\int_0^{\infty}dM \,\phi(M|\bar{M},\alpha, \beta) \\
    &\quad\quad \quad\quad\quad\quad \times \int_{E_1}^{\infty}dE \, \ell_{\gamma } (E) \, \sigma (E) \, E\frac{d^2N^{(1)}_{\gamma}}{dEdt} \\
    &= n_{ {\rm PBH}, i} \int_0^\infty dM \, \phi (M \vert \bar{M} , \alpha, \beta) \int_{E_1}^\infty dE \, E \frac{ d^2 N^{(1)}_{\gamma}}{dE dt} .
    \label{pHawk}
    \end{split}
\end{equation}
This integral is computed starting from \texttt{BlackHawk} \cite{arbeyBlackHawkV20Public2019, arbeyPhysicsStandardModel2021a} spectra and performing a series of interpolations and numerical integrals. Across the PBH mass range of interest, only the primary Hawking emission spectrum contributes.  

To compare the deposited energy from both mechanisms, we define the ratio
\begin{equation}
    R_p(\Lambda_b, \Lambda_M,  \bar{M}, \alpha, \beta)\equiv\frac{p_{\rm GI}(\Lambda_b, \Lambda_M, \bar{M}, \alpha, \beta)}{p_{\rm Hawk}(\bar{M}, \alpha, \beta)}.
    \label{Rp}
\end{equation}
Note that the ratio $R_p$ is independent of the (unknown) initial comoving PBH number density $n_{{\rm PBH}, i}$.

We first compare the deposited power per comoving volume by gravitational scattering that causes ionization to Hawking emission of photons energetic enough to photoionize the medium. We define the \emph{ionizing power ratio} as
\begin{equation}
    R_p^{\rm ion}(\bar{M}, \alpha, \beta)=R_p(b_{\rm th}^{\rm H}(M), M_{\rm max}^{\rm H}, \bar{M}, \alpha, \beta) ,
    \label{RpIon}
\end{equation}
where we set $\Lambda_b = b_{\rm th}^{\rm H}(M)$ and $\Lambda_M=M_{\rm max}^{\rm H}$ so as to only consider gravitational scattering events that are each energetic enough to ionize the neutral hydrogen. See Fig.~\ref{fig:ParamSpace} for a plot of $R_p^{\rm ion}(\bar{M})$ for various values of $\alpha$ and $\beta$. We find that $R_p^{\rm ion}\ll 1$ for all $\bar{M}$ within the asteroid-mass range, even as we vary $\alpha$ and $\beta$. We conclude that there is no region of parameter space, even with large $\alpha$ and $\beta$, within which gravitational ionization dominates photoionization by Hawking radiation around the time $t_{\rm rec}$.

Next we compare the \emph{total} rate of energy deposition from gravitational scattering to the power deposited in the medium by Hawking emission. Unlike for $R_p^{\rm ion}$, we now include scattering events with $b>b_{\rm th}^{\rm H}$, which would only impart center-of-mass motion to the hydrogen atoms rather than ionizing them. For this scenario, we define the cutoff $\Lambda_b$ such that
\begin{equation}
    \varepsilon(M,b,v_{\rm rec}) \equiv \frac{2 m_p v_{\rm rel}^2}{1 + (\Lambda_b/b_\perp)^2}=T_b,
\end{equation}
where $T_{\rm rec}=0.256 \, {\rm eV}$ is the background temperature at $t_{\rm rec}$.
This gives 
\begin{equation}
    \Lambda_b(M) = b_{\perp}\sqrt{\frac{2 m_p v_{\rm rec}^2}{T_{\rm rec}}-1}= 8.51 \, b_{\perp}.
\end{equation}
Using the fact that $b_{\perp}(M)/r_s(M) = 1/(2 v_{\rm rel}^2) \gg 1$, we can approximate Eq.~(\ref{dEdt1}) as:
\begin{equation}
    \dot{E} \simeq 26.9\frac{n_{\rm H}m_p G^2 M^2}{v_{\rm rec}}
    \label{dEdt2}
\end{equation}
And thus, by setting $\Lambda_b=8.51 b_{\perp}$ and $\Lambda_M = \infty$ in Eq.~(\ref{pGI}), we can express the total volumetric power from gravitational scattering analytically as
\begin{equation}
    \begin{split}
    &p_{\rm GI}^{\rm tot}(\bar{M},\alpha, \beta) \\
    &\>\> = p_{\rm GI}(8.51b_{\perp}, 
    \infty, \bar{M},\alpha, \beta)\\
    & \>\>\simeq  \,n_{\rm PBH, i} \frac{26.9\, n_{\rm H}\, m_p G^2}{v_{\rm rec}} \bar{M}^2 \frac{\Gamma\left(\frac{ \alpha+2}{\beta}\right)}{\Gamma(\alpha/\beta)}\left( \frac{\beta}{\alpha-1}\right)^{\frac{2}{\beta}}.
    \end{split}
    \label{pGItot}
\end{equation}

Meanwhile, as noted above, to calculate the power deposited by Hawking-emitted photons, we consider only the range $E \geq E_1$, since photons below this energy will not lose energy by scattering in the medium. Hence our expression for $p_{\rm Hawk}$ in Eq.~(\ref{pHawk}) remains unchanged. 

Thus, the \emph{total power ratio} may be written as
\begin{equation}
    R_p^{\rm tot}(\bar{M}, \alpha, \beta) = R_p(8.51 b_{\perp}, 
    \infty, 
    \bar{M}, \alpha, \beta) . 
    \label{RpTotAnalytical}
\end{equation}
We find that $R_p^{\rm tot} > 1$ for $\bar{M} \gtrsim M_{E_1}$, for a large region of realistic $\{ \alpha, \beta \}$ model parameter space. See Figure~\ref{fig:ParamSpace}. We consider ``realistic'' values of $\alpha$ and $\beta$ to be close to the standard critical collapse values $\alpha=\beta=\nu^{-1}\simeq 2.78$. Larger $\alpha$ and $\beta$ values force the distribution to be sharply peaked and thus to approach an unphysical monochromatic distribution.

To summarize, although gravitational ionization by PBHs remains highly subdominant compared to Hawking radiation photoionization ($R_p^{\rm ion} \ll 1$), the total energy deposited within the medium via gravitational scattering by PBHs can easily dominate the total power deposited via Hawking emission ($R_p^{\rm tot} \gg 1$). Energy deposited by gravitational scattering events with impact parameters $b \gg b_{\rm th}^{\rm H}$ would not ionize the medium---in such encounters, the gravitational tidal force on the hydrogen atom would be much too small to cause ionization. Rather, such encounters would transfer energy to {\it entire atoms}, thereby causing local heating of the medium.

Current PBH constraints arising from CMB measurements have focused on such phenomena as spectral distortions from the injection of energetic photons via Hawking emission \cite{carrConstraintsPrimordialBlack2016,carrPrimordialBlackHoles2020} and changes to the typical ionization fraction due to photoionization by Hawking radiation \cite{Koivu:2025add}. Whether the {\it non-ionizing} deposition of power via gravitational scattering by PBHs identified here---with its attendant local heating of the medium---could yield an observable signal, and hence perhaps an improved constraint on PBHs within the asteroid-mass window, remains the subject of further research.

\section{Gravitationally-Induced Nuclear Reactions}
\label{sec:GravNuclear}

In previous sections we considered gravitational ionization of neutral atoms via strong tidal forces from a transiting asteroid-mass PBH. This same basic process could also induce gravitational dissociation of nuclear matter, such as deuterons, by PBHs with correspondingly smaller Schwarzschild radii and hence steeper gravitational field gradients. In Sec.~\ref{sec:Deuteron} we consider the deuteron gravitational dissociation rates caused by PBH transits during big bang nucleosynthesis (BBN), and compare those rates with expected photodissociation rates from Hawking radiation emitted by the PBH. Unlike the case of ionizing neutral hydrogen atoms, we find that gravitational dissociation \emph{can} dominate photodissociation for an interesting range of PBH masses, though this depends on the relative velocity between dark matter and baryons in the early universe. Next, in Sec.~\ref{sec:Fission}, we consider gravitational effects on larger nuclei. In particular, we find scenarios in which the tidal force from a transiting PBH can deform heavy, unstable nuclei, such as Uranium-235, thereby inducing fission. 

\subsection{Deuteron Dissociation}
\label{sec:Deuteron}

Modern effective treatments of bound deuteron states use the ``Morse potential'' \cite{Khachi:2023iqp},
\begin{equation}
V_M (r) = V_0 \left[ e^{- 2 (r - r_m) / a_m} - 2 e^{-(r - r_m)/a_m} \right] ,
\label{MorsePotential}
\end{equation}
with recent best-fit values $V_0 = 114.153 \, {\rm MeV}$, $r_m = 0.841 \, {\rm fm}$, and $a_m = 0.350 \, {\rm fm}$ for the so-called ``global optimization algorithm'' or GOA method (see Table~I of Ref.~\cite{Khachi:2023iqp}). These parameters reproduce a charge radius $r_c = 2.13 \, {\rm fm}$ and binding energy $E_{\rm bind} = 2.225 \, {\rm MeV}$ for the deuteron, in close agreement with experimental values. If we equate $F_{\rm D} (r_c) \equiv -V_{M} (r_c) / r_c$ to the gravitational tidal force on the nucleons from a PBH passing within an impact parameter $b$ of one of the nucleons, we find the threshold impact parameter for gravitational dissociation of the deuteron:
\begin{equation}
\begin{split}
    b_{\rm th}^{\rm D} (M)& = \left[ -\frac{ 2 G M m_p r_{c}^2 }{V_M(r_c)} \right]^{1/3}\\ 
    & = \left[\frac{ 2 G M m_p r_{c}^2 }{V_0} \frac{e^{(2 r_c-r_m)/a_m}}{\left( 2 e^{r_c/a_m} -e^{r_m / a_m}   \right)} \right]^{1/3} \\
    & = 1.04 \times 10^{-4} \, {\rm fm} \left( \frac{ M}{1 \, {\rm g}} \right)^{1/3} .
    \end{split}
    \label{bMorsealt1}
\end{equation}
Again, setting $b_{\rm th}^{\rm D} (M_{\rm max}) = r_s (M_{\rm max})$ yields the maximum PBH mass for which gravitational tidal forces could dissociate a deuteron:
\begin{equation}
    \begin{split}
    M_{\rm max}^{\rm D} 
    & = \frac{r_c}{2 G}\sqrt{\frac{m_p \,e^{(2 r_c-r_m)/a_m}}{V_0 \left( 2 e^{r_{c}/a_m} - e^{r_m/a_m} \right)}} \\
    & = 1.84\times10^{16}\, {\rm g.}
    \end{split}
    \label{MmaxMorsealt2}
\end{equation}
The gravitationally-induced deuteron dissociation rate 
\begin{equation}
\begin{split}
    \Gamma_{\rm GI}^{\rm D}(M) & = n_{\rm D} \sigma \,v_{\rm rel} \\
    & = n_{\rm D}\pi \left[b_{\rm th}^{\rm D}(M)^2 - r_s(M)^2\right]v_{\rm rel}
    \end{split}
\end{equation}
will be maximized at $M_{\rm peak}^{\rm D} =  3^{-3/4} M_{\rm max}^{\rm D}$, as in Eq.~(\ref{Msigma}). For $M_{\rm max}^{\rm D}$ in Eq.~(\ref{MmaxMorsealt2}), this yields
\begin{equation}
    M_{\rm peak}^{\rm D} = 8.06 \times 10^{15} \, {\rm g} ,
    \label{MpeakD}
\end{equation}
with Schwarzschild radius $r_s (M_{\rm peak}^{\rm D}) = 1.2\times10^{-14} \, {\rm m}$. 

The interaction timescale $\tau = \Delta T (b_{\rm max}^{\rm D})$ over which the energy transfer to one of the nucleons would occur due to gravitational scattering from the passing PBH again takes the form of Eq.~(\ref{tau}), now using $b_{\rm th}^{\rm D} (M)$ from Eq.~(\ref{bMorsealt1}) along with $\langle \Delta r \rangle \rightarrow r_c = 2.13 \, {\rm fm}$ and $E_1 \rightarrow E_{\rm bind} = 2.225 \, {\rm MeV}$ for the deuteron. It is not clear what to expect for $v_{\rm rel}$ between dark matter particles (including PBHs) and baryons around the time of BBN. In our analysis, we consider two fiducial values: $v_{\rm rel} = 30\, {\rm km/s}$ (akin to the value around cosmic recombination), and $v_{\rm rel} = 1 \, {\rm km/s}$. For $v_{\rm rel} = 30 \, {\rm km/s}$, the ratio $\tau / E_{\rm bind}^{-1} \lesssim 0.2$ for $M < 3.98 \times 10^{18}  \, {\rm g}$, and hence for all $M < M^{\rm D}_{\rm max}$; for $v_{\rm rel} = 1 \, {\rm km/s}$, the ratio $\tau / E_{\rm bind}^{-1} < 1$ for $M < 6.62 \times 10^{15} \, {\rm g}$. 

\begin{figure}[t]
    \centering
    \includegraphics[width=1.0\linewidth]{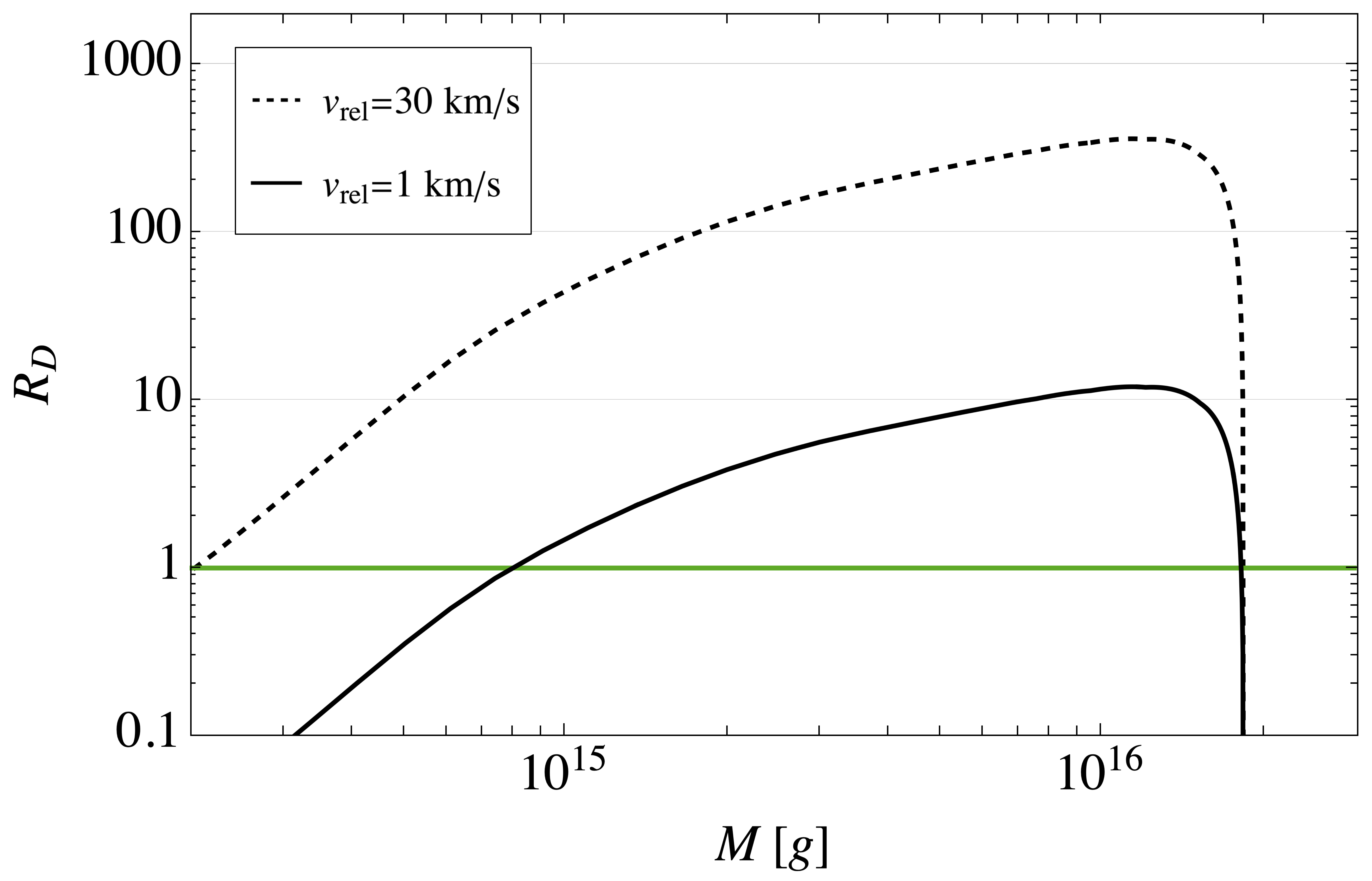}
    \caption{\justifying  The deuteron dissociation ratio $R_{\rm D}$ from Eq.~(\ref{RDratio}). For $v_{\rm rel}=30 \, {\rm km/s}$ (dashed), we find that $R_{\rm D}(M)>1$ for $2.0\times10^{14} \, {\rm g} \leq M \leq 1.8\times10^{16}\,{\rm g}$. For $v_{\rm rel}=1 \, {\rm km/s}$ (solid), we find that $R_{\rm D}(M)>1$ for $7.8\times10^{14} \, {\rm g} \leq M \leq 1.8\times10^{16}\,{\rm g}$. There will exist regimes in which deuteron dissociation by sub-asteroid-mass PBHs is dominated by gravitational dissociation rather than Hawking photodissociation for relative velocities above $v_{\rm rel}^{\rm min}=  0.084 \, {\rm km/s} $.}
    \label{fig:RDratio}
\end{figure}

Our goal is to determine whether there exists a range of PBH masses $M$ such that the gravitational deuteron dissociation rate $\Gamma_{\rm GI}^{\rm D}(M)$ exceeds the deuteron dissociation rate from emitted Hawking radiation. Unlike Section~\ref{sec:Comparison}, which considered a similar comparison between gravitational ionization and photon Hawking radiation power from a \emph{PBH population} at the epoch of recombination, here we consider a \emph{single PBH} at the time of maximum deuteron density during BBN. Note that because $M_{\rm max}^{\rm D} < 10^{17} \, {\rm g}$, all PBHs capable of gravitationally dissociating deuterons must fall below the asteroid-mass range and will therefore be hot enough to emit particles other than photons and neutrinos (which we ignore throughout). To avoid the complexities of hadronic interactions, we will only consider PBHs with mass $2\times10^{14} \, {\rm g} \lesssim M < M_{\rm max}^{\rm D}$, which have temperatures $T_H\lesssim140/\beta_{0} \, {\rm MeV} = 53 \, {\rm MeV}$, below the cutoff scale for pion emission. Such PBHs can only emit photons, neutrinos, electrons, positrons, and muons and their emission spectra are best captured by \emph{secondary spectra} as discussed in Section~\ref{sec:HawkingFormalism}. 

We note that the secondary $e^{\pm}$ emission spectra for PBHs in this mass range are significant and actually exceed the secondary photon spectra. Nonetheless, we neglect deuteron dissociation by charged leptons for two reasons. First, strictly \emph{elastic} electron-deuteron scattering has been measured for electron energies as high as $E_e = 300\, {\rm MeV}$ \cite{Simon:1981br}, well above the deuteron binding energy. Second, accounting for deuteron dissociation by charged leptons depends sensitively upon the deuteron structure functions (see, e.g., Ref.~\cite{JLABt20:2000qyq}), which remains beyond the scope of this work. Therefore we only focus on comparing gravitational dissociation rates to photodissociation rates in order to parallel the previous section on ionization.

The deuteron photodissociation rate by Hawking radiation from a single transiting PBH when the universe has background temperature $T_b$ is given by
\begin{equation}
    \begin{split}
    \Gamma&_{\rm PD}^{\rm D}(M) \\ &\approx  n_{\rm D}(T_b) \ell_{\rm D} (T_b) \int_0^{\infty} dE \, \sigma_{\rm PD} (E) \, f_{\rm D}\frac{ d^2 N^{(2)}_\gamma(M)}{dt dE} ,
    \end{split}
\end{equation}
where $n_{\rm D}$ is the deuteron number density, $\ell_{\rm D}$ is the mean free path for a photon to encounter a deuteron, $\sigma_{\rm PD}$ is the energy-dependent deuteron photodissociation cross-section, $dN_{\gamma}^{(2)}/dtdE$ is the secondary Hawking spectrum for photons, and $f_{\rm D}\equiv n_{\rm D}/n_{\rm ch}$ is the fraction of charged particles that are deuterons. For photon energies $E > 300 \, {\rm MeV}$, $\sigma_{\rm PD} (E)$ falls steeply as a power-law, $\sigma_{\rm PD} \sim E^{-p}$, with $p > 3$ \cite{Klein:2003bz,Belz:1995ge,Schulte:2001se}.

We estimate $\ell_{\rm D} (T_b)$ in terms of the comoving number density of charged particles $n_{\rm ch} (T_b)$ within the medium:
\begin{equation}
    \ell_{\rm D} (T_b) \simeq \left[ f_{\rm D}n_{\rm ch} (T_b)\right]^{-1/3} ,
    \label{lambdagammadef}
\end{equation}
where 
\begin{equation}
    n_{\rm ch} = \left(\frac{3}{4} \right) \frac{ \zeta (3)}{\pi^2} g_{\rm ch} (T_b) \, T_b^3 ,
    \label{nchdef}
\end{equation}
and $g_{\rm ch}$ is the number of degrees of freedom of relativistic charged fermions in the medium.

The number density of deuterons reaches its maximum fairly early in the process of BBN, for $T_b=T_{\rm D} \simeq 0.07\, {\rm MeV}$, at $t=t_{\rm D} \simeq 120 \, {\rm s}$ \cite{Pospelov:2010hj}. Around that time, $g_{\rm ch} \sim 4$, with two spin states each for free electrons and free protons in the plasma, and $f_{\rm D} = n_{\rm D} / (n_p + n_e) \simeq 3.9\times10^{-3}$ (based on Fig.~1 in Ref.~\cite{Pospelov:2010hj}), which yields
\begin{equation}
    \ell_{\rm D} (0.07 \, {\rm MeV}) \simeq 
   3.15 \times 10^{-11} \, {\rm m} .
\end{equation}

We fix $T_b=T_{\rm D}$ in order to evaluate $\Gamma_{\rm PD}^{\rm D}(M)$ in the most optimistic scenario. As indicated in Table~\ref{tab:EmissionMasses}, PBHs with $M < M_c (T_{\rm D}) = 4.21 \times 10^{18} \, {\rm g}$ will be net Hawking emitters in a background medium at temperature $T_{\rm D}$, which includes all PBHs that could also induce gravitational dissociation, given $M_{\rm max}^{\rm D} < M_c (T_{\rm D})$. When computing the Hawking photodissociation rate $\Gamma^{\rm D}_{\rm PD} (M)$, we take $\sigma_{\rm PD} (E)$ based on Fig.~1 in Ref.~\cite{Klein:2003bz}, and conservatively extrapolate the high-energy tail for $E > 1.2 \, {\rm GeV}$ with the same exponent that holds within the range $500 < E < 1.2 \, {\rm GeV}$. 

For a PBH of mass $M$, gravitational dissociation of deuterons will dominate photodissociation via Hawking emission if
\begin{equation}
    R_{\rm D}(M)\equiv \frac{\Gamma_{\rm GI}^{\rm D}(M)}{\Gamma_{\rm PD}^{\rm D}(M)} >1.
    \label{RDratio}
\end{equation}
See Fig.~\ref{fig:RDratio} for a plot of $R_{\rm D}(M)$.  For $v_{\rm rel}=30 \, {\rm km/s}$, we find that $R_{\rm D}(M)>1$ for $2.0\times10^{14} \, {\rm g} \leq M \leq 1.8\times10^{16}\,{\rm g}$. For $v_{\rm rel}=1 \, {\rm km/s}$, we find that $R_{\rm D}(M)>1$ for $7.8\times10^{14} \, {\rm g} \leq M \leq 1.8\times10^{16}\,{\rm g}$. The relative velocity $v_{\rm rel}^{\rm min}$ such that $R_D(M)=1$ at its maximum is: $v_{\rm rel}^{\rm min}=0.084 \, {\rm km/s}$.

Previous constraints on PBHs from BBN have focused on energy injection into the plasma via Hawking emission, including from the late-stage evaporation of PBHs, which occurs for PBHs with initial mass $M_i \lesssim {\cal O} (10^{10} \, {\rm g})$ \cite{carrNewCosmologicalConstraints2010b,Wang:2025pum}. Whether the new effect identified here---gravitationally-induced dissociation of deuterons via tidal forces from transiting PBHs---could be used to further constrain PBHs with masses $M \gtrsim 10^{14} \, {\rm g}$ remains an exciting possibility, and the subject of further research.

\subsection{Nuclear Fission}
\label{sec:Fission}

In this section, we consider whether the steep gravitational gradients from a PBH transit can induce fission of heavy elements via tidal deformation of a nucleus. Qualitatively, we consider a heavy nucleus, such as $\ce{^{235}_{92}U}$, some distance $b$ from a PBH of mass $M$. Steep gradients in the PBH gravitational potential exert non-uniform forces on the nucleus and deform it from its initial ground state into an oblong-shaped isomer. (See Refs.~\cite{bohrMechanismNuclearFission1939, plessetClassicalModelNuclear1941,strutinskySymmetricalShapesEquilibrium1963,myersNuclearMassesDeformations1966,bjornholmDoublehumpedFissionBarrier1980, ivanyukOptimalShapesFission2009, Kowal2023, walkerNuclearIsomers2023} for discussion of nuclear deformation and isomeric fission of heavy elements, and particularly of actinides like uranium.) Nuclear deformation is often modeled as that of an incompressible fluid by the liquid drop model, which allows calculation of nuclear binding energy due to contributions from volume, surface tension, Coulomb force, and other terms related to nucleon asymmetry and spin-coupling \cite{presentLiquidDropModel1991, Bertulani2007}. 

For heavy nuclei with mass number $A \gtrsim 150$, the fission activation energy is typically $E_A\sim \mathcal{O}(5 \, {\rm MeV})$, which is less than the individual nucleon separation energy $E_b/A$ or alpha particle separation energy $E_{\alpha \, {\rm sep}}$ \cite{WongBook}. Generally, for a nucleus such as ${\rm U}^{235}$, we have $E_A < E_b/A < E_{\alpha \, {\rm sep}}$, so the least energetically costly way to disrupt the nucleus is to induce fission. In comparison, a framework to model the splitting of light nuclei like carbon or oxygen by tidal forces---a process which would not release energy---should be geared around nucleon or $\alpha$ emission, which have comparable separation energies for $A\lesssim 50$ \cite{WongBook}.

We first model the axially symmetric deformation of a heavy nucleus with atomic number $Z$ and mass number $A$ to lowest order as \cite{bohrMechanismNuclearFission1939, myersNuclearMassesDeformations1966}:
\begin{equation}
    R(\theta) = R_0\left[1 + \epsilon P_2(\cos\theta) + ... \right],
    \label{NucDef}
\end{equation}
where the ground state radius for $A\gtrsim 20$ is given by
\begin{equation}
    R_0(A) \simeq 1.128 \,A^{1/3} \, {\rm fm}.
\end{equation}

We express the total energy of the nucleus as a function of the dimensionless deformation parameter $\epsilon$, which quantifies the amplitude of the ellipsoidal deformation. We then compute the threshold impact parameter $b_{\rm th}^{A}(M)$ such that tidal forces from a PBH of mass $M$ can deform the nucleus enough to cross the fission energy barrier $E_A$ in this potential. Once the barrier is crossed, competition between the Coulomb force and the nuclear `surface tension' causes the nucleus to undergo fission---splitting into two asymmetric daughter fragments and releasing prompt neutrons, $\gamma$-rays, and about $200 \, {\rm MeV}$ of energy.

\begin{figure}[t]
    \centering
    \includegraphics[width=1.0\linewidth]{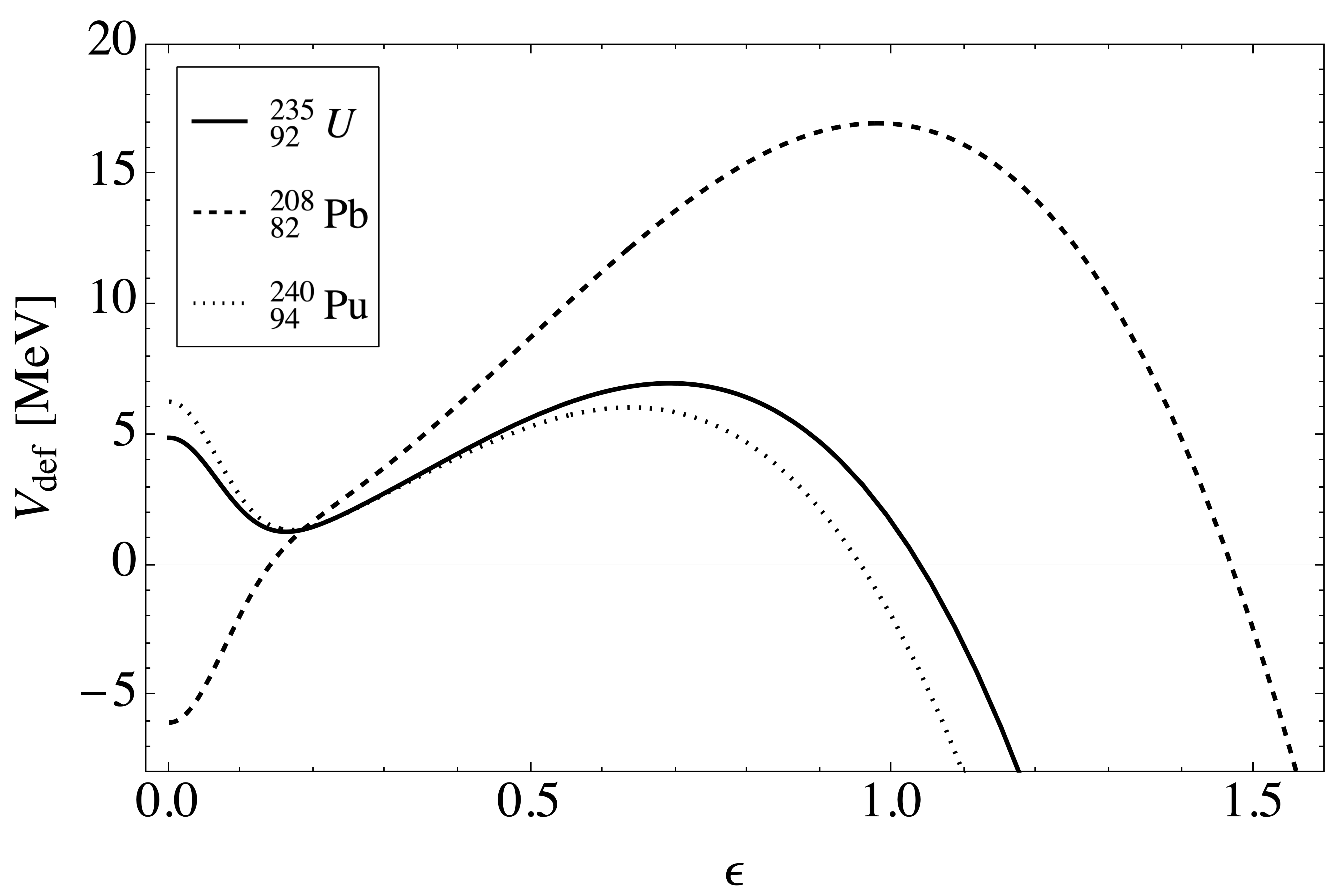}
    \caption{\justifying  Rescaled nuclear deformation potentials $\mathcal{E}_{Z,A}V_{\rm def}$ given by Eq.~(\ref{Vdef}) for three heavy isotopes. $V_{\rm def}$ is a function of the dimensionless deformation parameter $\epsilon$ from Eq.~(\ref{NucDef}). Note that $Z=82$ for lead is a magic number and thus the ground state is predicted to be spherical with $\epsilon=0$ at the minimum. The uranium and plutonium isotopes, by contrast, have low fission barriers and deformed ground states. Values of $E_A(Z,A)$ in $\mathcal{E}_{Z,A}$ are taken from Refs.~\cite{oberstedtExploringFissionBarrier2021a, myersNuclearMassesDeformations1966}.}
    \label{fig:Vdef}
\end{figure}

Following Ref.~\cite{myersNuclearMassesDeformations1966}, we parameterize the nuclear mass as
\begin{equation}
    \begin{split}
    \mathcal{M}(\epsilon,Z,A) = m_n&(A-Z) + m_p A \\
    &- E_b(Z,A) + \mathcal{E}_{Z,A}\,V_{\rm def}(\epsilon, Z,A)
    \label{NucMass}
    \end{split}
\end{equation}
where $E_b$ is the binding energy calculated from the liquid drop model,
\begin{equation}
    E_b(Z,A) = c_1 A - c_2 A^{2/3} - c_3 \frac{Z^2}{A^{1/3}}  + c_4 \frac{Z^2}{A} + \delta ,
\end{equation}
and $V_{\rm def}$ is the deformation potential, parameterized via
\begin{equation}
    \begin{split}
    V_{\rm def}(\epsilon, Z, A) = E&(Z, A)\epsilon^2 -F(Z,A)\epsilon^3 \\
    & + S(Z,A)\exp\left[\frac{-\epsilon^2}{\alpha_0(A)^2}\right].
    \end{split}
    \label{Vdef}
\end{equation}
See Eq.~(10) of Ref.~\cite{myersNuclearMassesDeformations1966} for the explicit forms of the functions $E(Z,A)$ and $F(Z,A)$ and note that we take the shape parameter $\gamma$=0 in their parameterization of $\mathcal{M}$. The third term of Eq.~(\ref{Vdef}) is a shell correction (see Sec.~3 of Ref.~\cite{myersNuclearMassesDeformations1966}), in which the shell function $S(Z,A)$ leads to spherical ground state configurations for magic number nuclei ($Z=$ 2, 8, 20, 28, 50, 82, 126...) and deformed ground states for non-magic nuclei. Fig.~\ref{fig:Vdef} plots the deformation potentials for $\ce{^{235}_{92}U}$, $\ce{^{208}_{82}Pb}$, and $\ce{^{240}_{94}Pu}$.

Because the ground states of non-magic nuclei are not spherically symmetric and have $\epsilon_{\rm min}\neq 0$, we define the threshold deformation as
\begin{equation}
    \epsilon_{\rm th}(Z,A) = \epsilon_{\rm max}(Z,A)-\epsilon_{\rm min}(Z,A),
\end{equation}
and the fission potential barrier from the model is
\begin{equation}
    E_A^{\rm model}(Z,A) = V_{\rm def}(\epsilon_{\rm max}, Z,A)-V_{\rm def}(\epsilon_{\rm min}, Z,A),
\end{equation}
where the values $\epsilon_{\rm max}$ and $\epsilon_{\rm min}$ that maximize and minimize the potential are found numerically from Eq.~(\ref{Vdef}). We introduce the parameter $\mathcal{E}_{Z,A}$ in Eq.~(\ref{NucMass}), defined as
\begin{equation}
    \mathcal{E}_{Z,A} = \frac{E_A(Z,A)}{E_A^{\rm model}(Z, A)},
\end{equation}
where $E_A(Z,A)$ is the reported experimental value for the fission barrier, in order to tune the potential to reproduce the accurate energy barrier. The overall rescaling of $V_{\rm def}$ by $\mathcal{E}_{Z,A}$ does not effect the values of $\epsilon_{\min}$ and $\epsilon_{\min}$.

Note that more complex modern models of isomeric fission give rise to potentials $V(\epsilon)$ with multiple wells \cite{bjornholmDoublehumpedFissionBarrier1980, brackFunnyHillsShellCorrection1972, lynnTheoreticalEvaluationsFission2003, wangNewCalculationsFivedimensional2019}, which admit metastable isomeric states and agree with experimental observations of shape isomers \cite{csigeExploringMultihumpedFission2013, oberstedtIdentificationShapeIsomer2007}. Here we assume a simpler potential $V(\epsilon)$ with just one minimum and tune the energy barrier to match reported experimental values via the parameter $\mathcal{E}_{Z,A}$.

The threshold impact parameter for a PBH transit capable of deforming a nucleus enough to induce fission must satisfy
\begin{equation}
    \frac{E_A(Z,A)}{2\epsilon_{\rm th}R_0} \simeq \frac{2 G M \mathcal{M}(\epsilon_{\rm max}, Z, A)}{(b_{\rm th}^{A})^3}\epsilon_{\rm th}R_0,
\end{equation}
which gives
\begin{equation}
    b_{\rm th}^{\rm A}(M) = \left(\frac{4 G M \mathcal{M}(\epsilon_{\rm max}, Z, A)}{E_A(Z,A)}\epsilon_{\rm th}^2R_0^2\right)^{1/3}.
\end{equation}

As in the previous sections, the maximum PBH mass capable of gravitationally-induced fission will satisfy
\begin{equation}
    b_{\rm th}^{\rm A} (M_{\rm max}^{\rm A}) = r_s(M_{\rm max}^{\rm A}).
\end{equation}
Specializing to \ce{^{235}_{92}U}, with reported fission activation energy $E_{\rm U}=(5.7 \pm 0.6) \, {\rm MeV}$ \cite{oberstedtExploringFissionBarrier2021a}, we find $\epsilon_{\rm min}=0.16$, $\epsilon_{\rm max}=0.69$. See Fig~\ref{fig:Vdef}. The maximum PBH mass that can induce fission of \ce{^{235}_{92}U} is therefore
\begin{equation}
    M_{\rm max}^{\rm U} = 7.31\times10^{17} \, {\rm g},
\end{equation}
with a Hawking temperature of $T_H = 14.4 \, {\rm keV}$ and a typical photon emission energy of $E_{\rm peak}= 87.0 \, {\rm keV}$.

\begin{figure}[t]
    \centering
    \includegraphics[width=1.0\linewidth]{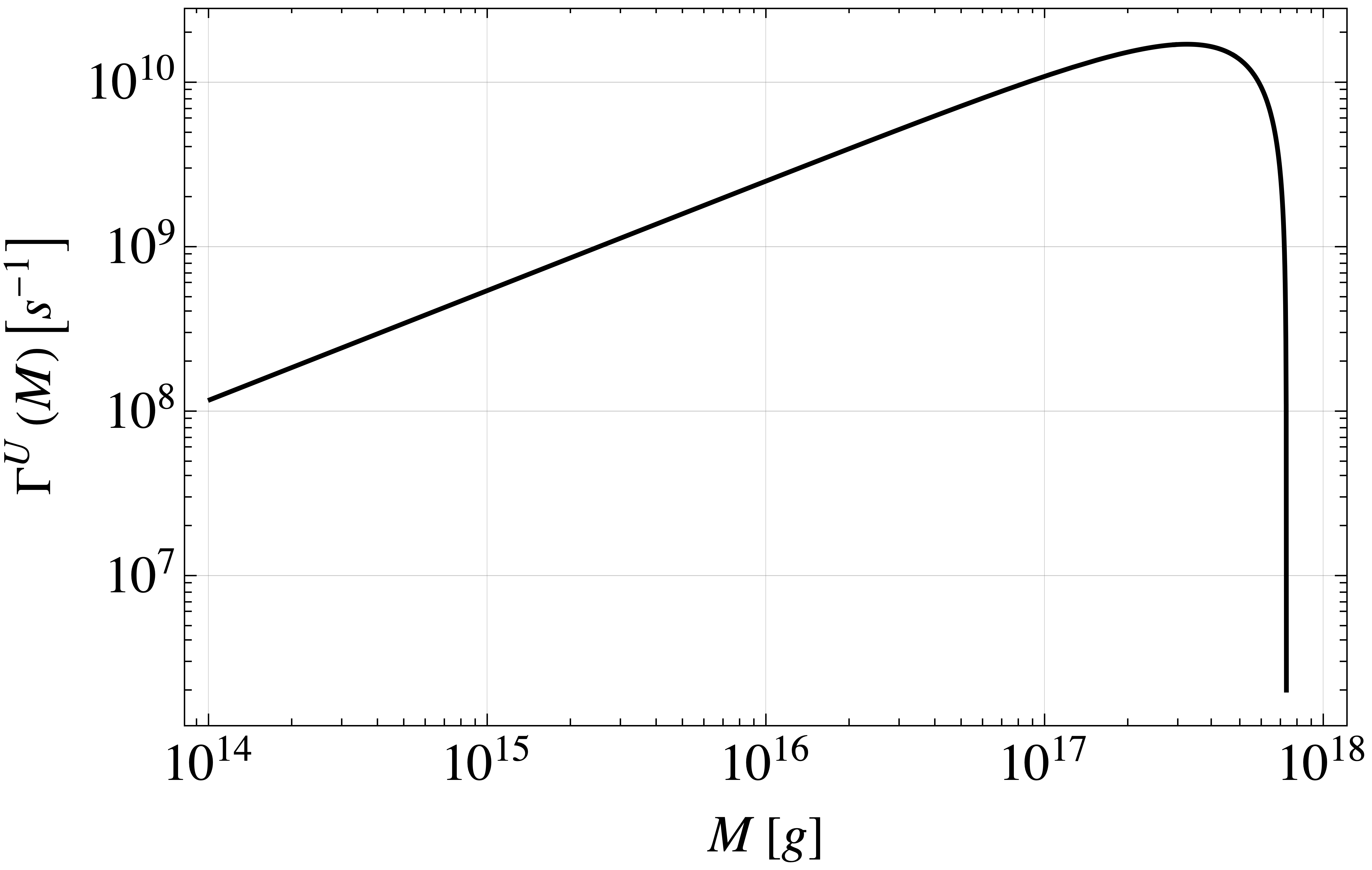}
    \caption{\justifying  Rate of gravitationally-induced nuclear fission $\Gamma^{\rm U}(M)$ from Eq.~(\ref{GammaU}) for a PBH incident upon a volume of \ce{^{235}_{92}U} with present-day velocity $v_{\rm rel} = 246 \, {\rm km/s}$. The rate is maximized for an asteroid-mass PBH of mass $M_{\rm peak}^{\rm U}=4.18\times10^{17} \, {\rm g}$. }
    \label{fig:Urate}
\end{figure}

For a PBH transiting through a volume of \ce{^{235}_{92}U} with density $\rho_{\rm U}=19.1 \, {\rm g \, cm}^{-3}$, we expect a fission rate of 
\begin{equation}
    \Gamma^{\rm U}(M) = \pi n_{\rm U} \, v_{\rm rel} \big[ b_{\rm th}^{\rm U}(M)^2-r_s(M)^2 \big],
    \label{GammaU}
\end{equation}
where $n_{\rm U} = \rho_{\rm U}/\mathcal{M}(\epsilon_{\rm min}, 92, 235)$. See Fig.~\ref{fig:Urate} for a plot of $\Gamma^{\rm U} (M)$ evaluated with $v_{\rm rel} = 246\,{\rm km/s}$ for typical present-day PBH velocities near Earth. The PBH mass which maximizes $\Gamma^{\rm U}(M)$ is
\begin{equation}
    M_{\rm peak}^{\rm U} = 3.21\times10^{17} \, {\rm g},
\end{equation}
with a Hawking temperature of $T_H = 32.8 \, {\rm keV}$ and a typical photon emission energy of $E_{\rm peak}= 198 \, {\rm keV}$. Notably, $M_{\rm peak}^{\rm U}$ falls within the unconstrained asteroid-mass range and is cold enough to only emit photons and neutrinos at energies well below the scale of $E_A$. 

The maximum fission rate for $v_{\rm rel} = 246 \, {\rm km/s}$ is 
\begin{equation}
    \Gamma^{\rm U}(M_{\rm peak}^{\rm U}) = 1.7\times10^{10} \, {\rm s}^{-1},
\end{equation}
which corresponds to an energy release rate of about $3.4\times10^{9} \, {\rm GeV s}^{-1}$ and the generation of $4.3\times10^{10}$ prompt neutrons per second (assuming an average of 2.5 neutrons per \ce{^{235}_{92}U} fission \cite{Bertulani2007}), which could induce further fission of the material in a chain reaction. For a PBH traveling roughly $10 \, {\rm cm}$ through a lump of enriched uranium, this corresponds to around $700$ fission events and the release of about
$140 \, {\rm GeV}$ of energy and a few thousand prompt neutrons.

The case of a PBH traversing a volume of Uranium and causing it to potentially explode is an interesting exercise, but a very unlikely event \cite{WheelerBook}. 
However, PBHs in the galaxy would have non-trivial impact rates with stars, including white dwarfs and neutron stars \cite{Capela:2012jz,Capela:2013yf,Lehmann:2020yxb,Genolini:2020ejw,Lehmann:2022vdt,Esser:2022owk,Caplan:2023ddo,Santarelli:2024uqx,Baumgarte:2024mei,Baumgarte:2024buu,Baumgarte:2024iby,DeLorenci:2024xez,DeLorenci:2025wbn}. A recent analysis even suggests that white dwarfs might be triggered to explode as supernovae via spontaneous fission of uranium within their stellar cores \cite{Horowitz:2021nlr}. Whether such reactions might themselves be catalyzed by passage of a PBH remains an intriguing possibility.

\section{Discussion}
\label{sec:Discussion}

Asteroid-mass PBHs are interesting as theoretical objects in their own right, for their potential role in the early universe, and as a candidate which could comprise most or all of the dark matter today. It is difficult to directly detect such PBHs, with masses $10^{17} \, {\rm g} \lesssim M \lesssim 10^{23} \, {\rm g}$, due to their cold temperatures and small radii. Constraining asteroid-mass PBHs will therefore require the development of new observables. The effects of gravitational tidal forces from Schwarzschild PBHs on matter provide intriguing potential observables that would be \emph{unique} to PBHs and distinguish them from macroscopic objects of the same mass. In this work we have focused on Schwarzschild PBHs, because asteroid-mass PBHs are unlikely to retain any spin or Standard Model charge for long timescales after formation. We investigated three phenomena caused by tidal forces during a PBH transit through a medium: ionization of neutral hydrogen, dissociation of deuterons, and fission of heavy nuclei. 

We find that Schwarzschild PBHs with masses $M\leq M_{\rm max}^{\rm H}=2.43\times10^{22}\,{\rm g}$ are capable of gravitationally ionizing neutral hydrogen atoms. Lyman-$\alpha$ photon emission rates from present-day gravitational ionization of the IPM or ISM by a PBH transit are far too low to be detectable. Furthermore, even after recombination ends at $z\simeq1090$, when neutral hydrogen density is maximized, gravitational ionization rates for realistic PBHs populations would be dominated by Hawking emission photoionization rates. We do find however, that \emph{total energy deposition} to the neutral hydrogen medium after recombination by a population of PBHs with typical mass $\bar{M}\gtrsim 5\times10^{21} \, {\rm g}$ is dominated by gravitational scattering rather than Hawking emission, across a wide range of PBH population model parameters $\{ \alpha, \beta \}$. This implies that gravitational interactions by PBHs, even those which do not ionize the medium, would be the main source of heating of neutral hydrogen after cosmic recombination for such populations. Whether such heating of the medium could be used to place new constraints on the PBH dark matter fraction in the asteroid-mass range remains the subject of further research.

We further demonstrate that gravitational dissociation of deuterons by PBHs is possible for $M\leq M_{\rm max}^{\rm D}=1.84\times10^{16}\, {\rm g}$, a regime which lies close to but below the asteroid-mass range. We focus on deuteron dissociation during BBN around $T_b=0.07 \, {\rm MeV}$ when deuteron densities were highest. We find that, for relative velocities between the PBHs and the baryonic matter such that $v_{\rm rel}> 0.084 \, {\rm km/s}$, gravitational dissociation rates can significantly dominate photodissociation rates from Hawking radiation. Deuteron density is a sensitive probe of the dynamics and contents of the universe during BBN, so possible implications for the disruption of deuteron formation by PBHs remains a topic for future study. 

The last phenomenon we investigate is fission of heavy elements due to tidal deformations induced by a PBH transit. Using a model of the nuclear potential as a function of deformation parameter $\epsilon$, we find that PBHs with $M\leq M_{\rm max}^{\rm U} = 7.31\times10^{17}\, {\rm g}$ could tidally deform a \ce{^{235}_{92}U} nucleus enough to cross the fission energy barrier during a transit at typical speeds expected of PBHs near Earth today. Though encounters between asteroid-mass PBHs and material on Earth would be exceedingly rare, nuclear dissociation by PBHs has interesting implications for collisions between PBHs and stars and white dwarfs, which are expected to occur throughout our galaxy.

\section*{Acknowledgments}

It is a pleasure to thank Michael Baker, Thomas Baumgarte, Peter Fisher, Joaquim Iguaz Juan, Benjamin Lehmann, Stefano Profumo, Aidan Symons, Andrea Thamm, and Rainer Weiss for helpful discussions. Portions of this research were conducted in MIT's Center for Theoretical Physics --- A Leinweber Institute and supported by the Office of High Energy Physics within the Office of Science of the U.S.~Department of Energy under grant Contract Number DE-SC0012567. We also gratefully acknowledge support from the Amar G.~Bose Research Grant Program at MIT.


\begin{thebibliography}{122}%
\makeatletter
\providecommand \@ifxundefined [1]{%
 \@ifx{#1\undefined}
}%
\providecommand \@ifnum [1]{%
 \ifnum #1\expandafter \@firstoftwo
 \else \expandafter \@secondoftwo
 \fi
}%
\providecommand \@ifx [1]{%
 \ifx #1\expandafter \@firstoftwo
 \else \expandafter \@secondoftwo
 \fi
}%
\providecommand \natexlab [1]{#1}%
\providecommand \enquote  [1]{``#1''}%
\providecommand \bibnamefont  [1]{#1}%
\providecommand \bibfnamefont [1]{#1}%
\providecommand \citenamefont [1]{#1}%
\providecommand \href@noop [0]{\@secondoftwo}%
\providecommand \href [0]{\begingroup \@sanitize@url \@href}%
\providecommand \@href[1]{\@@startlink{#1}\@@href}%
\providecommand \@@href[1]{\endgroup#1\@@endlink}%
\providecommand \@sanitize@url [0]{\catcode `\\12\catcode `\$12\catcode `\&12\catcode `\#12\catcode `\^12\catcode `\_12\catcode `\%12\relax}%
\providecommand \@@startlink[1]{}%
\providecommand \@@endlink[0]{}%
\providecommand \url  [0]{\begingroup\@sanitize@url \@url }%
\providecommand \@url [1]{\endgroup\@href {#1}{\urlprefix }}%
\providecommand \urlprefix  [0]{URL }%
\providecommand \Eprint [0]{\href }%
\providecommand \doibase [0]{http://dx.doi.org/}%
\providecommand \selectlanguage [0]{\@gobble}%
\providecommand \bibinfo  [0]{\@secondoftwo}%
\providecommand \bibfield  [0]{\@secondoftwo}%
\providecommand \translation [1]{[#1]}%
\providecommand \BibitemOpen [0]{}%
\providecommand \bibitemStop [0]{}%
\providecommand \bibitemNoStop [0]{.\EOS\space}%
\providecommand \EOS [0]{\spacefactor3000\relax}%
\providecommand \BibitemShut  [1]{\csname bibitem#1\endcsname}%
\let\auto@bib@innerbib\@empty
\bibitem [{\citenamefont {Zel'dovich}\ and\ \citenamefont {Novikov}(1966)}]{zeldovichHypothesisCoresRetarded1966}%
  \BibitemOpen
  \bibfield  {author} {\bibinfo {author} {\bibfnamefont {{\relax Ya}.~B.}\ \bibnamefont {Zel'dovich}}\ and\ \bibinfo {author} {\bibfnamefont {I.~D.}\ \bibnamefont {Novikov}},\ }\bibfield  {title} {\enquote {\bibinfo {title} {The {{Hypothesis}} of {{Cores Retarded}} during {{Expansion}} and the {{Hot Cosmological Model}}},}\ }\href@noop {} {\bibfield  {journal} {\bibinfo  {journal} {Astron. Zhurnal}\ }\textbf {\bibinfo {volume} {43}},\ \bibinfo {pages} {758} (\bibinfo {year} {1966})}\BibitemShut {NoStop}%
\bibitem [{\citenamefont {Hawking}(1971)}]{hawkingGravitationallyCollapsedObjects1971}%
  \BibitemOpen
  \bibfield  {author} {\bibinfo {author} {\bibfnamefont {Stephen}\ \bibnamefont {Hawking}},\ }\bibfield  {title} {\enquote {\bibinfo {title} {Gravitationally {{Collapsed Objects}} of {{Very Low Mass}}},}\ }\href {\doibase 10.1093/mnras/152.1.75} {\bibfield  {journal} {\bibinfo  {journal} {Mon. Not. R. Astron. Soc.}\ }\textbf {\bibinfo {volume} {152}},\ \bibinfo {pages} {75--78} (\bibinfo {year} {1971})}\BibitemShut {NoStop}%
\bibitem [{\citenamefont {Carr}\ and\ \citenamefont {Hawking}(1974)}]{carrBlackHolesEarly1974}%
  \BibitemOpen
  \bibfield  {author} {\bibinfo {author} {\bibfnamefont {B.~J.}\ \bibnamefont {Carr}}\ and\ \bibinfo {author} {\bibfnamefont {S.~W.}\ \bibnamefont {Hawking}},\ }\bibfield  {title} {\enquote {\bibinfo {title} {Black {{Holes}} in the {{Early Universe}}},}\ }\href {\doibase 10.1093/mnras/168.2.399} {\bibfield  {journal} {\bibinfo  {journal} {Mon. Not. R. Astron. Soc.}\ }\textbf {\bibinfo {volume} {168}},\ \bibinfo {pages} {399--415} (\bibinfo {year} {1974})}\BibitemShut {NoStop}%
\bibitem [{\citenamefont {Khlopov}(2010)}]{Khlopov:2008qy}%
  \BibitemOpen
  \bibfield  {author} {\bibinfo {author} {\bibfnamefont {Maxim~Yu.}\ \bibnamefont {Khlopov}},\ }\bibfield  {title} {\enquote {\bibinfo {title} {{Primordial Black Holes}},}\ }\href {\doibase 10.1088/1674-4527/10/6/001} {\bibfield  {journal} {\bibinfo  {journal} {Res. Astron. Astrophys.}\ }\textbf {\bibinfo {volume} {10}},\ \bibinfo {pages} {495--528} (\bibinfo {year} {2010})},\ \Eprint {http://arxiv.org/abs/0801.0116} {arXiv:0801.0116 [astro-ph]} \BibitemShut {NoStop}%
\bibitem [{\citenamefont {Carr}\ and\ \citenamefont {K{\"u}hnel}(2020)}]{carrPrimordialBlackHoles2020}%
  \BibitemOpen
  \bibfield  {author} {\bibinfo {author} {\bibfnamefont {Bernard}\ \bibnamefont {Carr}}\ and\ \bibinfo {author} {\bibfnamefont {Florian}\ \bibnamefont {K{\"u}hnel}},\ }\bibfield  {title} {\enquote {\bibinfo {title} {Primordial {{Black Holes}} as {{Dark Matter}}: {{Recent Developments}}},}\ }\href {\doibase 10.1146/annurev-nucl-050520-125911} {\bibfield  {journal} {\bibinfo  {journal} {Annu. Rev. Nucl. Part. Sci.}\ }\textbf {\bibinfo {volume} {70}},\ \bibinfo {pages} {355--394} (\bibinfo {year} {2020})}\BibitemShut {NoStop}%
\bibitem [{\citenamefont {Green}\ and\ \citenamefont {Kavanagh}(2021)}]{greenPrimordialBlackHoles2021}%
  \BibitemOpen
  \bibfield  {author} {\bibinfo {author} {\bibfnamefont {Anne~M.}\ \bibnamefont {Green}}\ and\ \bibinfo {author} {\bibfnamefont {Bradley~J.}\ \bibnamefont {Kavanagh}},\ }\bibfield  {title} {\enquote {\bibinfo {title} {Primordial black holes as a dark matter candidate},}\ }\href {\doibase 10.1088/1361-6471/abc534} {\bibfield  {journal} {\bibinfo  {journal} {J. Phys. G: Nucl. Part. Phys.}\ }\textbf {\bibinfo {volume} {48}},\ \bibinfo {pages} {043001} (\bibinfo {year} {2021})}\BibitemShut {NoStop}%
\bibitem [{\citenamefont {Carr}\ \emph {et~al.}(2024)\citenamefont {Carr}, \citenamefont {Clesse}, \citenamefont {{Garcia-Bellido}}, \citenamefont {Hawkins},\ and\ \citenamefont {Kuhnel}}]{carrObservationalEvidencePrimordial2024}%
  \BibitemOpen
  \bibfield  {author} {\bibinfo {author} {\bibfnamefont {Bernard}\ \bibnamefont {Carr}}, \bibinfo {author} {\bibfnamefont {Sebastien}\ \bibnamefont {Clesse}}, \bibinfo {author} {\bibfnamefont {Juan}\ \bibnamefont {{Garcia-Bellido}}}, \bibinfo {author} {\bibfnamefont {Michael}\ \bibnamefont {Hawkins}}, \ and\ \bibinfo {author} {\bibfnamefont {Florian}\ \bibnamefont {Kuhnel}},\ }\bibfield  {title} {\enquote {\bibinfo {title} {Observational {{Evidence}} for {{Primordial Black Holes}}: {{A Positivist Perspective}}},}\ }\href {\doibase 10.1016/j.physrep.2023.11.005} {\bibfield  {journal} {\bibinfo  {journal} {Phys. Rep.}\ }\textbf {\bibinfo {volume} {1054}},\ \bibinfo {pages} {1--68} (\bibinfo {year} {2024})},\ \Eprint {http://arxiv.org/abs/2306.03903} {arXiv:2306.03903 [astro-ph, physics:gr-qc, physics:hep-ph]} \BibitemShut {NoStop}%
\bibitem [{\citenamefont {Escriv{\`a}}\ \emph {et~al.}(2024)\citenamefont {Escriv{\`a}}, \citenamefont {Kuhnel},\ and\ \citenamefont {Tada}}]{escrivaPrimordialBlackHoles2024}%
  \BibitemOpen
  \bibfield  {author} {\bibinfo {author} {\bibfnamefont {Albert}\ \bibnamefont {Escriv{\`a}}}, \bibinfo {author} {\bibfnamefont {Florian}\ \bibnamefont {Kuhnel}}, \ and\ \bibinfo {author} {\bibfnamefont {Yuichiro}\ \bibnamefont {Tada}},\ }\bibfield  {title} {\enquote {\bibinfo {title} {Primordial {{Black Holes}}},}\ \ }(\bibinfo {year} {2024})\ pp.\ \bibinfo {pages} {261--377},\ \Eprint {http://arxiv.org/abs/2211.05767} {arXiv:2211.05767 [astro-ph, physics:gr-qc, physics:hep-ph, physics:hep-th]} \BibitemShut {NoStop}%
\bibitem [{\citenamefont {Klipfel}\ \emph {et~al.}(2025)\citenamefont {Klipfel}, \citenamefont {Fisher},\ and\ \citenamefont {Kaiser}}]{Klipfel:2025bvh}%
  \BibitemOpen
  \bibfield  {author} {\bibinfo {author} {\bibfnamefont {Alexandra~P.}\ \bibnamefont {Klipfel}}, \bibinfo {author} {\bibfnamefont {Peter}\ \bibnamefont {Fisher}}, \ and\ \bibinfo {author} {\bibfnamefont {David~I.}\ \bibnamefont {Kaiser}},\ }\bibfield  {title} {\enquote {\bibinfo {title} {{Hawking radiation signatures from primordial black holes transiting the inner Solar System: Prospects for detection}},}\ }\href {\doibase 10.1103/9jyp-24sw} {\bibfield  {journal} {\bibinfo  {journal} {Phys. Rev. D}\ }\textbf {\bibinfo {volume} {112}},\ \bibinfo {pages} {103007} (\bibinfo {year} {2025})},\ \Eprint {http://arxiv.org/abs/2506.14041} {arXiv:2506.14041 [astro-ph.CO]} \BibitemShut {NoStop}%
\bibitem [{\citenamefont {Carr}\ \emph {et~al.}(2016)\citenamefont {Carr}, \citenamefont {Kohri}, \citenamefont {Sendouda},\ and\ \citenamefont {Yokoyama}}]{carrConstraintsPrimordialBlack2016}%
  \BibitemOpen
  \bibfield  {author} {\bibinfo {author} {\bibfnamefont {B.~J.}\ \bibnamefont {Carr}}, \bibinfo {author} {\bibfnamefont {Kazunori}\ \bibnamefont {Kohri}}, \bibinfo {author} {\bibfnamefont {Yuuiti}\ \bibnamefont {Sendouda}}, \ and\ \bibinfo {author} {\bibfnamefont {Jun'ichi}\ \bibnamefont {Yokoyama}},\ }\bibfield  {title} {\enquote {\bibinfo {title} {Constraints on primordial black holes from {{Galactic}} gamma-ray background},}\ }\href {\doibase 10.1103/PhysRevD.94.044029} {\bibfield  {journal} {\bibinfo  {journal} {Phys. Rev. D}\ }\textbf {\bibinfo {volume} {94}},\ \bibinfo {pages} {044029} (\bibinfo {year} {2016})},\ \Eprint {http://arxiv.org/abs/1604.05349} {arXiv:1604.05349 [astro-ph]} \BibitemShut {NoStop}%
\bibitem [{\citenamefont {Smyth}\ \emph {et~al.}(2020)\citenamefont {Smyth}, \citenamefont {Profumo}, \citenamefont {English}, \citenamefont {Jeltema}, \citenamefont {McKinnon},\ and\ \citenamefont {Guhathakurta}}]{Smyth:2019whb}%
  \BibitemOpen
  \bibfield  {author} {\bibinfo {author} {\bibfnamefont {Nolan}\ \bibnamefont {Smyth}}, \bibinfo {author} {\bibfnamefont {Stefano}\ \bibnamefont {Profumo}}, \bibinfo {author} {\bibfnamefont {Samuel}\ \bibnamefont {English}}, \bibinfo {author} {\bibfnamefont {Tesla}\ \bibnamefont {Jeltema}}, \bibinfo {author} {\bibfnamefont {Kevin}\ \bibnamefont {McKinnon}}, \ and\ \bibinfo {author} {\bibfnamefont {Puragra}\ \bibnamefont {Guhathakurta}},\ }\bibfield  {title} {\enquote {\bibinfo {title} {{Updated Constraints on Asteroid-Mass Primordial Black Holes as Dark Matter}},}\ }\href {\doibase 10.1103/PhysRevD.101.063005} {\bibfield  {journal} {\bibinfo  {journal} {Phys. Rev. D}\ }\textbf {\bibinfo {volume} {101}},\ \bibinfo {pages} {063005} (\bibinfo {year} {2020})},\ \Eprint {http://arxiv.org/abs/1910.01285} {arXiv:1910.01285 [astro-ph.CO]} \BibitemShut {NoStop}%
\bibitem [{\citenamefont {Carr}\ \emph {et~al.}(2021)\citenamefont {Carr}, \citenamefont {Kohri}, \citenamefont {Sendouda},\ and\ \citenamefont {Yokoyama}}]{carrConstraintsPrimordialBlack2021}%
  \BibitemOpen
  \bibfield  {author} {\bibinfo {author} {\bibfnamefont {Bernard}\ \bibnamefont {Carr}}, \bibinfo {author} {\bibfnamefont {Kazunori}\ \bibnamefont {Kohri}}, \bibinfo {author} {\bibfnamefont {Yuuiti}\ \bibnamefont {Sendouda}}, \ and\ \bibinfo {author} {\bibfnamefont {Jun'ichi}\ \bibnamefont {Yokoyama}},\ }\bibfield  {title} {\enquote {\bibinfo {title} {Constraints on primordial black holes},}\ }\href {\doibase 10.1088/1361-6633/ac1e31} {\bibfield  {journal} {\bibinfo  {journal} {Rep. Prog. Phys.}\ }\textbf {\bibinfo {volume} {84}},\ \bibinfo {pages} {116902} (\bibinfo {year} {2021})}\BibitemShut {NoStop}%
\bibitem [{\citenamefont {Gorton}\ and\ \citenamefont {Green}(2024)}]{gortonHowOpenAsteroidmass2024}%
  \BibitemOpen
  \bibfield  {author} {\bibinfo {author} {\bibfnamefont {Matthew}\ \bibnamefont {Gorton}}\ and\ \bibinfo {author} {\bibfnamefont {Anne~M.}\ \bibnamefont {Green}},\ }\href@noop {} {\enquote {\bibinfo {title} {How open is the asteroid-mass primordial black hole window?}}\ } (\bibinfo {year} {2024}),\ \Eprint {http://arxiv.org/abs/2403.03839} {arXiv:2403.03839 [astro-ph, physics:hep-ph]} \BibitemShut {NoStop}%
\bibitem [{\citenamefont {De~la Torre~Luque}\ \emph {et~al.}(2024)\citenamefont {De~la Torre~Luque}, \citenamefont {Koechler},\ and\ \citenamefont {Balaji}}]{DelaTorreLuque:2024qms}%
  \BibitemOpen
  \bibfield  {author} {\bibinfo {author} {\bibfnamefont {Pedro}\ \bibnamefont {De~la Torre~Luque}}, \bibinfo {author} {\bibfnamefont {Jordan}\ \bibnamefont {Koechler}}, \ and\ \bibinfo {author} {\bibfnamefont {Shyam}\ \bibnamefont {Balaji}},\ }\bibfield  {title} {\enquote {\bibinfo {title} {{Refining Galactic primordial black hole evaporation constraints}},}\ }\href {\doibase 10.1103/PhysRevD.110.123022} {\bibfield  {journal} {\bibinfo  {journal} {Phys. Rev. D}\ }\textbf {\bibinfo {volume} {110}},\ \bibinfo {pages} {123022} (\bibinfo {year} {2024})},\ \bibinfo {note} {[Erratum: Phys.Rev.D 112, 109904 (2025)]},\ \Eprint {http://arxiv.org/abs/2406.11949} {arXiv:2406.11949 [astro-ph.HE]} \BibitemShut {NoStop}%
\bibitem [{\citenamefont {Dror}\ \emph {et~al.}(2019)\citenamefont {Dror}, \citenamefont {Ramani}, \citenamefont {Trickle},\ and\ \citenamefont {Zurek}}]{Dror:2019twh}%
  \BibitemOpen
  \bibfield  {author} {\bibinfo {author} {\bibfnamefont {Jeff~A.}\ \bibnamefont {Dror}}, \bibinfo {author} {\bibfnamefont {Harikrishnan}\ \bibnamefont {Ramani}}, \bibinfo {author} {\bibfnamefont {Tanner}\ \bibnamefont {Trickle}}, \ and\ \bibinfo {author} {\bibfnamefont {Kathryn~M.}\ \bibnamefont {Zurek}},\ }\bibfield  {title} {\enquote {\bibinfo {title} {{Pulsar Timing Probes of Primordial Black Holes and Subhalos}},}\ }\href {\doibase 10.1103/PhysRevD.100.023003} {\bibfield  {journal} {\bibinfo  {journal} {Phys. Rev. D}\ }\textbf {\bibinfo {volume} {100}},\ \bibinfo {pages} {023003} (\bibinfo {year} {2019})},\ \Eprint {http://arxiv.org/abs/1901.04490} {arXiv:1901.04490 [astro-ph.CO]} \BibitemShut {NoStop}%
\bibitem [{\citenamefont {Li}\ \emph {et~al.}(2023)\citenamefont {Li}, \citenamefont {Huang}, \citenamefont {Huang},\ and\ \citenamefont {Shu}}]{Li:2022oqo}%
  \BibitemOpen
  \bibfield  {author} {\bibinfo {author} {\bibfnamefont {Ya-Ling}\ \bibnamefont {Li}}, \bibinfo {author} {\bibfnamefont {Guo-Qing}\ \bibnamefont {Huang}}, \bibinfo {author} {\bibfnamefont {Zong-Qiang}\ \bibnamefont {Huang}}, \ and\ \bibinfo {author} {\bibfnamefont {Fu-Wen}\ \bibnamefont {Shu}},\ }\bibfield  {title} {\enquote {\bibinfo {title} {{Detecting sublunar-mass primordial black holes with the Earth-Moon binary system}},}\ }\href {\doibase 10.1103/PhysRevD.107.084019} {\bibfield  {journal} {\bibinfo  {journal} {Phys. Rev. D}\ }\textbf {\bibinfo {volume} {107}},\ \bibinfo {pages} {084019} (\bibinfo {year} {2023})},\ \Eprint {http://arxiv.org/abs/2209.12415} {arXiv:2209.12415 [gr-qc]} \BibitemShut {NoStop}%
\bibitem [{\citenamefont {Tran}\ \emph {et~al.}(2024)\citenamefont {Tran}, \citenamefont {Geller}, \citenamefont {Lehmann},\ and\ \citenamefont {Kaiser}}]{tranCloseEncountersPrimordial2024}%
  \BibitemOpen
  \bibfield  {author} {\bibinfo {author} {\bibfnamefont {Tung~X.}\ \bibnamefont {Tran}}, \bibinfo {author} {\bibfnamefont {Sarah~R.}\ \bibnamefont {Geller}}, \bibinfo {author} {\bibfnamefont {Benjamin~V.}\ \bibnamefont {Lehmann}}, \ and\ \bibinfo {author} {\bibfnamefont {David~I.}\ \bibnamefont {Kaiser}},\ }\bibfield  {title} {\enquote {\bibinfo {title} {Close encounters of the primordial kind: A new observable for primordial black holes as dark matter},}\ }\href {\doibase 10.1103/PhysRevD.110.063533} {\bibfield  {journal} {\bibinfo  {journal} {Phys. Rev. D}\ }\textbf {\bibinfo {volume} {110}},\ \bibinfo {pages} {063533} (\bibinfo {year} {2024})},\ \Eprint {http://arxiv.org/abs/2312.17217} {arXiv:2312.17217 [astro-ph]} \BibitemShut {NoStop}%
\bibitem [{\citenamefont {{Cuadrat-Grzybowski}}\ \emph {et~al.}(2024)\citenamefont {{Cuadrat-Grzybowski}}, \citenamefont {Clesse}, \citenamefont {Defraigne}, \citenamefont {Van~Camp},\ and\ \citenamefont {Bertrand}}]{cuadrat-grzybowskiProbingPrimordialBlack2024}%
  \BibitemOpen
  \bibfield  {author} {\bibinfo {author} {\bibfnamefont {Michal}\ \bibnamefont {{Cuadrat-Grzybowski}}}, \bibinfo {author} {\bibfnamefont {S{\'e}bastien}\ \bibnamefont {Clesse}}, \bibinfo {author} {\bibfnamefont {Pascale}\ \bibnamefont {Defraigne}}, \bibinfo {author} {\bibfnamefont {Michel}\ \bibnamefont {Van~Camp}}, \ and\ \bibinfo {author} {\bibfnamefont {Bruno}\ \bibnamefont {Bertrand}},\ }\href@noop {} {\enquote {\bibinfo {title} {Probing {{Primordial Black Holes}} and {{Dark Matter Clumps}} in the {{Solar System}} with {{Gravimeter}} and {{GNSS Networks}}},}\ } (\bibinfo {year} {2024}),\ \Eprint {http://arxiv.org/abs/2403.14397} {arXiv:2403.14397 [astro-ph, physics:gr-qc, physics:hep-ph]} \BibitemShut {NoStop}%
\bibitem [{\citenamefont {Brown}\ \emph {et~al.}(2025)\citenamefont {Brown}, \citenamefont {He},\ and\ \citenamefont {Unwin}}]{Brown:2025awt}%
  \BibitemOpen
  \bibfield  {author} {\bibinfo {author} {\bibfnamefont {Garett}\ \bibnamefont {Brown}}, \bibinfo {author} {\bibfnamefont {Linda}\ \bibnamefont {He}}, \ and\ \bibinfo {author} {\bibfnamefont {James}\ \bibnamefont {Unwin}},\ }\bibfield  {title} {\enquote {\bibinfo {title} {{The Potential Impact of Primordial Black Holes on Exoplanet Systems}},}\ }\href {\doibase 10.33232/001c.146689} {\bibfield  {journal} {\bibinfo  {journal} {Open J. Astrophys.}\ }\textbf {\bibinfo {volume} {8}},\ \bibinfo {pages} {2025} (\bibinfo {year} {2025})},\ \Eprint {http://arxiv.org/abs/2507.05389} {arXiv:2507.05389 [astro-ph.GA]} \BibitemShut {NoStop}%
\bibitem [{\citenamefont {De~Lorenci}\ \emph {et~al.}(2025{\natexlab{a}})\citenamefont {De~Lorenci}, \citenamefont {Kaiser}, \citenamefont {Peter}, \citenamefont {Ruiz},\ and\ \citenamefont {Wolfe}}]{DeLorenci:2025wbn}%
  \BibitemOpen
  \bibfield  {author} {\bibinfo {author} {\bibfnamefont {Vitorio~A.}\ \bibnamefont {De~Lorenci}}, \bibinfo {author} {\bibfnamefont {David~I.}\ \bibnamefont {Kaiser}}, \bibinfo {author} {\bibfnamefont {Patrick}\ \bibnamefont {Peter}}, \bibinfo {author} {\bibfnamefont {Lucas~S.}\ \bibnamefont {Ruiz}}, \ and\ \bibinfo {author} {\bibfnamefont {Noah~E.}\ \bibnamefont {Wolfe}},\ }\bibfield  {title} {\enquote {\bibinfo {title} {{Gravitational wave signals from primordial black holes orbiting solar-type stars}},}\ }\href {\doibase 10.1103/294z-nfj4} {\bibfield  {journal} {\bibinfo  {journal} {Phys. Rev. D}\ }\textbf {\bibinfo {volume} {112}},\ \bibinfo {pages} {063063} (\bibinfo {year} {2025}{\natexlab{a}})},\ \Eprint {http://arxiv.org/abs/2504.07517} {arXiv:2504.07517 [gr-qc]} \BibitemShut {NoStop}%
\bibitem [{\citenamefont {Thoss}\ and\ \citenamefont {Loeb}(2025)}]{Thoss:2025yht}%
  \BibitemOpen
  \bibfield  {author} {\bibinfo {author} {\bibfnamefont {Valentin}\ \bibnamefont {Thoss}}\ and\ \bibinfo {author} {\bibfnamefont {Abraham}\ \bibnamefont {Loeb}},\ }\bibfield  {title} {\enquote {\bibinfo {title} {{Detecting dark objects in the Solar System with gravitational wave observatories}},}\ }\href {\doibase 10.1103/g3wh-hg4x} {\bibfield  {journal} {\bibinfo  {journal} {Phys. Rev. D}\ }\textbf {\bibinfo {volume} {112}},\ \bibinfo {pages} {083050} (\bibinfo {year} {2025})},\ \Eprint {http://arxiv.org/abs/2507.19577} {arXiv:2507.19577 [gr-qc]} \BibitemShut {NoStop}%
\bibitem [{\citenamefont {{Alonso-Monsalve}}\ and\ \citenamefont {Kaiser}(2024)}]{alonso-monsalvePrimordialBlackHoles2024}%
  \BibitemOpen
  \bibfield  {author} {\bibinfo {author} {\bibfnamefont {Elba}\ \bibnamefont {{Alonso-Monsalve}}}\ and\ \bibinfo {author} {\bibfnamefont {David~I.}\ \bibnamefont {Kaiser}},\ }\bibfield  {title} {\enquote {\bibinfo {title} {Primordial {{Black Holes}} with {{QCD Color Charge}}},}\ }\href {\doibase 10.1103/PhysRevLett.132.231402} {\bibfield  {journal} {\bibinfo  {journal} {Phys. Rev. Lett.}\ }\textbf {\bibinfo {volume} {132}},\ \bibinfo {pages} {231402} (\bibinfo {year} {2024})},\ \Eprint {http://arxiv.org/abs/2310.16877} {arXiv:2310.16877 [astro-ph, physics:gr-qc, physics:hep-ph, physics:hep-th]} \BibitemShut {NoStop}%
\bibitem [{\citenamefont {Baker}\ \emph {et~al.}(2025{\natexlab{a}})\citenamefont {Baker}, \citenamefont {Iguaz~Juan}, \citenamefont {Symons},\ and\ \citenamefont {Thamm}}]{Baker:2025zxm}%
  \BibitemOpen
  \bibfield  {author} {\bibinfo {author} {\bibfnamefont {Michael~J.}\ \bibnamefont {Baker}}, \bibinfo {author} {\bibfnamefont {Joaquim}\ \bibnamefont {Iguaz~Juan}}, \bibinfo {author} {\bibfnamefont {Aidan}\ \bibnamefont {Symons}}, \ and\ \bibinfo {author} {\bibfnamefont {Andrea}\ \bibnamefont {Thamm}},\ }\bibfield  {title} {\enquote {\bibinfo {title} {{Could We Observe an Exploding Black Hole in the Near Future?}}}\ }\href {\doibase 10.1103/nwgd-g3zl} {\bibfield  {journal} {\bibinfo  {journal} {Phys. Rev. Lett.}\ }\textbf {\bibinfo {volume} {135}},\ \bibinfo {pages} {111002} (\bibinfo {year} {2025}{\natexlab{a}})},\ \Eprint {http://arxiv.org/abs/2503.10755} {arXiv:2503.10755 [hep-ph]} \BibitemShut {NoStop}%
\bibitem [{\citenamefont {Santiago}\ \emph {et~al.}(2025)\citenamefont {Santiago}, \citenamefont {Feng}, \citenamefont {Schuster},\ and\ \citenamefont {Visser}}]{Santiago:2025rzb}%
  \BibitemOpen
  \bibfield  {author} {\bibinfo {author} {\bibfnamefont {Jessica}\ \bibnamefont {Santiago}}, \bibinfo {author} {\bibfnamefont {Justin}\ \bibnamefont {Feng}}, \bibinfo {author} {\bibfnamefont {Sebastian}\ \bibnamefont {Schuster}}, \ and\ \bibinfo {author} {\bibfnamefont {Matt}\ \bibnamefont {Visser}},\ }\bibfield  {title} {\enquote {\bibinfo {title} {{Immortality through the dark forces: Dark-charge primordial black holes as dark matter candidates}},}\ }\href@noop {} {\  (\bibinfo {year} {2025})},\ \Eprint {http://arxiv.org/abs/2503.20696} {arXiv:2503.20696 [gr-qc]} \BibitemShut {NoStop}%
\bibitem [{\citenamefont {Carter}(1974)}]{carterChargeParticleConservation1974}%
  \BibitemOpen
  \bibfield  {author} {\bibinfo {author} {\bibfnamefont {B.}~\bibnamefont {Carter}},\ }\bibfield  {title} {\enquote {\bibinfo {title} {Charge and {{Particle Conservation}} in {{Black-Hole Decay}}},}\ }\href {\doibase 10.1103/PhysRevLett.33.558} {\bibfield  {journal} {\bibinfo  {journal} {Phys. Rev. Lett.}\ }\textbf {\bibinfo {volume} {33}},\ \bibinfo {pages} {558--561} (\bibinfo {year} {1974})}\BibitemShut {NoStop}%
\bibitem [{\citenamefont {Gibbons}(1975)}]{gibbonsVacuumPolarizationSpontaneous1975}%
  \BibitemOpen
  \bibfield  {author} {\bibinfo {author} {\bibfnamefont {G.~W.}\ \bibnamefont {Gibbons}},\ }\bibfield  {title} {\enquote {\bibinfo {title} {Vacuum polarization and the spontaneous loss of charge by black holes},}\ }\href {\doibase 10.1007/BF01609829} {\bibfield  {journal} {\bibinfo  {journal} {Commun.Math. Phys.}\ }\textbf {\bibinfo {volume} {44}},\ \bibinfo {pages} {245--264} (\bibinfo {year} {1975})}\BibitemShut {NoStop}%
\bibitem [{\citenamefont {Page}(1976{\natexlab{a}})}]{pageParticleEmissionRates1976a}%
  \BibitemOpen
  \bibfield  {author} {\bibinfo {author} {\bibfnamefont {Don~N.}\ \bibnamefont {Page}},\ }\bibfield  {title} {\enquote {\bibinfo {title} {Particle emission rates from a black hole. {{II}}. {{Massless}} particles from a rotating hole},}\ }\href {\doibase 10.1103/PhysRevD.14.3260} {\bibfield  {journal} {\bibinfo  {journal} {Phys. Rev. D}\ }\textbf {\bibinfo {volume} {14}},\ \bibinfo {pages} {3260--3273} (\bibinfo {year} {1976}{\natexlab{a}})}\BibitemShut {NoStop}%
\bibitem [{\citenamefont {Chiba}\ and\ \citenamefont {Yokoyama}(2017)}]{chibaSpinDistributionPrimordial2017}%
  \BibitemOpen
  \bibfield  {author} {\bibinfo {author} {\bibfnamefont {Takeshi}\ \bibnamefont {Chiba}}\ and\ \bibinfo {author} {\bibfnamefont {Shuichiro}\ \bibnamefont {Yokoyama}},\ }\bibfield  {title} {\enquote {\bibinfo {title} {Spin {{Distribution}} of {{Primordial Black Holes}}},}\ }\href {\doibase 10.1093/ptep/ptx087} {\bibfield  {journal} {\bibinfo  {journal} {Prog. Theor. Exp. Phys.}\ }\textbf {\bibinfo {volume} {2017}} (\bibinfo {year} {2017}),\ 10.1093/ptep/ptx087},\ \Eprint {http://arxiv.org/abs/1704.06573} {arXiv:1704.06573 [astro-ph, physics:gr-qc]} \BibitemShut {NoStop}%
\bibitem [{\citenamefont {Jaraba}\ and\ \citenamefont {Garcia-Bellido}(2021)}]{Jaraba:2021ces}%
  \BibitemOpen
  \bibfield  {author} {\bibinfo {author} {\bibfnamefont {Santiago}\ \bibnamefont {Jaraba}}\ and\ \bibinfo {author} {\bibfnamefont {Juan}\ \bibnamefont {Garcia-Bellido}},\ }\bibfield  {title} {\enquote {\bibinfo {title} {{Black hole induced spins from hyperbolic encounters in dense clusters}},}\ }\href {\doibase 10.1016/j.dark.2021.100882} {\bibfield  {journal} {\bibinfo  {journal} {Phys. Dark Univ.}\ }\textbf {\bibinfo {volume} {34}},\ \bibinfo {pages} {100882} (\bibinfo {year} {2021})},\ \Eprint {http://arxiv.org/abs/2106.01436} {arXiv:2106.01436 [gr-qc]} \BibitemShut {NoStop}%
\bibitem [{\citenamefont {De~Luca}\ \emph {et~al.}(2020)\citenamefont {De~Luca}, \citenamefont {Franciolini}, \citenamefont {Pani},\ and\ \citenamefont {Riotto}}]{delucaEvolutionPrimordialBlack2020}%
  \BibitemOpen
  \bibfield  {author} {\bibinfo {author} {\bibfnamefont {V.}~\bibnamefont {De~Luca}}, \bibinfo {author} {\bibfnamefont {G.}~\bibnamefont {Franciolini}}, \bibinfo {author} {\bibfnamefont {P.}~\bibnamefont {Pani}}, \ and\ \bibinfo {author} {\bibfnamefont {A.}~\bibnamefont {Riotto}},\ }\bibfield  {title} {\enquote {\bibinfo {title} {The {{Evolution}} of {{Primordial Black Holes}} and their {{Final Observable Spins}}},}\ }\href {\doibase 10.1088/1475-7516/2020/04/052} {\bibfield  {journal} {\bibinfo  {journal} {J. Cosmol. Astropart. Phys.}\ }\textbf {\bibinfo {volume} {2020}},\ \bibinfo {pages} {052--052} (\bibinfo {year} {2020})},\ \Eprint {http://arxiv.org/abs/2003.02778} {arXiv:2003.02778 [astro-ph, physics:gr-qc, physics:hep-ph]} \BibitemShut {NoStop}%
\bibitem [{\citenamefont {Chongchitnan}\ and\ \citenamefont {Silk}(2021)}]{chongchitnanExtremeValueStatisticsSpin2021}%
  \BibitemOpen
  \bibfield  {author} {\bibinfo {author} {\bibfnamefont {Siri}\ \bibnamefont {Chongchitnan}}\ and\ \bibinfo {author} {\bibfnamefont {Joseph}\ \bibnamefont {Silk}},\ }\bibfield  {title} {\enquote {\bibinfo {title} {Extreme-{{Value Statistics}} of the {{Spin}} of {{Primordial Black Holes}}},}\ }\href {\doibase 10.1103/PhysRevD.104.083018} {\bibfield  {journal} {\bibinfo  {journal} {Phys. Rev. D}\ }\textbf {\bibinfo {volume} {104}},\ \bibinfo {pages} {083018} (\bibinfo {year} {2021})},\ \Eprint {http://arxiv.org/abs/2109.12268} {arXiv:2109.12268 [astro-ph]} \BibitemShut {NoStop}%
\bibitem [{\citenamefont {Rice}\ and\ \citenamefont {Zhang}(2017)}]{Rice:2017avg}%
  \BibitemOpen
  \bibfield  {author} {\bibinfo {author} {\bibfnamefont {Jared~R.}\ \bibnamefont {Rice}}\ and\ \bibinfo {author} {\bibfnamefont {Bing}\ \bibnamefont {Zhang}},\ }\bibfield  {title} {\enquote {\bibinfo {title} {{Cosmological evolution of primordial black holes}},}\ }\href {\doibase 10.1016/j.jheap.2017.02.002} {\bibfield  {journal} {\bibinfo  {journal} {JHEAp}\ }\textbf {\bibinfo {volume} {13-14}},\ \bibinfo {pages} {22--31} (\bibinfo {year} {2017})},\ \Eprint {http://arxiv.org/abs/1702.08069} {arXiv:1702.08069 [astro-ph.HE]} \BibitemShut {NoStop}%
\bibitem [{\citenamefont {Tseliakhovich}\ and\ \citenamefont {Hirata}(2010)}]{Tseliakhovich_2010}%
  \BibitemOpen
  \bibfield  {author} {\bibinfo {author} {\bibfnamefont {Dmitriy}\ \bibnamefont {Tseliakhovich}}\ and\ \bibinfo {author} {\bibfnamefont {Christopher}\ \bibnamefont {Hirata}},\ }\bibfield  {title} {\enquote {\bibinfo {title} {Relative velocity of dark matter and baryonic fluids and the formation of the first structures},}\ }\href {\doibase 10.1103/physrevd.82.083520} {\bibfield  {journal} {\bibinfo  {journal} {Phys.~Rev.~D}\ }\textbf {\bibinfo {volume} {82}} (\bibinfo {year} {2010}),\ 10.1103/physrevd.82.083520},\ \Eprint {http://arxiv.org/abs/1005.2416} {arXiv:1005.2416 [astro-ph]} \BibitemShut {NoStop}%
\bibitem [{\citenamefont {Jackson}(1962)}]{Jackson}%
  \BibitemOpen
  \bibfield  {author} {\bibinfo {author} {\bibfnamefont {J.~D.}\ \bibnamefont {Jackson}},\ }\href@noop {} {\emph {\bibinfo {title} {Classical Electrodynamics}}}\ (\bibinfo  {publisher} {Wiley},\ \bibinfo {address} {New York},\ \bibinfo {year} {1962})\BibitemShut {NoStop}%
\bibitem [{\citenamefont {Landau}(1944)}]{Landau:216256}%
  \BibitemOpen
  \bibfield  {author} {\bibinfo {author} {\bibfnamefont {Lev~Davidovich}\ \bibnamefont {Landau}},\ }\bibfield  {title} {\enquote {\bibinfo {title} {{On the energy loss of fast particles by ionization}},}\ }\href {https://cds.cern.ch/record/216256} {\bibfield  {journal} {\bibinfo  {journal} {J. Phys.}\ }\textbf {\bibinfo {volume} {8}},\ \bibinfo {pages} {201--205} (\bibinfo {year} {1944})}\BibitemShut {NoStop}%
\bibitem [{\citenamefont {Vavilov}(1957)}]{osti_4311507}%
  \BibitemOpen
  \bibfield  {author} {\bibinfo {author} {\bibfnamefont {P~V}\ \bibnamefont {Vavilov}},\ }\bibfield  {title} {\enquote {\bibinfo {title} {Ionization losses of high-energy heavy particles},}\ }\href {https://www.osti.gov/biblio/4311507} {\bibfield  {journal} {\bibinfo  {journal} {Soviet Phys. JETP}\ }\textbf {\bibinfo {volume} {Vol: 5}} (\bibinfo {year} {1957})}\BibitemShut {NoStop}%
\bibitem [{\citenamefont {Zwiebach}(2022)}]{Zwiebach}%
  \BibitemOpen
  \bibfield  {author} {\bibinfo {author} {\bibfnamefont {Barton}\ \bibnamefont {Zwiebach}},\ }\href@noop {} {\emph {\bibinfo {title} {Mastering Quantum Mechanics: Essentials, Theory, Applications}}}\ (\bibinfo  {publisher} {MIT Press},\ \bibinfo {address} {Cambridge, MA},\ \bibinfo {year} {2022})\BibitemShut {NoStop}%
\bibitem [{\citenamefont {Holzer}(1977)}]{holzerNeutralHydrogenInterplanetary1977}%
  \BibitemOpen
  \bibfield  {author} {\bibinfo {author} {\bibfnamefont {Thomas~E.}\ \bibnamefont {Holzer}},\ }\bibfield  {title} {\enquote {\bibinfo {title} {Neutral hydrogen in interplanetary space},}\ }\href {\doibase 10.1029/RG015i004p00467} {\bibfield  {journal} {\bibinfo  {journal} {Rev. Geophys.}\ }\textbf {\bibinfo {volume} {15}},\ \bibinfo {pages} {467--490} (\bibinfo {year} {1977})}\BibitemShut {NoStop}%
\bibitem [{\citenamefont {Swaczyna}\ \emph {et~al.}(2024)\citenamefont {Swaczyna}, \citenamefont {Bzowski}, \citenamefont {Dialynas},\ and\ \citenamefont {{et al}}}]{swaczynaInterstellarNeutralHydrogen2024}%
  \BibitemOpen
  \bibfield  {author} {\bibinfo {author} {\bibfnamefont {P.}~\bibnamefont {Swaczyna}}, \bibinfo {author} {\bibfnamefont {M.}~\bibnamefont {Bzowski}}, \bibinfo {author} {\bibfnamefont {K.}~\bibnamefont {Dialynas}}, \ and\ \bibinfo {author} {\bibnamefont {{et al}}},\ }\bibfield  {title} {\enquote {\bibinfo {title} {Interstellar {{Neutral Hydrogen}} in the {{Heliosphere}}: {{New Horizons Observations}} in the {{Context}} of {{Models}}},}\ }\href {\doibase 10.3847/2041-8213/ad5832} {\bibfield  {journal} {\bibinfo  {journal} {Astrophys. J., Lett.}\ }\textbf {\bibinfo {volume} {969}},\ \bibinfo {pages} {L20} (\bibinfo {year} {2024})}\BibitemShut {NoStop}%
\bibitem [{\citenamefont {Hayakawa}\ \emph {et~al.}(1961)\citenamefont {Hayakawa}, \citenamefont {Nishimura},\ and\ \citenamefont {Takayanagi}}]{hayakawaRadiationInterstellarHydrogen1961}%
  \BibitemOpen
  \bibfield  {author} {\bibinfo {author} {\bibfnamefont {Satio}\ \bibnamefont {Hayakawa}}, \bibinfo {author} {\bibfnamefont {Shiro}\ \bibnamefont {Nishimura}}, \ and\ \bibinfo {author} {\bibfnamefont {Kazuo}\ \bibnamefont {Takayanagi}},\ }\bibfield  {title} {\enquote {\bibinfo {title} {Radiation from the {{Interstellar Hydrogen Atoms}}},}\ }\href {\doibase https://ui.adsabs.harvard.edu/abs/1961PASJ...13..184H} {\bibfield  {journal} {\bibinfo  {journal} {Publ. Astron. Soc. Jpn.}\ }\textbf {\bibinfo {volume} {13}},\ \bibinfo {pages} {184--206} (\bibinfo {year} {1961})}\BibitemShut {NoStop}%
\bibitem [{\citenamefont {Swaczyna}\ \emph {et~al.}(2020)\citenamefont {Swaczyna}, \citenamefont {McComas}, \citenamefont {Zirnstein},\ and\ \citenamefont {{et al}}}]{swaczynaDensityNeutralHydrogen2020}%
  \BibitemOpen
  \bibfield  {author} {\bibinfo {author} {\bibfnamefont {P.}~\bibnamefont {Swaczyna}}, \bibinfo {author} {\bibfnamefont {D.~J.}\ \bibnamefont {McComas}}, \bibinfo {author} {\bibfnamefont {E.~J.}\ \bibnamefont {Zirnstein}}, \ and\ \bibinfo {author} {\bibnamefont {{et al}}},\ }\bibfield  {title} {\enquote {\bibinfo {title} {Density of {{Neutral Hydrogen}} in the {{Sun}}'s {{Interstellar Neighborhood}}},}\ }\href {\doibase 10.3847/1538-4357/abb80a} {\bibfield  {journal} {\bibinfo  {journal} {Astrophys. J.}\ }\textbf {\bibinfo {volume} {903}},\ \bibinfo {pages} {48} (\bibinfo {year} {2020})}\BibitemShut {NoStop}%
\bibitem [{\citenamefont {Lynch}\ \emph {et~al.}(2024)\citenamefont {Lynch}, \citenamefont {Knox},\ and\ \citenamefont {Chluba}}]{Lynch:2024gmp}%
  \BibitemOpen
  \bibfield  {author} {\bibinfo {author} {\bibfnamefont {Gabriel~P.}\ \bibnamefont {Lynch}}, \bibinfo {author} {\bibfnamefont {Lloyd}\ \bibnamefont {Knox}}, \ and\ \bibinfo {author} {\bibfnamefont {Jens}\ \bibnamefont {Chluba}},\ }\bibfield  {title} {\enquote {\bibinfo {title} {{Reconstructing the recombination history by combining early and late cosmological probes}},}\ }\href {\doibase 10.1103/PhysRevD.110.063518} {\bibfield  {journal} {\bibinfo  {journal} {Phys. Rev. D}\ }\textbf {\bibinfo {volume} {110}},\ \bibinfo {pages} {063518} (\bibinfo {year} {2024})},\ \Eprint {http://arxiv.org/abs/2404.05715} {arXiv:2404.05715 [astro-ph.CO]} \BibitemShut {NoStop}%
\bibitem [{\citenamefont {Collaboration}\ \emph {et~al.}(2020)\citenamefont {Collaboration}, \citenamefont {Aghanim}, \citenamefont {Akrami},\ and\ \citenamefont {{et al}}}]{planckcollaborationPlanck2018Results2020}%
  \BibitemOpen
  \bibfield  {author} {\bibinfo {author} {\bibfnamefont {Planck}\ \bibnamefont {Collaboration}}, \bibinfo {author} {\bibfnamefont {N.}~\bibnamefont {Aghanim}}, \bibinfo {author} {\bibfnamefont {Y.}~\bibnamefont {Akrami}}, \ and\ \bibinfo {author} {\bibnamefont {{et al}}},\ }\bibfield  {title} {\enquote {\bibinfo {title} {Planck 2018 results. {{VI}}. {{Cosmological}} parameters},}\ }\href {\doibase 10.1051/0004-6361/201833910} {\bibfield  {journal} {\bibinfo  {journal} {Astron. Astrophys.}\ }\textbf {\bibinfo {volume} {641}},\ \bibinfo {pages} {A6} (\bibinfo {year} {2020})},\ \Eprint {http://arxiv.org/abs/1807.06209} {arXiv:1807.06209 [astro-ph]} \BibitemShut {NoStop}%
\bibitem [{\citenamefont {Cerde{\~n}o}\ and\ \citenamefont {Greene}(2010)}]{cerdenoParticleDarkMatter2010}%
  \BibitemOpen
  \bibfield  {author} {\bibinfo {author} {\bibfnamefont {D.G.}\ \bibnamefont {Cerde{\~n}o}}\ and\ \bibinfo {author} {\bibfnamefont {A.M.}\ \bibnamefont {Greene}},\ }\href@noop {} {\emph {\bibinfo {title} {Particle {{Dark Matter}}: {{Observations}}, {{Models}} and {{Searches}}}}}\ (\bibinfo  {publisher} {Cambridge University Press},\ \bibinfo {address} {Cambridge, UK; New York},\ \bibinfo {year} {2010})\BibitemShut {NoStop}%
\bibitem [{\citenamefont {Choi}\ \emph {et~al.}(2014)\citenamefont {Choi}, \citenamefont {Rott},\ and\ \citenamefont {Itow}}]{choiImpactDarkMatter2014}%
  \BibitemOpen
  \bibfield  {author} {\bibinfo {author} {\bibfnamefont {Koun}\ \bibnamefont {Choi}}, \bibinfo {author} {\bibfnamefont {Carsten}\ \bibnamefont {Rott}}, \ and\ \bibinfo {author} {\bibfnamefont {Yoshitaka}\ \bibnamefont {Itow}},\ }\bibfield  {title} {\enquote {\bibinfo {title} {Impact of {{Dark Matter Velocity Distributions}} on {{Capture Rates}} in the {{Sun}}},}\ }\href {\doibase 10.1088/1475-7516/2014/05/049} {\bibfield  {journal} {\bibinfo  {journal} {J. Cosmol. Astropart. Phys.}\ }\textbf {\bibinfo {volume} {2014}},\ \bibinfo {pages} {049--049} (\bibinfo {year} {2014})},\ \Eprint {http://arxiv.org/abs/1312.0273} {arXiv:1312.0273 [astro-ph]} \BibitemShut {NoStop}%
\bibitem [{\citenamefont {{Slater}}(1964)}]{SlaterRadii}%
  \BibitemOpen
  \bibfield  {author} {\bibinfo {author} {\bibfnamefont {J.~C.}\ \bibnamefont {{Slater}}},\ }\bibfield  {title} {\enquote {\bibinfo {title} {{Atomic Radii in Crystals}},}\ }\href {\doibase 10.1063/1.1725697} {\bibfield  {journal} {\bibinfo  {journal} {\jcp}\ }\textbf {\bibinfo {volume} {41}},\ \bibinfo {pages} {3199--3204} (\bibinfo {year} {1964})}\BibitemShut {NoStop}%
\bibitem [{\citenamefont {Pospelov}\ and\ \citenamefont {Pradler}(2010)}]{Pospelov:2010hj}%
  \BibitemOpen
  \bibfield  {author} {\bibinfo {author} {\bibfnamefont {Maxim}\ \bibnamefont {Pospelov}}\ and\ \bibinfo {author} {\bibfnamefont {Josef}\ \bibnamefont {Pradler}},\ }\bibfield  {title} {\enquote {\bibinfo {title} {{Big Bang Nucleosynthesis as a Probe of New Physics}},}\ }\href {\doibase 10.1146/annurev.nucl.012809.104521} {\bibfield  {journal} {\bibinfo  {journal} {Ann. Rev. Nucl. Part. Sci.}\ }\textbf {\bibinfo {volume} {60}},\ \bibinfo {pages} {539--568} (\bibinfo {year} {2010})},\ \Eprint {http://arxiv.org/abs/1011.1054} {arXiv:1011.1054 [hep-ph]} \BibitemShut {NoStop}%
\bibitem [{\citenamefont {Aghanim}\ \emph {et~al.}(2020)\citenamefont {Aghanim} \emph {et~al.}}]{Planck:2018vyg}%
  \BibitemOpen
  \bibfield  {author} {\bibinfo {author} {\bibfnamefont {N.}~\bibnamefont {Aghanim}} \emph {et~al.} (\bibinfo {collaboration} {Planck}),\ }\bibfield  {title} {\enquote {\bibinfo {title} {{Planck 2018 results. VI. Cosmological parameters}},}\ }\href {\doibase 10.1051/0004-6361/201833910} {\bibfield  {journal} {\bibinfo  {journal} {Astron. Astrophys.}\ }\textbf {\bibinfo {volume} {641}},\ \bibinfo {pages} {A6} (\bibinfo {year} {2020})},\ \bibinfo {note} {[Erratum: Astron.Astrophys. 652, C4 (2021)]},\ \Eprint {http://arxiv.org/abs/1807.06209} {arXiv:1807.06209 [astro-ph.CO]} \BibitemShut {NoStop}%
\bibitem [{\citenamefont {{Basu}}\ \emph {et~al.}(2022)\citenamefont {{Basu}}, \citenamefont {{Roy}}, \citenamefont {{Beuther}}, \citenamefont {{Syed}}, \citenamefont {{Ott}}, \citenamefont {{Soler}}, \citenamefont {{Stil}},\ and\ \citenamefont {{Rugel}}}]{BasuISM}%
  \BibitemOpen
  \bibfield  {author} {\bibinfo {author} {\bibfnamefont {Arghyadeep}\ \bibnamefont {{Basu}}}, \bibinfo {author} {\bibfnamefont {Nirupam}\ \bibnamefont {{Roy}}}, \bibinfo {author} {\bibfnamefont {Henrik}\ \bibnamefont {{Beuther}}}, \bibinfo {author} {\bibfnamefont {Jonas}\ \bibnamefont {{Syed}}}, \bibinfo {author} {\bibfnamefont {J{\"u}rgen}\ \bibnamefont {{Ott}}}, \bibinfo {author} {\bibfnamefont {Juan~D.}\ \bibnamefont {{Soler}}}, \bibinfo {author} {\bibfnamefont {Jeroen}\ \bibnamefont {{Stil}}}, \ and\ \bibinfo {author} {\bibfnamefont {Michael~R.}\ \bibnamefont {{Rugel}}},\ }\bibfield  {title} {\enquote {\bibinfo {title} {{Properties of atomic hydrogen gas in the Galactic plane from THOR 21-cm absorption spectra: a comparison with the high latitude gas}},}\ }\href {\doibase 10.1093/mnras/stac3043} {\bibfield  {journal} {\bibinfo  {journal} {{M.~Not.~Roy.~Astron.~Soc.}}\ }\textbf {\bibinfo {volume} {517}},\ \bibinfo {pages} {5063--5068} (\bibinfo {year} {2022})},\ \Eprint {http://arxiv.org/abs/2210.11551}
  {arXiv:2210.11551 [astro-ph.GA]} \BibitemShut {NoStop}%
\bibitem [{\citenamefont {{Patra}}\ and\ \citenamefont {{Roy}}(2024)}]{PatraISM}%
  \BibitemOpen
  \bibfield  {author} {\bibinfo {author} {\bibfnamefont {Narendra~Nath}\ \bibnamefont {{Patra}}}\ and\ \bibinfo {author} {\bibfnamefont {Nirupam}\ \bibnamefont {{Roy}}},\ }\bibfield  {title} {\enquote {\bibinfo {title} {{The temperature of the neutral interstellar medium in the Galaxy}},}\ }\href {\doibase 10.1093/mnras/stae771} {\bibfield  {journal} {\bibinfo  {journal} {{M.~Not.~Roy.~Astron.~Soc.}}\ }\textbf {\bibinfo {volume} {529}},\ \bibinfo {pages} {4037--4049} (\bibinfo {year} {2024})},\ \Eprint {http://arxiv.org/abs/2403.11653} {arXiv:2403.11653 [astro-ph.GA]} \BibitemShut {NoStop}%
\bibitem [{\citenamefont {{Fixsen}}(2009)}]{FixsenCMBT0}%
  \BibitemOpen
  \bibfield  {author} {\bibinfo {author} {\bibfnamefont {D.~J.}\ \bibnamefont {{Fixsen}}},\ }\bibfield  {title} {\enquote {\bibinfo {title} {{The Temperature of the Cosmic Microwave Background}},}\ }\href {\doibase 10.1088/0004-637X/707/2/916} {\bibfield  {journal} {\bibinfo  {journal} {{Ap.~J.}}\ }\textbf {\bibinfo {volume} {707}},\ \bibinfo {pages} {916--920} (\bibinfo {year} {2009})},\ \Eprint {http://arxiv.org/abs/0911.1955} {arXiv:0911.1955 [astro-ph.CO]} \BibitemShut {NoStop}%
\bibitem [{\citenamefont {{Alexandra P.~Klipfel and David I.~Kaiser}}()}]{KlipfelEMsignatures}%
  \BibitemOpen
  \bibfield  {author} {\bibinfo {author} {\bibnamefont {{Alexandra P.~Klipfel and David I.~Kaiser}}},\ }\bibfield  {title} {\enquote {\bibinfo {title} {{Electromagnetic Signatures of Galactic Primordial Black Holes}},}\ }\href@noop {} {\bibinfo  {journal} {{in prep}}\ }\BibitemShut {NoStop}%
\bibitem [{\citenamefont {Baker}\ and\ \citenamefont {Thamm}(2022)}]{Baker:2021btk}%
  \BibitemOpen
\bibfield  {journal} {  }\bibfield  {author} {\bibinfo {author} {\bibfnamefont {Michael~J.}\ \bibnamefont {Baker}}\ and\ \bibinfo {author} {\bibfnamefont {Andrea}\ \bibnamefont {Thamm}},\ }\bibfield  {title} {\enquote {\bibinfo {title} {{Probing the particle spectrum of nature with evaporating black holes}},}\ }\href {\doibase 10.21468/SciPostPhys.12.5.150} {\bibfield  {journal} {\bibinfo  {journal} {SciPost Phys.}\ }\textbf {\bibinfo {volume} {12}},\ \bibinfo {pages} {150} (\bibinfo {year} {2022})},\ \Eprint {http://arxiv.org/abs/2105.10506} {arXiv:2105.10506 [hep-ph]} \BibitemShut {NoStop}%
\bibitem [{\citenamefont {Baker}\ and\ \citenamefont {Thamm}(2023)}]{Baker:2022rkn}%
  \BibitemOpen
  \bibfield  {author} {\bibinfo {author} {\bibfnamefont {Michael~J.}\ \bibnamefont {Baker}}\ and\ \bibinfo {author} {\bibfnamefont {Andrea}\ \bibnamefont {Thamm}},\ }\bibfield  {title} {\enquote {\bibinfo {title} {{Black hole evaporation beyond the Standard Model of particle physics}},}\ }\href {\doibase 10.1007/JHEP01(2023)063} {\bibfield  {journal} {\bibinfo  {journal} {JHEP}\ }\textbf {\bibinfo {volume} {01}},\ \bibinfo {pages} {063} (\bibinfo {year} {2023})},\ \Eprint {http://arxiv.org/abs/2210.02805} {arXiv:2210.02805 [hep-ph]} \BibitemShut {NoStop}%
\bibitem [{\citenamefont {Baker}\ \emph {et~al.}(2025{\natexlab{b}})\citenamefont {Baker}, \citenamefont {Iguaz~Juan}, \citenamefont {Symons},\ and\ \citenamefont {Thamm}}]{Baker:2025ffi}%
  \BibitemOpen
  \bibfield  {author} {\bibinfo {author} {\bibfnamefont {Michael~J.}\ \bibnamefont {Baker}}, \bibinfo {author} {\bibfnamefont {Joaquim}\ \bibnamefont {Iguaz~Juan}}, \bibinfo {author} {\bibfnamefont {Aidan}\ \bibnamefont {Symons}}, \ and\ \bibinfo {author} {\bibfnamefont {Andrea}\ \bibnamefont {Thamm}},\ }\bibfield  {title} {\enquote {\bibinfo {title} {{Probing Dark Sectors with Exploding Black Holes: Gamma Rays}},}\ }\href@noop {} {\  (\bibinfo {year} {2025}{\natexlab{b}})},\ \Eprint {http://arxiv.org/abs/2512.19603} {arXiv:2512.19603 [hep-ph]} \BibitemShut {NoStop}%
\bibitem [{\citenamefont {Barrau}\ \emph {et~al.}(2022)\citenamefont {Barrau}, \citenamefont {Martineau},\ and\ \citenamefont {Renevey}}]{Barrau:2022bfg}%
  \BibitemOpen
  \bibfield  {author} {\bibinfo {author} {\bibfnamefont {Aur{\'e}lien}\ \bibnamefont {Barrau}}, \bibinfo {author} {\bibfnamefont {Killian}\ \bibnamefont {Martineau}}, \ and\ \bibinfo {author} {\bibfnamefont {Cyril}\ \bibnamefont {Renevey}},\ }\bibfield  {title} {\enquote {\bibinfo {title} {{Catastrophic fate of Schwarzschild black holes in a thermal bath}},}\ }\href {\doibase 10.1103/PhysRevD.106.023509} {\bibfield  {journal} {\bibinfo  {journal} {Phys. Rev. D}\ }\textbf {\bibinfo {volume} {106}},\ \bibinfo {pages} {023509} (\bibinfo {year} {2022})},\ \Eprint {http://arxiv.org/abs/2203.13297} {arXiv:2203.13297 [gr-qc]} \BibitemShut {NoStop}%
\bibitem [{\citenamefont {Loeb}(2024)}]{Loeb:2024gga}%
  \BibitemOpen
  \bibfield  {author} {\bibinfo {author} {\bibfnamefont {Abraham}\ \bibnamefont {Loeb}},\ }\bibfield  {title} {\enquote {\bibinfo {title} {{Quantum-mechanical Suppression of Accretion by Primordial Black Holes}},}\ }\href {\doibase 10.3847/2041-8213/ad887d} {\bibfield  {journal} {\bibinfo  {journal} {Astrophys. J. Lett.}\ }\textbf {\bibinfo {volume} {975}},\ \bibinfo {pages} {L15} (\bibinfo {year} {2024})},\ \Eprint {http://arxiv.org/abs/2409.09081} {arXiv:2409.09081 [astro-ph.HE]} \BibitemShut {NoStop}%
\bibitem [{\citenamefont {Chatterjee}\ \emph {et~al.}(2025)\citenamefont {Chatterjee}, \citenamefont {Kalita},\ and\ \citenamefont {Maity}}]{Chatterjee:2025wnt}%
  \BibitemOpen
  \bibfield  {author} {\bibinfo {author} {\bibfnamefont {Ayan}\ \bibnamefont {Chatterjee}}, \bibinfo {author} {\bibfnamefont {Jitumani}\ \bibnamefont {Kalita}}, \ and\ \bibinfo {author} {\bibfnamefont {Debaprasad}\ \bibnamefont {Maity}},\ }\bibfield  {title} {\enquote {\bibinfo {title} {{Evaporation of Primordial Black Holes in a Thermal Universe: A Thermofield Dynamics Approach}},}\ }\href@noop {} {\  (\bibinfo {year} {2025})},\ \Eprint {http://arxiv.org/abs/2512.07284} {arXiv:2512.07284 [hep-th]} \BibitemShut {NoStop}%
\bibitem [{\citenamefont {Page}(1976{\natexlab{b}})}]{pageParticleEmissionRates1976}%
  \BibitemOpen
  \bibfield  {author} {\bibinfo {author} {\bibfnamefont {Don~N.}\ \bibnamefont {Page}},\ }\bibfield  {title} {\enquote {\bibinfo {title} {Particle emission rates from a black hole: {{Massless}} particles from an uncharged, nonrotating hole},}\ }\href {\doibase 10.1103/PhysRevD.13.198} {\bibfield  {journal} {\bibinfo  {journal} {Phys. Rev. D}\ }\textbf {\bibinfo {volume} {13}},\ \bibinfo {pages} {198--206} (\bibinfo {year} {1976}{\natexlab{b}})}\BibitemShut {NoStop}%
\bibitem [{\citenamefont {MacGibbon}\ and\ \citenamefont {Webber}(1990)}]{macgibbonQuarkGluonjetEmission1990}%
  \BibitemOpen
  \bibfield  {author} {\bibinfo {author} {\bibfnamefont {Jane~H.}\ \bibnamefont {MacGibbon}}\ and\ \bibinfo {author} {\bibfnamefont {B.~R.}\ \bibnamefont {Webber}},\ }\bibfield  {title} {\enquote {\bibinfo {title} {Quark- and gluon-jet emission from primordial black holes: {{The}} instantaneous spectra},}\ }\href {\doibase 10.1103/PhysRevD.41.3052} {\bibfield  {journal} {\bibinfo  {journal} {Phys. Rev. D}\ }\textbf {\bibinfo {volume} {41}},\ \bibinfo {pages} {3052--3079} (\bibinfo {year} {1990})}\BibitemShut {NoStop}%
\bibitem [{\citenamefont {Teukolsky}(1973)}]{teukolskyPerturbationsRotatingBlack1973}%
  \BibitemOpen
  \bibfield  {author} {\bibinfo {author} {\bibfnamefont {Saul~A.}\ \bibnamefont {Teukolsky}},\ }\bibfield  {title} {\enquote {\bibinfo {title} {Perturbations of a {{Rotating Black Hole}}. {{I}}. {{Fundamental Equations}} for {{Gravitational}}, {{Electromagnetic}}, and {{Neutrino-Field Perturbations}}},}\ }\href {\doibase 10.1086/152444} {\bibfield  {journal} {\bibinfo  {journal} {Astrophys. J.}\ }\textbf {\bibinfo {volume} {185}},\ \bibinfo {pages} {635} (\bibinfo {year} {1973})}\BibitemShut {NoStop}%
\bibitem [{\citenamefont {Teukolsky}\ and\ \citenamefont {Press}(1974)}]{teukolskyPerturbationsRotatingBlack1974}%
  \BibitemOpen
  \bibfield  {author} {\bibinfo {author} {\bibfnamefont {S.~A.}\ \bibnamefont {Teukolsky}}\ and\ \bibinfo {author} {\bibfnamefont {W.~H.}\ \bibnamefont {Press}},\ }\bibfield  {title} {\enquote {\bibinfo {title} {Perturbations of a rotating black hole. {{III}}. {{Interaction}} of the hole with gravitational and electromagnetic radiation.}}\ }\href {\doibase 10.1086/153180} {\bibfield  {journal} {\bibinfo  {journal} {Astrophys. J.}\ }\textbf {\bibinfo {volume} {193}},\ \bibinfo {pages} {443--461} (\bibinfo {year} {1974})}\BibitemShut {NoStop}%
\bibitem [{\citenamefont {Arbey}\ and\ \citenamefont {Auffinger}(2019)}]{arbeyBlackHawkV20Public2019}%
  \BibitemOpen
  \bibfield  {author} {\bibinfo {author} {\bibfnamefont {Alexandre}\ \bibnamefont {Arbey}}\ and\ \bibinfo {author} {\bibfnamefont {J{\'e}r{\'e}my}\ \bibnamefont {Auffinger}},\ }\bibfield  {title} {\enquote {\bibinfo {title} {{{BlackHawk}} v2.0: {{A}} public code for calculating the {{Hawking}} evaporation spectra of any black hole distribution},}\ }\href {\doibase 10.1140/epjc/s10052-019-7161-1} {\bibfield  {journal} {\bibinfo  {journal} {Eur. Phys. J. C}\ }\textbf {\bibinfo {volume} {79}},\ \bibinfo {pages} {693} (\bibinfo {year} {2019})},\ \Eprint {http://arxiv.org/abs/1905.04268} {arXiv:1905.04268 [astro-ph, physics:gr-qc, physics:hep-ph]} \BibitemShut {NoStop}%
\bibitem [{\citenamefont {Arbey}\ and\ \citenamefont {Auffinger}(2021)}]{arbeyPhysicsStandardModel2021a}%
  \BibitemOpen
  \bibfield  {author} {\bibinfo {author} {\bibfnamefont {Alexandre}\ \bibnamefont {Arbey}}\ and\ \bibinfo {author} {\bibfnamefont {J{\'e}r{\'e}my}\ \bibnamefont {Auffinger}},\ }\bibfield  {title} {\enquote {\bibinfo {title} {Physics {{Beyond}} the {{Standard Model}} with {{BlackHawk}} v2.0},}\ }\href {\doibase 10.1140/epjc/s10052-021-09702-8} {\bibfield  {journal} {\bibinfo  {journal} {Eur. Phys. J. C}\ }\textbf {\bibinfo {volume} {81}},\ \bibinfo {pages} {910} (\bibinfo {year} {2021})},\ \Eprint {http://arxiv.org/abs/2108.02737} {arXiv:2108.02737 [gr-qc]} \BibitemShut {NoStop}%
\bibitem [{\citenamefont {Gray}\ and\ \citenamefont {Visser}(2018)}]{grayGreybodyFactorsSchwarzschild2018}%
  \BibitemOpen
  \bibfield  {author} {\bibinfo {author} {\bibfnamefont {Finnian}\ \bibnamefont {Gray}}\ and\ \bibinfo {author} {\bibfnamefont {Matt}\ \bibnamefont {Visser}},\ }\bibfield  {title} {\enquote {\bibinfo {title} {Greybody {{Factors}} for {{Schwarzschild Black Holes}}: {{Path-Ordered Exponentials}} and {{Product Integrals}}},}\ }\href {\doibase 10.3390/universe4090093} {\bibfield  {journal} {\bibinfo  {journal} {Universe}\ }\textbf {\bibinfo {volume} {4}},\ \bibinfo {pages} {93} (\bibinfo {year} {2018})}\BibitemShut {NoStop}%
\bibitem [{\citenamefont {MacGibbon}(1991)}]{macgibbonQuarkGluonjetEmission1991}%
  \BibitemOpen
  \bibfield  {author} {\bibinfo {author} {\bibfnamefont {Jane~H.}\ \bibnamefont {MacGibbon}},\ }\bibfield  {title} {\enquote {\bibinfo {title} {Quark- and gluon-jet emission from primordial black holes. {{II}}. {{The}} emission over the black-hole lifetime},}\ }\href {\doibase 10.1103/PhysRevD.44.376} {\bibfield  {journal} {\bibinfo  {journal} {Phys. Rev. D}\ }\textbf {\bibinfo {volume} {44}},\ \bibinfo {pages} {376--392} (\bibinfo {year} {1991})}\BibitemShut {NoStop}%
\bibitem [{\citenamefont {Sj{\"o}strand}\ \emph {et~al.}(2015)\citenamefont {Sj{\"o}strand}, \citenamefont {Ask}, \citenamefont {Christiansen}, \citenamefont {Corke}, \citenamefont {Desai}, \citenamefont {Ilten}, \citenamefont {Mrenna}, \citenamefont {Prestel}, \citenamefont {Rasmussen},\ and\ \citenamefont {Skands}}]{Sjostrand:2014zea}%
  \BibitemOpen
  \bibfield  {author} {\bibinfo {author} {\bibfnamefont {Torbj{\"o}rn}\ \bibnamefont {Sj{\"o}strand}}, \bibinfo {author} {\bibfnamefont {Stefan}\ \bibnamefont {Ask}}, \bibinfo {author} {\bibfnamefont {Jesper~R.}\ \bibnamefont {Christiansen}}, \bibinfo {author} {\bibfnamefont {Richard}\ \bibnamefont {Corke}}, \bibinfo {author} {\bibfnamefont {Nishita}\ \bibnamefont {Desai}}, \bibinfo {author} {\bibfnamefont {Philip}\ \bibnamefont {Ilten}}, \bibinfo {author} {\bibfnamefont {Stephen}\ \bibnamefont {Mrenna}}, \bibinfo {author} {\bibfnamefont {Stefan}\ \bibnamefont {Prestel}}, \bibinfo {author} {\bibfnamefont {Christine~O.}\ \bibnamefont {Rasmussen}}, \ and\ \bibinfo {author} {\bibfnamefont {Peter~Z.}\ \bibnamefont {Skands}},\ }\bibfield  {title} {\enquote {\bibinfo {title} {{An introduction to PYTHIA 8.2}},}\ }\href {\doibase 10.1016/j.cpc.2015.01.024} {\bibfield  {journal} {\bibinfo  {journal} {Comput. Phys. Commun.}\ }\textbf {\bibinfo {volume} {191}},\ \bibinfo {pages} {159--177} (\bibinfo {year} {2015})},\
  \Eprint {http://arxiv.org/abs/1410.3012} {arXiv:1410.3012 [hep-ph]} \BibitemShut {NoStop}%
\bibitem [{\citenamefont {Coogan}\ \emph {et~al.}(2020)\citenamefont {Coogan}, \citenamefont {Morrison},\ and\ \citenamefont {Profumo}}]{Coogan:2019qpu}%
  \BibitemOpen
  \bibfield  {author} {\bibinfo {author} {\bibfnamefont {Adam}\ \bibnamefont {Coogan}}, \bibinfo {author} {\bibfnamefont {Logan}\ \bibnamefont {Morrison}}, \ and\ \bibinfo {author} {\bibfnamefont {Stefano}\ \bibnamefont {Profumo}},\ }\bibfield  {title} {\enquote {\bibinfo {title} {{Hazma: A Python Toolkit for Studying Indirect Detection of Sub-GeV Dark Matter}},}\ }\href {\doibase 10.1088/1475-7516/2020/01/056} {\bibfield  {journal} {\bibinfo  {journal} {JCAP}\ }\textbf {\bibinfo {volume} {01}},\ \bibinfo {pages} {056} (\bibinfo {year} {2020})},\ \Eprint {http://arxiv.org/abs/1907.11846} {arXiv:1907.11846 [hep-ph]} \BibitemShut {NoStop}%
\bibitem [{\citenamefont {Mosbech}\ and\ \citenamefont {Picker}(2022)}]{mosbechEffectsHawkingEvaporation2022}%
  \BibitemOpen
  \bibfield  {author} {\bibinfo {author} {\bibfnamefont {Markus}\ \bibnamefont {Mosbech}}\ and\ \bibinfo {author} {\bibfnamefont {Zachary}\ \bibnamefont {Picker}},\ }\bibfield  {title} {\enquote {\bibinfo {title} {Effects of {{Hawking}} evaporation on {{PBH}} distributions},}\ }\href {\doibase 10.21468/SciPostPhys.13.4.100} {\bibfield  {journal} {\bibinfo  {journal} {SciPost Phys.}\ }\textbf {\bibinfo {volume} {13}},\ \bibinfo {pages} {100} (\bibinfo {year} {2022})}\BibitemShut {NoStop}%
\bibitem [{\citenamefont {Cang}\ \emph {et~al.}(2022)\citenamefont {Cang}, \citenamefont {Gao},\ and\ \citenamefont {Ma}}]{cang21cmConstraintsSpinning2022}%
  \BibitemOpen
  \bibfield  {author} {\bibinfo {author} {\bibfnamefont {Junsong}\ \bibnamefont {Cang}}, \bibinfo {author} {\bibfnamefont {Yu}~\bibnamefont {Gao}}, \ and\ \bibinfo {author} {\bibfnamefont {Yin-Zhe}\ \bibnamefont {Ma}},\ }\bibfield  {title} {\enquote {\bibinfo {title} {21-cm constraints on spinning primordial black holes},}\ }\href {\doibase 10.1088/1475-7516/2022/03/012} {\bibfield  {journal} {\bibinfo  {journal} {J. Cosmol. Astropart. Phys.}\ }\textbf {\bibinfo {volume} {2022}},\ \bibinfo {pages} {012} (\bibinfo {year} {2022})}\BibitemShut {NoStop}%
\bibitem [{\citenamefont {Choptuik}(1993)}]{Choptuik:1992jv}%
  \BibitemOpen
  \bibfield  {author} {\bibinfo {author} {\bibfnamefont {Matthew~W.}\ \bibnamefont {Choptuik}},\ }\bibfield  {title} {\enquote {\bibinfo {title} {{Universality and scaling in gravitational collapse of a massless scalar field}},}\ }\href {\doibase 10.1103/PhysRevLett.70.9} {\bibfield  {journal} {\bibinfo  {journal} {Phys. Rev. Lett.}\ }\textbf {\bibinfo {volume} {70}},\ \bibinfo {pages} {9--12} (\bibinfo {year} {1993})}\BibitemShut {NoStop}%
\bibitem [{\citenamefont {Evans}\ and\ \citenamefont {Coleman}(1994)}]{Evans:1994pj}%
  \BibitemOpen
  \bibfield  {author} {\bibinfo {author} {\bibfnamefont {Charles~R.}\ \bibnamefont {Evans}}\ and\ \bibinfo {author} {\bibfnamefont {Jason~S.}\ \bibnamefont {Coleman}},\ }\bibfield  {title} {\enquote {\bibinfo {title} {{Observation of critical phenomena and selfsimilarity in the gravitational collapse of radiation fluid}},}\ }\href {\doibase 10.1103/PhysRevLett.72.1782} {\bibfield  {journal} {\bibinfo  {journal} {Phys. Rev. Lett.}\ }\textbf {\bibinfo {volume} {72}},\ \bibinfo {pages} {1782--1785} (\bibinfo {year} {1994})},\ \Eprint {http://arxiv.org/abs/gr-qc/9402041} {arXiv:gr-qc/9402041} \BibitemShut {NoStop}%
\bibitem [{\citenamefont {Niemeyer}\ and\ \citenamefont {Jedamzik}(1999{\natexlab{a}})}]{Niemeyer:1999ak}%
  \BibitemOpen
  \bibfield  {author} {\bibinfo {author} {\bibfnamefont {Jens~C.}\ \bibnamefont {Niemeyer}}\ and\ \bibinfo {author} {\bibfnamefont {K.}~\bibnamefont {Jedamzik}},\ }\bibfield  {title} {\enquote {\bibinfo {title} {{Dynamics of primordial black hole formation}},}\ }\href {\doibase 10.1103/PhysRevD.59.124013} {\bibfield  {journal} {\bibinfo  {journal} {Phys. Rev. D}\ }\textbf {\bibinfo {volume} {59}},\ \bibinfo {pages} {124013} (\bibinfo {year} {1999}{\natexlab{a}})},\ \Eprint {http://arxiv.org/abs/astro-ph/9901292} {arXiv:astro-ph/9901292} \BibitemShut {NoStop}%
\bibitem [{\citenamefont {Gundlach}\ and\ \citenamefont {Martin-Garcia}(2007)}]{Gundlach:2007gc}%
  \BibitemOpen
  \bibfield  {author} {\bibinfo {author} {\bibfnamefont {Carsten}\ \bibnamefont {Gundlach}}\ and\ \bibinfo {author} {\bibfnamefont {Jose~M.}\ \bibnamefont {Martin-Garcia}},\ }\bibfield  {title} {\enquote {\bibinfo {title} {{Critical phenomena in gravitational collapse}},}\ }\href {\doibase 10.12942/lrr-2007-5} {\bibfield  {journal} {\bibinfo  {journal} {Living Rev. Rel.}\ }\textbf {\bibinfo {volume} {10}},\ \bibinfo {pages} {5} (\bibinfo {year} {2007})},\ \Eprint {http://arxiv.org/abs/0711.4620} {arXiv:0711.4620 [gr-qc]} \BibitemShut {NoStop}%
\bibitem [{\citenamefont {Musco}\ \emph {et~al.}(2009)\citenamefont {Musco}, \citenamefont {Miller},\ and\ \citenamefont {Polnarev}}]{Musco:2008hv}%
  \BibitemOpen
  \bibfield  {author} {\bibinfo {author} {\bibfnamefont {Ilia}\ \bibnamefont {Musco}}, \bibinfo {author} {\bibfnamefont {John~C.}\ \bibnamefont {Miller}}, \ and\ \bibinfo {author} {\bibfnamefont {Alexander~G.}\ \bibnamefont {Polnarev}},\ }\bibfield  {title} {\enquote {\bibinfo {title} {{Primordial black hole formation in the radiative era: Investigation of the critical nature of the collapse}},}\ }\href {\doibase 10.1088/0264-9381/26/23/235001} {\bibfield  {journal} {\bibinfo  {journal} {Class. Quant. Grav.}\ }\textbf {\bibinfo {volume} {26}},\ \bibinfo {pages} {235001} (\bibinfo {year} {2009})},\ \Eprint {http://arxiv.org/abs/0811.1452} {arXiv:0811.1452 [gr-qc]} \BibitemShut {NoStop}%
\bibitem [{\citenamefont {Gow}\ \emph {et~al.}(2022)\citenamefont {Gow}, \citenamefont {Byrnes},\ and\ \citenamefont {Hall}}]{Gow:2020cou}%
  \BibitemOpen
  \bibfield  {author} {\bibinfo {author} {\bibfnamefont {Andrew~D.}\ \bibnamefont {Gow}}, \bibinfo {author} {\bibfnamefont {Christian~T.}\ \bibnamefont {Byrnes}}, \ and\ \bibinfo {author} {\bibfnamefont {Alex}\ \bibnamefont {Hall}},\ }\bibfield  {title} {\enquote {\bibinfo {title} {{Accurate model for the primordial black hole mass distribution from a peak in the power spectrum}},}\ }\href {\doibase 10.1103/PhysRevD.105.023503} {\bibfield  {journal} {\bibinfo  {journal} {Phys. Rev. D}\ }\textbf {\bibinfo {volume} {105}},\ \bibinfo {pages} {023503} (\bibinfo {year} {2022})},\ \Eprint {http://arxiv.org/abs/2009.03204} {arXiv:2009.03204 [astro-ph.CO]} \BibitemShut {NoStop}%
\bibitem [{\citenamefont {Niemeyer}\ and\ \citenamefont {Jedamzik}(1998)}]{Niemeyer:1997mt}%
  \BibitemOpen
  \bibfield  {author} {\bibinfo {author} {\bibfnamefont {Jens~C.}\ \bibnamefont {Niemeyer}}\ and\ \bibinfo {author} {\bibfnamefont {K.}~\bibnamefont {Jedamzik}},\ }\bibfield  {title} {\enquote {\bibinfo {title} {{Near-critical gravitational collapse and the initial mass function of primordial black holes}},}\ }\href {\doibase 10.1103/PhysRevLett.80.5481} {\bibfield  {journal} {\bibinfo  {journal} {Phys. Rev. Lett.}\ }\textbf {\bibinfo {volume} {80}},\ \bibinfo {pages} {5481--5484} (\bibinfo {year} {1998})},\ \Eprint {http://arxiv.org/abs/astro-ph/9709072} {arXiv:astro-ph/9709072} \BibitemShut {NoStop}%
\bibitem [{\citenamefont {Green}\ and\ \citenamefont {Liddle}(1999)}]{Green:1999xm}%
  \BibitemOpen
  \bibfield  {author} {\bibinfo {author} {\bibfnamefont {Anne~M.}\ \bibnamefont {Green}}\ and\ \bibinfo {author} {\bibfnamefont {Andrew~R.}\ \bibnamefont {Liddle}},\ }\bibfield  {title} {\enquote {\bibinfo {title} {{Critical collapse and the primordial black hole initial mass function}},}\ }\href {\doibase 10.1103/PhysRevD.60.063509} {\bibfield  {journal} {\bibinfo  {journal} {Phys. Rev. D}\ }\textbf {\bibinfo {volume} {60}},\ \bibinfo {pages} {063509} (\bibinfo {year} {1999})},\ \Eprint {http://arxiv.org/abs/astro-ph/9901268} {arXiv:astro-ph/9901268} \BibitemShut {NoStop}%
\bibitem [{\citenamefont {Niemeyer}\ and\ \citenamefont {Jedamzik}(1999{\natexlab{b}})}]{niemeyerDynamicsPrimordialBlack1999}%
  \BibitemOpen
  \bibfield  {author} {\bibinfo {author} {\bibfnamefont {J.~C.}\ \bibnamefont {Niemeyer}}\ and\ \bibinfo {author} {\bibfnamefont {K.}~\bibnamefont {Jedamzik}},\ }\bibfield  {title} {\enquote {\bibinfo {title} {Dynamics of {{Primordial Black Hole Formation}}},}\ }\href {\doibase 10.1103/PhysRevD.59.124013} {\bibfield  {journal} {\bibinfo  {journal} {Phys. Rev. D}\ }\textbf {\bibinfo {volume} {59}},\ \bibinfo {pages} {124013} (\bibinfo {year} {1999}{\natexlab{b}})},\ \Eprint {http://arxiv.org/abs/astro-ph/9901292} {arXiv:astro-ph/9901292} \BibitemShut {NoStop}%
\bibitem [{\citenamefont {Klipfel}\ and\ \citenamefont {Kaiser}(2025)}]{Klipfel:2025jql}%
  \BibitemOpen
  \bibfield  {author} {\bibinfo {author} {\bibfnamefont {Alexandra~P.}\ \bibnamefont {Klipfel}}\ and\ \bibinfo {author} {\bibfnamefont {David~I.}\ \bibnamefont {Kaiser}},\ }\bibfield  {title} {\enquote {\bibinfo {title} {{Ultrahigh-Energy Neutrinos from Primordial Black Holes}},}\ }\href {\doibase 10.1103/vnm4-7wdc} {\bibfield  {journal} {\bibinfo  {journal} {Phys. Rev. Lett.}\ }\textbf {\bibinfo {volume} {135}},\ \bibinfo {pages} {121003} (\bibinfo {year} {2025})},\ \Eprint {http://arxiv.org/abs/2503.19227} {arXiv:2503.19227 [hep-ph]} \BibitemShut {NoStop}%
\bibitem [{\citenamefont {Carr}(1975)}]{Carr:1975qj}%
  \BibitemOpen
  \bibfield  {author} {\bibinfo {author} {\bibfnamefont {Bernard~J.}\ \bibnamefont {Carr}},\ }\bibfield  {title} {\enquote {\bibinfo {title} {{The Primordial black hole mass spectrum}},}\ }\href {\doibase 10.1086/153853} {\bibfield  {journal} {\bibinfo  {journal} {Astrophys. J.}\ }\textbf {\bibinfo {volume} {201}},\ \bibinfo {pages} {1--19} (\bibinfo {year} {1975})}\BibitemShut {NoStop}%
\bibitem [{\citenamefont {Kuhnel}\ \emph {et~al.}(2016)\citenamefont {Kuhnel}, \citenamefont {Rampf},\ and\ \citenamefont {Sandstad}}]{Kuhnel:2015vtw}%
  \BibitemOpen
  \bibfield  {author} {\bibinfo {author} {\bibfnamefont {Florian}\ \bibnamefont {Kuhnel}}, \bibinfo {author} {\bibfnamefont {Cornelius}\ \bibnamefont {Rampf}}, \ and\ \bibinfo {author} {\bibfnamefont {Marit}\ \bibnamefont {Sandstad}},\ }\bibfield  {title} {\enquote {\bibinfo {title} {{Effects of Critical Collapse on Primordial Black-Hole Mass Spectra}},}\ }\href {\doibase 10.1140/epjc/s10052-016-3945-8} {\bibfield  {journal} {\bibinfo  {journal} {Eur. Phys. J. C}\ }\textbf {\bibinfo {volume} {76}},\ \bibinfo {pages} {93} (\bibinfo {year} {2016})},\ \Eprint {http://arxiv.org/abs/1512.00488} {arXiv:1512.00488 [astro-ph.CO]} \BibitemShut {NoStop}%
\bibitem [{\citenamefont {Koivu}\ \emph {et~al.}(2025)\citenamefont {Koivu}, \citenamefont {Gnedin},\ and\ \citenamefont {Hirata}}]{Koivu:2025add}%
  \BibitemOpen
  \bibfield  {author} {\bibinfo {author} {\bibfnamefont {Emily}\ \bibnamefont {Koivu}}, \bibinfo {author} {\bibfnamefont {Nickolay~Y.}\ \bibnamefont {Gnedin}}, \ and\ \bibinfo {author} {\bibfnamefont {Christopher~M.}\ \bibnamefont {Hirata}},\ }\bibfield  {title} {\enquote {\bibinfo {title} {{Effects of Primordial Black Holes on IGM History}},}\ }\href@noop {} {\  (\bibinfo {year} {2025})},\ \Eprint {http://arxiv.org/abs/2510.00246} {arXiv:2510.00246 [astro-ph.CO]} \BibitemShut {NoStop}%
\bibitem [{\citenamefont {Khachi}\ \emph {et~al.}(2023)\citenamefont {Khachi}, \citenamefont {Kumar}, \citenamefont {Kumar},\ and\ \citenamefont {Sastri}}]{Khachi:2023iqp}%
  \BibitemOpen
  \bibfield  {author} {\bibinfo {author} {\bibfnamefont {Anil}\ \bibnamefont {Khachi}}, \bibinfo {author} {\bibfnamefont {Lalit}\ \bibnamefont {Kumar}}, \bibinfo {author} {\bibfnamefont {M.~R.~Ganesh}\ \bibnamefont {Kumar}}, \ and\ \bibinfo {author} {\bibfnamefont {O.~S. K.~S.}\ \bibnamefont {Sastri}},\ }\bibfield  {title} {\enquote {\bibinfo {title} {{Deuteron structure and form factors: Using an inverse potential approach}},}\ }\href {\doibase 10.1103/PhysRevC.107.064002} {\bibfield  {journal} {\bibinfo  {journal} {Phys. Rev. C}\ }\textbf {\bibinfo {volume} {107}},\ \bibinfo {pages} {064002} (\bibinfo {year} {2023})},\ \Eprint {http://arxiv.org/abs/2209.03575} {arXiv:2209.03575 [nucl-th]} \BibitemShut {NoStop}%
\bibitem [{\citenamefont {Simon}\ \emph {et~al.}(1981)\citenamefont {Simon}, \citenamefont {Schmitt},\ and\ \citenamefont {Walther}}]{Simon:1981br}%
  \BibitemOpen
  \bibfield  {author} {\bibinfo {author} {\bibfnamefont {G.~G.}\ \bibnamefont {Simon}}, \bibinfo {author} {\bibfnamefont {C.}~\bibnamefont {Schmitt}}, \ and\ \bibinfo {author} {\bibfnamefont {V.~H.}\ \bibnamefont {Walther}},\ }\bibfield  {title} {\enquote {\bibinfo {title} {{Elastic Electric and Magnetic $e D$ Scattering at Low Momentum Transfer}},}\ }\href {\doibase 10.1016/0375-9474(81)90572-8} {\bibfield  {journal} {\bibinfo  {journal} {Nucl. Phys. A}\ }\textbf {\bibinfo {volume} {364}},\ \bibinfo {pages} {285--296} (\bibinfo {year} {1981})}\BibitemShut {NoStop}%
\bibitem [{\citenamefont {Abbott}\ \emph {et~al.}(2000)\citenamefont {Abbott} \emph {et~al.}}]{JLABt20:2000qyq}%
  \BibitemOpen
  \bibfield  {author} {\bibinfo {author} {\bibfnamefont {D.}~\bibnamefont {Abbott}} \emph {et~al.} (\bibinfo {collaboration} {JLAB t20}),\ }\bibfield  {title} {\enquote {\bibinfo {title} {{Phenomenology of the deuteron electromagnetic form-factors}},}\ }\href {\doibase 10.1007/PL00013629} {\bibfield  {journal} {\bibinfo  {journal} {Eur. Phys. J. A}\ }\textbf {\bibinfo {volume} {7}},\ \bibinfo {pages} {421--427} (\bibinfo {year} {2000})},\ \Eprint {http://arxiv.org/abs/nucl-ex/0002003} {arXiv:nucl-ex/0002003} \BibitemShut {NoStop}%
\bibitem [{\citenamefont {Klein}\ and\ \citenamefont {Vogt}(2003)}]{Klein:2003bz}%
  \BibitemOpen
  \bibfield  {author} {\bibinfo {author} {\bibfnamefont {Spencer}\ \bibnamefont {Klein}}\ and\ \bibinfo {author} {\bibfnamefont {Ramona}\ \bibnamefont {Vogt}},\ }\bibfield  {title} {\enquote {\bibinfo {title} {{Deuteron photodissociation in ultraperipheral relativistic heavy ion on deuteron collisions}},}\ }\href {\doibase 10.1103/PhysRevC.68.017902} {\bibfield  {journal} {\bibinfo  {journal} {Phys. Rev. C}\ }\textbf {\bibinfo {volume} {68}},\ \bibinfo {pages} {017902} (\bibinfo {year} {2003})},\ \Eprint {http://arxiv.org/abs/nucl-ex/0303013} {arXiv:nucl-ex/0303013} \BibitemShut {NoStop}%
\bibitem [{\citenamefont {Belz}\ \emph {et~al.}(1995)\citenamefont {Belz} \emph {et~al.}}]{Belz:1995ge}%
  \BibitemOpen
  \bibfield  {author} {\bibinfo {author} {\bibfnamefont {J.~E.}\ \bibnamefont {Belz}} \emph {et~al.},\ }\bibfield  {title} {\enquote {\bibinfo {title} {{Two body photodisintegration of the deuteron up to 2.8-GeV}},}\ }\href {\doibase 10.1103/PhysRevLett.74.646} {\bibfield  {journal} {\bibinfo  {journal} {Phys. Rev. Lett.}\ }\textbf {\bibinfo {volume} {74}},\ \bibinfo {pages} {646--649} (\bibinfo {year} {1995})}\BibitemShut {NoStop}%
\bibitem [{\citenamefont {Schulte}\ \emph {et~al.}(2001)\citenamefont {Schulte} \emph {et~al.}}]{Schulte:2001se}%
  \BibitemOpen
  \bibfield  {author} {\bibinfo {author} {\bibfnamefont {E.~C.}\ \bibnamefont {Schulte}} \emph {et~al.},\ }\bibfield  {title} {\enquote {\bibinfo {title} {{Measurement of the high energy two-body deuteron photodisintegration differential cross-section}},}\ }\href {\doibase 10.1103/PhysRevLett.87.102302} {\bibfield  {journal} {\bibinfo  {journal} {Phys. Rev. Lett.}\ }\textbf {\bibinfo {volume} {87}},\ \bibinfo {pages} {102302} (\bibinfo {year} {2001})}\BibitemShut {NoStop}%
\bibitem [{\citenamefont {Carr}\ \emph {et~al.}(2010)\citenamefont {Carr}, \citenamefont {Kohri}, \citenamefont {Sendouda},\ and\ \citenamefont {Yokoyama}}]{carrNewCosmologicalConstraints2010b}%
  \BibitemOpen
  \bibfield  {author} {\bibinfo {author} {\bibfnamefont {B.~J.}\ \bibnamefont {Carr}}, \bibinfo {author} {\bibfnamefont {Kazunori}\ \bibnamefont {Kohri}}, \bibinfo {author} {\bibfnamefont {Yuuiti}\ \bibnamefont {Sendouda}}, \ and\ \bibinfo {author} {\bibfnamefont {Jun'ichi}\ \bibnamefont {Yokoyama}},\ }\bibfield  {title} {\enquote {\bibinfo {title} {New cosmological constraints on primordial black holes},}\ }\href {\doibase 10.1103/PhysRevD.81.104019} {\bibfield  {journal} {\bibinfo  {journal} {Phys. Rev. D}\ }\textbf {\bibinfo {volume} {81}},\ \bibinfo {pages} {104019} (\bibinfo {year} {2010})},\ \Eprint {http://arxiv.org/abs/0912.5297} {arXiv:0912.5297 [astro-ph]} \BibitemShut {NoStop}%
\bibitem [{\citenamefont {Wang}\ \emph {et~al.}(2025)\citenamefont {Wang}, \citenamefont {Grohs},\ and\ \citenamefont {Mersini-Houghton}}]{Wang:2025pum}%
  \BibitemOpen
  \bibfield  {author} {\bibinfo {author} {\bibfnamefont {Tianning}\ \bibnamefont {Wang}}, \bibinfo {author} {\bibfnamefont {Evan}\ \bibnamefont {Grohs}}, \ and\ \bibinfo {author} {\bibfnamefont {Laura}\ \bibnamefont {Mersini-Houghton}},\ }\bibfield  {title} {\enquote {\bibinfo {title} {{How Primordial Black Holes Change BBN}},}\ }\href@noop {} {\  (\bibinfo {year} {2025})},\ \Eprint {http://arxiv.org/abs/2511.18646} {arXiv:2511.18646 [astro-ph.CO]} \BibitemShut {NoStop}%
\bibitem [{\citenamefont {Bohr}\ and\ \citenamefont {Wheeler}(1939)}]{bohrMechanismNuclearFission1939}%
  \BibitemOpen
  \bibfield  {author} {\bibinfo {author} {\bibfnamefont {Niels}\ \bibnamefont {Bohr}}\ and\ \bibinfo {author} {\bibfnamefont {John~Archibald}\ \bibnamefont {Wheeler}},\ }\bibfield  {title} {\enquote {\bibinfo {title} {The {{Mechanism}} of {{Nuclear Fission}}},}\ }\href {\doibase 10.1103/PhysRev.56.426} {\bibfield  {journal} {\bibinfo  {journal} {Phys. Rev.}\ }\textbf {\bibinfo {volume} {56}},\ \bibinfo {pages} {426--450} (\bibinfo {year} {1939})}\BibitemShut {NoStop}%
\bibitem [{\citenamefont {Plesset}(1941)}]{plessetClassicalModelNuclear1941}%
  \BibitemOpen
  \bibfield  {author} {\bibinfo {author} {\bibfnamefont {M.~S.}\ \bibnamefont {Plesset}},\ }\bibfield  {title} {\enquote {\bibinfo {title} {On the {{Classical Model}} of {{Nuclear Fission}}},}\ }\href {\doibase 10.1119/1.1991623} {\bibfield  {journal} {\bibinfo  {journal} {Am. J. Phys.}\ }\textbf {\bibinfo {volume} {9}},\ \bibinfo {pages} {1--10} (\bibinfo {year} {1941})}\BibitemShut {NoStop}%
\bibitem [{\citenamefont {Strutinsky}\ \emph {et~al.}(1963)\citenamefont {Strutinsky}, \citenamefont {Lyashchenko},\ and\ \citenamefont {Popov}}]{strutinskySymmetricalShapesEquilibrium1963}%
  \BibitemOpen
  \bibfield  {author} {\bibinfo {author} {\bibfnamefont {V.~M.}\ \bibnamefont {Strutinsky}}, \bibinfo {author} {\bibfnamefont {N.~{\relax Ya}.}\ \bibnamefont {Lyashchenko}}, \ and\ \bibinfo {author} {\bibfnamefont {N.~A.}\ \bibnamefont {Popov}},\ }\bibfield  {title} {\enquote {\bibinfo {title} {Symmetrical shapes of equilibrium for a liquid drop model},}\ }\href {\doibase 10.1016/0029-5582(63)90635-7} {\bibfield  {journal} {\bibinfo  {journal} {Nuclear Physics}\ }\textbf {\bibinfo {volume} {46}},\ \bibinfo {pages} {639--659} (\bibinfo {year} {1963})}\BibitemShut {NoStop}%
\bibitem [{\citenamefont {Myers}\ and\ \citenamefont {Swiatecki}(1966)}]{myersNuclearMassesDeformations1966}%
  \BibitemOpen
  \bibfield  {author} {\bibinfo {author} {\bibfnamefont {William~D.}\ \bibnamefont {Myers}}\ and\ \bibinfo {author} {\bibfnamefont {Wladyslaw~J.}\ \bibnamefont {Swiatecki}},\ }\bibfield  {title} {\enquote {\bibinfo {title} {Nuclear masses and deformations},}\ }\href {\doibase 10.1016/0029-5582(66)90639-0} {\bibfield  {journal} {\bibinfo  {journal} {Nuclear Physics}\ }\textbf {\bibinfo {volume} {81}},\ \bibinfo {pages} {1--60} (\bibinfo {year} {1966})}\BibitemShut {NoStop}%
\bibitem [{\citenamefont {Bj{\o}rnholm}\ and\ \citenamefont {Lynn}(1980)}]{bjornholmDoublehumpedFissionBarrier1980}%
  \BibitemOpen
  \bibfield  {author} {\bibinfo {author} {\bibfnamefont {S.}~\bibnamefont {Bj{\o}rnholm}}\ and\ \bibinfo {author} {\bibfnamefont {J.~E.}\ \bibnamefont {Lynn}},\ }\bibfield  {title} {\enquote {\bibinfo {title} {The double-humped fission barrier},}\ }\href {\doibase 10.1103/RevModPhys.52.725} {\bibfield  {journal} {\bibinfo  {journal} {Rev. Mod. Phys.}\ }\textbf {\bibinfo {volume} {52}},\ \bibinfo {pages} {725--931} (\bibinfo {year} {1980})}\BibitemShut {NoStop}%
\bibitem [{\citenamefont {Ivanyuk}\ and\ \citenamefont {Pomorski}(2009)}]{ivanyukOptimalShapesFission2009}%
  \BibitemOpen
  \bibfield  {author} {\bibinfo {author} {\bibfnamefont {F.~A.}\ \bibnamefont {Ivanyuk}}\ and\ \bibinfo {author} {\bibfnamefont {K.}~\bibnamefont {Pomorski}},\ }\bibfield  {title} {\enquote {\bibinfo {title} {Optimal shapes and fission barriers of nuclei within the liquid drop model},}\ }\href {\doibase 10.1103/PhysRevC.79.054327} {\bibfield  {journal} {\bibinfo  {journal} {Phys. Rev. C}\ }\textbf {\bibinfo {volume} {79}},\ \bibinfo {pages} {054327} (\bibinfo {year} {2009})}\BibitemShut {NoStop}%
\bibitem [{\citenamefont {Kowal}\ and\ \citenamefont {Skalski}(2023)}]{Kowal2023}%
  \BibitemOpen
  \bibfield  {author} {\bibinfo {author} {\bibfnamefont {Micha{\l}}\ \bibnamefont {Kowal}}\ and\ \bibinfo {author} {\bibfnamefont {Janusz}\ \bibnamefont {Skalski}},\ }\bibfield  {title} {\enquote {\bibinfo {title} {The fission barrier of heaviest nuclei from a macroscopic-microscopic perspective},}\ }in\ \href@noop {} {\emph {\bibinfo {booktitle} {Handbook of Nuclear Physics}}},\ \bibinfo {editor} {edited by\ \bibinfo {editor} {\bibfnamefont {Isao}\ \bibnamefont {Tanihata}}, \bibinfo {editor} {\bibfnamefont {Hiroshi}\ \bibnamefont {Toki}}, \ and\ \bibinfo {editor} {\bibfnamefont {Toshitaka}\ \bibnamefont {Kajino}}}\ (\bibinfo  {publisher} {Springer Nature Singapore},\ \bibinfo {address} {Singapore},\ \bibinfo {year} {2023})\ pp.\ \bibinfo {pages} {945--982}\BibitemShut {NoStop}%
\bibitem [{\citenamefont {Walker}\ and\ \citenamefont {Podoly{\'a}k}(2023)}]{walkerNuclearIsomers2023}%
  \BibitemOpen
  \bibfield  {author} {\bibinfo {author} {\bibfnamefont {Philip~M.}\ \bibnamefont {Walker}}\ and\ \bibinfo {author} {\bibfnamefont {Zsolt}\ \bibnamefont {Podoly{\'a}k}},\ }\bibfield  {title} {\enquote {\bibinfo {title} {Nuclear {{Isomers}}},}\ }in\ \href@noop {} {\emph {\bibinfo {booktitle} {Handbook of {{Nuclear Physics}}}}}\ (\bibinfo  {publisher} {Springer, Singapore},\ \bibinfo {year} {2023})\ pp.\ \bibinfo {pages} {487--523}\BibitemShut {NoStop}%
\bibitem [{\citenamefont {Present}\ \emph {et~al.}(1991)\citenamefont {Present}, \citenamefont {Reines},\ and\ \citenamefont {Knipp}}]{presentLiquidDropModel1991}%
  \BibitemOpen
  \bibfield  {author} {\bibinfo {author} {\bibfnamefont {R.~D.}\ \bibnamefont {Present}}, \bibinfo {author} {\bibfnamefont {F.}~\bibnamefont {Reines}}, \ and\ \bibinfo {author} {\bibfnamefont {J.~K.}\ \bibnamefont {Knipp}},\ }\bibfield  {title} {\enquote {\bibinfo {title} {The {{Liquid Drop Model}} for {{Nuclear Fission}}},}\ }in\ \href@noop {} {\emph {\bibinfo {booktitle} {Neutrinos and {{Other Matters}}}}}\ (\bibinfo  {publisher} {World Scientific},\ \bibinfo {address} {Signapore},\ \bibinfo {year} {1991})\ pp.\ \bibinfo {pages} {461--461}\BibitemShut {NoStop}%
\bibitem [{\citenamefont {Bertulani}(2007)}]{Bertulani2007}%
  \BibitemOpen
  \bibfield  {author} {\bibinfo {author} {\bibfnamefont {Carlos~A.}\ \bibnamefont {Bertulani}},\ }\href@noop {} {\emph {\bibinfo {title} {Nuclear Physics in a Nutshell}}}\ (\bibinfo  {publisher} {Princeton University Press},\ \bibinfo {year} {2007})\BibitemShut {NoStop}%
\bibitem [{\citenamefont {{Wong}}(1998)}]{WongBook}%
  \BibitemOpen
  \bibfield  {author} {\bibinfo {author} {\bibfnamefont {Samuel S.~M.}\ \bibnamefont {{Wong}}},\ }\href@noop {} {\emph {\bibinfo {title} {{Introductory Nuclear Physics, 2nd Edition}}}}\ (\bibinfo  {publisher} {{Wiley-VCH}},\ \bibinfo {year} {1998})\BibitemShut {NoStop}%
\bibitem [{\citenamefont {Oberstedt}\ and\ \citenamefont {Oberstedt}(2021)}]{oberstedtExploringFissionBarrier2021a}%
  \BibitemOpen
  \bibfield  {author} {\bibinfo {author} {\bibfnamefont {A.}~\bibnamefont {Oberstedt}}\ and\ \bibinfo {author} {\bibfnamefont {S.}~\bibnamefont {Oberstedt}},\ }\bibfield  {title} {\enquote {\bibinfo {title} {{Exploring the Fission Barrier of $^{235}{\rm U}$}},}\ }\href {\doibase 10.1103/PhysRevC.104.024611} {\bibfield  {journal} {\bibinfo  {journal} {Phys. Rev. C}\ }\textbf {\bibinfo {volume} {104}},\ \bibinfo {pages} {024611} (\bibinfo {year} {2021})}\BibitemShut {NoStop}%
\bibitem [{\citenamefont {Brack}\ \emph {et~al.}(1972)\citenamefont {Brack}, \citenamefont {Damgaard}, \citenamefont {Jensen}, \citenamefont {Pauli}, \citenamefont {Strutinsky},\ and\ \citenamefont {Wong}}]{brackFunnyHillsShellCorrection1972}%
  \BibitemOpen
  \bibfield  {author} {\bibinfo {author} {\bibfnamefont {M.}~\bibnamefont {Brack}}, \bibinfo {author} {\bibfnamefont {Jens}\ \bibnamefont {Damgaard}}, \bibinfo {author} {\bibfnamefont {A.~S.}\ \bibnamefont {Jensen}}, \bibinfo {author} {\bibfnamefont {H.~C.}\ \bibnamefont {Pauli}}, \bibinfo {author} {\bibfnamefont {V.~M.}\ \bibnamefont {Strutinsky}}, \ and\ \bibinfo {author} {\bibfnamefont {C.~Y.}\ \bibnamefont {Wong}},\ }\bibfield  {title} {\enquote {\bibinfo {title} {Funny {{Hills}}: {{The Shell-Correction Approach}} to {{Nuclear Shell Effects}} and {{Its Applications}} to the {{Fission Process}}},}\ }\href {\doibase 10.1103/RevModPhys.44.320} {\bibfield  {journal} {\bibinfo  {journal} {Rev. Mod. Phys.}\ }\textbf {\bibinfo {volume} {44}},\ \bibinfo {pages} {320--405} (\bibinfo {year} {1972})}\BibitemShut {NoStop}%
\bibitem [{\citenamefont {Lynn}\ and\ \citenamefont {Hayes}(2003)}]{lynnTheoreticalEvaluationsFission2003}%
  \BibitemOpen
  \bibfield  {author} {\bibinfo {author} {\bibfnamefont {J.~Eric}\ \bibnamefont {Lynn}}\ and\ \bibinfo {author} {\bibfnamefont {A.~C.}\ \bibnamefont {Hayes}},\ }\bibfield  {title} {\enquote {\bibinfo {title} {{Theoretical Evaluations of the Fission Cross Section of the 77 {{eV}} Isomer of $^{235}{\rm U}$}},}\ }\href {\doibase 10.1103/PhysRevC.67.014607} {\bibfield  {journal} {\bibinfo  {journal} {Phys. Rev. C}\ }\textbf {\bibinfo {volume} {67}},\ \bibinfo {pages} {014607} (\bibinfo {year} {2003})}\BibitemShut {NoStop}%
\bibitem [{\citenamefont {Wang}\ \emph {et~al.}(2019)\citenamefont {Wang}, \citenamefont {Zhu}, \citenamefont {Zhong},\ and\ \citenamefont {Fan}}]{wangNewCalculationsFivedimensional2019}%
  \BibitemOpen
  \bibfield  {author} {\bibinfo {author} {\bibfnamefont {Zhiming}\ \bibnamefont {Wang}}, \bibinfo {author} {\bibfnamefont {Wenjie}\ \bibnamefont {Zhu}}, \bibinfo {author} {\bibfnamefont {Chunlai}\ \bibnamefont {Zhong}}, \ and\ \bibinfo {author} {\bibfnamefont {Tieshuan}\ \bibnamefont {Fan}},\ }\bibfield  {title} {\enquote {\bibinfo {title} {New calculations of five-dimensional fission barriers for actinide nuclei},}\ }\href {\doibase 10.1016/j.nuclphysa.2019.05.014} {\bibfield  {journal} {\bibinfo  {journal} {Nuclear Physics A}\ }\textbf {\bibinfo {volume} {989}},\ \bibinfo {pages} {81--96} (\bibinfo {year} {2019})}\BibitemShut {NoStop}%
\bibitem [{\citenamefont {Csige}\ \emph {et~al.}(2013)\citenamefont {Csige}, \citenamefont {Filipescu}, \citenamefont {Glodariu},\ and\ \citenamefont {{et al}}}]{csigeExploringMultihumpedFission2013}%
  \BibitemOpen
  \bibfield  {author} {\bibinfo {author} {\bibfnamefont {L.}~\bibnamefont {Csige}}, \bibinfo {author} {\bibfnamefont {D.~M.}\ \bibnamefont {Filipescu}}, \bibinfo {author} {\bibfnamefont {T.}~\bibnamefont {Glodariu}}, \ and\ \bibinfo {author} {\bibnamefont {{et al}}},\ }\bibfield  {title} {\enquote {\bibinfo {title} {{Exploring the Multihumped Fission Barrier of $^{238}{\rm U}$ via Sub-Barrier Photofission}},}\ }\href {\doibase 10.1103/PhysRevC.87.044321} {\bibfield  {journal} {\bibinfo  {journal} {Phys. Rev. C}\ }\textbf {\bibinfo {volume} {87}},\ \bibinfo {pages} {044321} (\bibinfo {year} {2013})}\BibitemShut {NoStop}%
\bibitem [{\citenamefont {Oberstedt}\ \emph {et~al.}(2007)\citenamefont {Oberstedt}, \citenamefont {Oberstedt}, \citenamefont {Gawrys},\ and\ \citenamefont {Kornilov}}]{oberstedtIdentificationShapeIsomer2007}%
  \BibitemOpen
  \bibfield  {author} {\bibinfo {author} {\bibfnamefont {A.}~\bibnamefont {Oberstedt}}, \bibinfo {author} {\bibfnamefont {S.}~\bibnamefont {Oberstedt}}, \bibinfo {author} {\bibfnamefont {M.}~\bibnamefont {Gawrys}}, \ and\ \bibinfo {author} {\bibfnamefont {N.}~\bibnamefont {Kornilov}},\ }\bibfield  {title} {\enquote {\bibinfo {title} {{Identification of a {{Shape Isomer}} in $^{235}{\rm U}$}},}\ }\href {\doibase 10.1103/PhysRevLett.99.042502} {\bibfield  {journal} {\bibinfo  {journal} {Phys. Rev. Lett.}\ }\textbf {\bibinfo {volume} {99}},\ \bibinfo {pages} {042502} (\bibinfo {year} {2007})}\BibitemShut {NoStop}%
\bibitem [{\citenamefont {{J.~Craig Wheeler}}({1986})}]{WheelerBook}%
  \BibitemOpen
  \bibfield  {author} {\bibinfo {author} {\bibnamefont {{J.~Craig Wheeler}}},\ }\href@noop {} {\emph {\bibinfo {title} {{The Krone Experiment}}}}\ (\bibinfo  {publisher} {{Pressworks}},\ \bibinfo {address} {{Dallas, TX}},\ \bibinfo {year} {{1986}})\BibitemShut {NoStop}%
\bibitem [{\citenamefont {Capela}\ \emph {et~al.}(2013{\natexlab{a}})\citenamefont {Capela}, \citenamefont {Pshirkov},\ and\ \citenamefont {Tinyakov}}]{Capela:2012jz}%
  \BibitemOpen
  \bibfield  {author} {\bibinfo {author} {\bibfnamefont {Fabio}\ \bibnamefont {Capela}}, \bibinfo {author} {\bibfnamefont {Maxim}\ \bibnamefont {Pshirkov}}, \ and\ \bibinfo {author} {\bibfnamefont {Peter}\ \bibnamefont {Tinyakov}},\ }\bibfield  {title} {\enquote {\bibinfo {title} {{Constraints on Primordial Black Holes as Dark Matter Candidates from Star Formation}},}\ }\href {\doibase 10.1103/PhysRevD.87.023507} {\bibfield  {journal} {\bibinfo  {journal} {Phys. Rev. D}\ }\textbf {\bibinfo {volume} {87}},\ \bibinfo {pages} {023507} (\bibinfo {year} {2013}{\natexlab{a}})},\ \Eprint {http://arxiv.org/abs/1209.6021} {arXiv:1209.6021 [astro-ph.CO]} \BibitemShut {NoStop}%
\bibitem [{\citenamefont {Capela}\ \emph {et~al.}(2013{\natexlab{b}})\citenamefont {Capela}, \citenamefont {Pshirkov},\ and\ \citenamefont {Tinyakov}}]{Capela:2013yf}%
  \BibitemOpen
  \bibfield  {author} {\bibinfo {author} {\bibfnamefont {Fabio}\ \bibnamefont {Capela}}, \bibinfo {author} {\bibfnamefont {Maxim}\ \bibnamefont {Pshirkov}}, \ and\ \bibinfo {author} {\bibfnamefont {Peter}\ \bibnamefont {Tinyakov}},\ }\bibfield  {title} {\enquote {\bibinfo {title} {{Constraints on primordial black holes as dark matter candidates from capture by neutron stars}},}\ }\href {\doibase 10.1103/PhysRevD.87.123524} {\bibfield  {journal} {\bibinfo  {journal} {Phys. Rev. D}\ }\textbf {\bibinfo {volume} {87}},\ \bibinfo {pages} {123524} (\bibinfo {year} {2013}{\natexlab{b}})},\ \Eprint {http://arxiv.org/abs/1301.4984} {arXiv:1301.4984 [astro-ph.CO]} \BibitemShut {NoStop}%
\bibitem [{\citenamefont {Lehmann}\ \emph {et~al.}(2021)\citenamefont {Lehmann}, \citenamefont {Ross}, \citenamefont {Webber},\ and\ \citenamefont {Profumo}}]{Lehmann:2020yxb}%
  \BibitemOpen
  \bibfield  {author} {\bibinfo {author} {\bibfnamefont {Benjamin~V.}\ \bibnamefont {Lehmann}}, \bibinfo {author} {\bibfnamefont {Olivia~G.}\ \bibnamefont {Ross}}, \bibinfo {author} {\bibfnamefont {Ava}\ \bibnamefont {Webber}}, \ and\ \bibinfo {author} {\bibfnamefont {Stefano}\ \bibnamefont {Profumo}},\ }\bibfield  {title} {\enquote {\bibinfo {title} {{Three-body capture, ejection, and the demographics of bound objects in binary systems}},}\ }\href {\doibase 10.1093/mnras/stab1121} {\bibfield  {journal} {\bibinfo  {journal} {Mon. Not. Roy. Astron. Soc.}\ }\textbf {\bibinfo {volume} {505}},\ \bibinfo {pages} {1017--1028} (\bibinfo {year} {2021})},\ \Eprint {http://arxiv.org/abs/2012.05875} {arXiv:2012.05875 [astro-ph.SR]} \BibitemShut {NoStop}%
\bibitem [{\citenamefont {G{\'e}nolini}\ \emph {et~al.}(2020)\citenamefont {G{\'e}nolini}, \citenamefont {Serpico},\ and\ \citenamefont {Tinyakov}}]{Genolini:2020ejw}%
  \BibitemOpen
  \bibfield  {author} {\bibinfo {author} {\bibfnamefont {Yoann}\ \bibnamefont {G{\'e}nolini}}, \bibinfo {author} {\bibfnamefont {Pasquale}\ \bibnamefont {Serpico}}, \ and\ \bibinfo {author} {\bibfnamefont {Peter}\ \bibnamefont {Tinyakov}},\ }\bibfield  {title} {\enquote {\bibinfo {title} {{Revisiting primordial black hole capture into neutron stars}},}\ }\href {\doibase 10.1103/PhysRevD.102.083004} {\bibfield  {journal} {\bibinfo  {journal} {Phys. Rev. D}\ }\textbf {\bibinfo {volume} {102}},\ \bibinfo {pages} {083004} (\bibinfo {year} {2020})},\ \Eprint {http://arxiv.org/abs/2006.16975} {arXiv:2006.16975 [astro-ph.HE]} \BibitemShut {NoStop}%
\bibitem [{\citenamefont {Lehmann}\ \emph {et~al.}(2022)\citenamefont {Lehmann}, \citenamefont {Webber}, \citenamefont {Ross},\ and\ \citenamefont {Profumo}}]{Lehmann:2022vdt}%
  \BibitemOpen
  \bibfield  {author} {\bibinfo {author} {\bibfnamefont {Benjamin~V.}\ \bibnamefont {Lehmann}}, \bibinfo {author} {\bibfnamefont {Ava}\ \bibnamefont {Webber}}, \bibinfo {author} {\bibfnamefont {Olivia~G.}\ \bibnamefont {Ross}}, \ and\ \bibinfo {author} {\bibfnamefont {Stefano}\ \bibnamefont {Profumo}},\ }\bibfield  {title} {\enquote {\bibinfo {title} {{Capture of primordial black holes in extrasolar systems}},}\ }\href {\doibase 10.1088/1475-7516/2022/08/079} {\bibfield  {journal} {\bibinfo  {journal} {JCAP}\ }\textbf {\bibinfo {volume} {08}},\ \bibinfo {pages} {079} (\bibinfo {year} {2022})},\ \Eprint {http://arxiv.org/abs/2205.09756} {arXiv:2205.09756 [astro-ph.EP]} \BibitemShut {NoStop}%
\bibitem [{\citenamefont {Esser}\ and\ \citenamefont {Tinyakov}(2023)}]{Esser:2022owk}%
  \BibitemOpen
  \bibfield  {author} {\bibinfo {author} {\bibfnamefont {Nicolas}\ \bibnamefont {Esser}}\ and\ \bibinfo {author} {\bibfnamefont {Peter}\ \bibnamefont {Tinyakov}},\ }\bibfield  {title} {\enquote {\bibinfo {title} {{Constraints on primordial black holes from observation of stars in dwarf galaxies}},}\ }\href {\doibase 10.1103/PhysRevD.107.103052} {\bibfield  {journal} {\bibinfo  {journal} {Phys. Rev. D}\ }\textbf {\bibinfo {volume} {107}},\ \bibinfo {pages} {103052} (\bibinfo {year} {2023})},\ \Eprint {http://arxiv.org/abs/2207.07412} {arXiv:2207.07412 [astro-ph.HE]} \BibitemShut {NoStop}%
\bibitem [{\citenamefont {Caplan}\ \emph {et~al.}(2024)\citenamefont {Caplan}, \citenamefont {Bellinger},\ and\ \citenamefont {Santarelli}}]{Caplan:2023ddo}%
  \BibitemOpen
  \bibfield  {author} {\bibinfo {author} {\bibfnamefont {Matthew~E.}\ \bibnamefont {Caplan}}, \bibinfo {author} {\bibfnamefont {Earl~P.}\ \bibnamefont {Bellinger}}, \ and\ \bibinfo {author} {\bibfnamefont {Andrew~D.}\ \bibnamefont {Santarelli}},\ }\bibfield  {title} {\enquote {\bibinfo {title} {{Is there a black hole in the center of the Sun?}}}\ }\href {\doibase 10.1007/s10509-024-04270-1} {\bibfield  {journal} {\bibinfo  {journal} {Astrophys. Space Sci.}\ }\textbf {\bibinfo {volume} {369}},\ \bibinfo {pages} {8} (\bibinfo {year} {2024})},\ \Eprint {http://arxiv.org/abs/2312.07647} {arXiv:2312.07647 [astro-ph.SR]} \BibitemShut {NoStop}%
\bibitem [{\citenamefont {Santarelli}\ \emph {et~al.}(2024)\citenamefont {Santarelli}, \citenamefont {Caplan},\ and\ \citenamefont {Bellinger}}]{Santarelli:2024uqx}%
  \BibitemOpen
  \bibfield  {author} {\bibinfo {author} {\bibfnamefont {Andrew~D.}\ \bibnamefont {Santarelli}}, \bibinfo {author} {\bibfnamefont {Matthew~E.}\ \bibnamefont {Caplan}}, \ and\ \bibinfo {author} {\bibfnamefont {Earl~P.}\ \bibnamefont {Bellinger}},\ }\bibfield  {title} {\enquote {\bibinfo {title} {{Formation of Sub-Chandrasekhar-mass Black Holes and Red Stragglers via Hawking Stars in Ultrafaint Dwarf Galaxies}},}\ }\href {\doibase 10.3847/1538-4357/ad8ec0} {\bibfield  {journal} {\bibinfo  {journal} {Astrophys. J.}\ }\textbf {\bibinfo {volume} {977}},\ \bibinfo {pages} {145} (\bibinfo {year} {2024})},\ \Eprint {http://arxiv.org/abs/2406.17052} {arXiv:2406.17052 [astro-ph.GA]} \BibitemShut {NoStop}%
\bibitem [{\citenamefont {Baumgarte}\ and\ \citenamefont {Shapiro}(2024{\natexlab{a}})}]{Baumgarte:2024mei}%
  \BibitemOpen
  \bibfield  {author} {\bibinfo {author} {\bibfnamefont {Thomas~W.}\ \bibnamefont {Baumgarte}}\ and\ \bibinfo {author} {\bibfnamefont {Stuart~L.}\ \bibnamefont {Shapiro}},\ }\bibfield  {title} {\enquote {\bibinfo {title} {{Primordial black holes, gravitational wave beats, and the nuclear equation of state}},}\ }\href {\doibase 10.1103/PhysRevD.110.L021303} {\bibfield  {journal} {\bibinfo  {journal} {Phys. Rev. D}\ }\textbf {\bibinfo {volume} {110}},\ \bibinfo {pages} {L021303} (\bibinfo {year} {2024}{\natexlab{a}})},\ \Eprint {http://arxiv.org/abs/2402.01838} {arXiv:2402.01838 [gr-qc]} \BibitemShut {NoStop}%
\bibitem [{\citenamefont {Baumgarte}\ and\ \citenamefont {Shapiro}(2024{\natexlab{b}})}]{Baumgarte:2024buu}%
  \BibitemOpen
  \bibfield  {author} {\bibinfo {author} {\bibfnamefont {Thomas~W.}\ \bibnamefont {Baumgarte}}\ and\ \bibinfo {author} {\bibfnamefont {Stuart~L.}\ \bibnamefont {Shapiro}},\ }\bibfield  {title} {\enquote {\bibinfo {title} {{Could long-period transients be powered by primordial black hole capture?}}}\ }\href {\doibase 10.1103/PhysRevD.109.063004} {\bibfield  {journal} {\bibinfo  {journal} {Phys. Rev. D}\ }\textbf {\bibinfo {volume} {109}},\ \bibinfo {pages} {063004} (\bibinfo {year} {2024}{\natexlab{b}})},\ \Eprint {http://arxiv.org/abs/2402.11019} {arXiv:2402.11019 [astro-ph.HE]} \BibitemShut {NoStop}%
\bibitem [{\citenamefont {Baumgarte}\ and\ \citenamefont {Shapiro}(2024{\natexlab{c}})}]{Baumgarte:2024iby}%
  \BibitemOpen
  \bibfield  {author} {\bibinfo {author} {\bibfnamefont {Thomas~W.}\ \bibnamefont {Baumgarte}}\ and\ \bibinfo {author} {\bibfnamefont {Stuart~L.}\ \bibnamefont {Shapiro}},\ }\bibfield  {title} {\enquote {\bibinfo {title} {{Primordial black holes captured by neutron stars: Relativistic point-mass treatment}},}\ }\href {\doibase 10.1103/PhysRevD.109.123012} {\bibfield  {journal} {\bibinfo  {journal} {Phys. Rev. D}\ }\textbf {\bibinfo {volume} {109}},\ \bibinfo {pages} {123012} (\bibinfo {year} {2024}{\natexlab{c}})},\ \Eprint {http://arxiv.org/abs/2404.08735} {arXiv:2404.08735 [gr-qc]} \BibitemShut {NoStop}%
\bibitem [{\citenamefont {De~Lorenci}\ \emph {et~al.}(2025{\natexlab{b}})\citenamefont {De~Lorenci}, \citenamefont {Kaiser},\ and\ \citenamefont {Peter}}]{DeLorenci:2024xez}%
  \BibitemOpen
  \bibfield  {author} {\bibinfo {author} {\bibfnamefont {Vitorio~A.}\ \bibnamefont {De~Lorenci}}, \bibinfo {author} {\bibfnamefont {David~I.}\ \bibnamefont {Kaiser}}, \ and\ \bibinfo {author} {\bibfnamefont {Patrick}\ \bibnamefont {Peter}},\ }\bibfield  {title} {\enquote {\bibinfo {title} {{Orbital motion of primordial black holes crossing Sun-like stars}},}\ }\href {\doibase 10.1119/5.0256145} {\bibfield  {journal} {\bibinfo  {journal} {Am. J. Phys.}\ }\textbf {\bibinfo {volume} {93}},\ \bibinfo {pages} {943--950} (\bibinfo {year} {2025}{\natexlab{b}})},\ \Eprint {http://arxiv.org/abs/2405.08113} {arXiv:2405.08113 [astro-ph.CO]} \BibitemShut {NoStop}%
\bibitem [{\citenamefont {Horowitz}\ and\ \citenamefont {Caplan}(2021)}]{Horowitz:2021nlr}%
  \BibitemOpen
  \bibfield  {author} {\bibinfo {author} {\bibfnamefont {C.~J.}\ \bibnamefont {Horowitz}}\ and\ \bibinfo {author} {\bibfnamefont {M.~E.}\ \bibnamefont {Caplan}},\ }\bibfield  {title} {\enquote {\bibinfo {title} {{Actinide crystallization and fission reactions in cooling white dwarf stars}},}\ }\href {\doibase 10.1103/PhysRevLett.126.131101} {\bibfield  {journal} {\bibinfo  {journal} {Phys. Rev. Lett.}\ }\textbf {\bibinfo {volume} {126}},\ \bibinfo {pages} {131101} (\bibinfo {year} {2021})},\ \Eprint {http://arxiv.org/abs/2103.02122} {arXiv:2103.02122 [astro-ph.SR]} \BibitemShut {NoStop}%
\end{thebibliography}

%

\end{document}